\documentclass[aps,preprint,nofootinbib,hyperref]{revtex4}%
\usepackage{amsfonts}
\usepackage{amsmath}
\usepackage{amssymb}
\usepackage{graphicx}%
\providecommand{\U}[1]{\protect\rule{.1in}{.1in}}
\providecommand{\U}[1]{\protect\rule{.1in}{.1in}}
\providecommand{\U}[1]{\protect\rule{.1in}{.1in}}
\providecommand{\U}[1]{\protect\rule{.1in}{.1in}}
\providecommand{\U}[1]{\protect\rule{.1in}{.1in}}

\begin{document}
\title[Dualities 2T-Physics]{Generalized Dualities in 1T-Physics as \\Holographic Predictions from 2T-Physics}
\author{Ignacio J. Araya and Itzhak Bars}
\affiliation{Department of Physics and Astronomy, University of Southern California, Los
Angeles, CA 90089-0484}

\begin{abstract}
In the conventional formalism of physics, with 1-time, systems with different
Hamiltonians or Lagrangians have different physical interpretations and are
considered to be independent systems unrelated to each other. However, in this
paper we construct explicitly canonical maps in 1T phase space (including
timelike components, specifically the Hamiltonian) to show that it is
appropriate to regard various 1T-physics systems, with different Lagrangians
or Hamiltonians, as being duals of each other. This concept is similar in
spirit to dualities discovered in more complicated examples in field theory or
string theory. Our approach makes it evident that such generalized dualities
are widespread. This suggests that, as a general phenomenon, there are hidden
relations and hidden symmetries that conventional 1T-physics does not capture,
implying the existence of a more unified formulation of physics that naturally
supplies the hidden information. In fact, we show that 2T-physics in
(d+2)-dimensions is the generator of these dualities in 1T-physics in
d-dimensions by providing a holographic perspective that unifies all the dual
1T systems into one. The unifying ingredient is a gauge symmetry in phase
space. Via such dualities it is then possible to gain new insights toward new
physical predictions not suspected before, and suggest new methods of
computation that yield results not obtained before. As an illustration, we
will provide concrete examples of 1T-systems in classical mechanics that are
solved analytically for the first time via our dualities. These dualities in
classical mechanics have counterparts in quantum mechanics and field theory,
and in some simpler cases they have already been constructed in field theory.
We comment on the impact of our approach on the meaning of spacetime and on
the development of new computational methods based on dualities.

\end{abstract}
\maketitle
\tableofcontents

\section{Introduction}

Symmetry concepts and computational techniques that emerged from 2T-physics in
$4+2$ dimensions were successfully applied recently in $3+1$ dimensional
cosmology, to obtain for the first time analytically the full set of
homogeneous cosmological solutions of the standard model of particle physics
coupled to gravity \cite{BSTconformalSM}, and to propose a new cyclic
cosmology driven only by the Higgs field with no recourse to an inflaton
\cite{BSThiggsCosmo}, in a geodesically complete universe \cite{BSTsailing}.
The underlying $4+2$ dimensions predicts the presence of a local conformal
Weyl symmetry in $3+1$ dimensions with restrictions on how to couple the Higgs
field to gravity such that the new conformally invariant standard model is
geodesically complete through cosmological singularities in a cyclic universe.
This Weyl symmetry carries information and imposes properties related to the
extra $1+1$ space and time dimensions \cite{2Tgravity}. Unprecedented analytic
control in these computations emerged from some very simple duality concepts
that amounted to making Weyl gauge transformations between different fixed
Weyl gauges of the same conformal standard model. Such gauge transformations,
or dualities, amount to simple changes of the perspective of the $3+1$
dimensional phase space within the $4+2$ dimensional phase space, which is
what we will study more generally in this paper.

Two crucial observations in M-theory in 1995-1996 provided the initial hints
for constructing 2T-physics in 1998 based on phase space gauge symmetry
\cite{2T-BDA}. These were: $\left(  i\right)  $ U-dualities in M-theory
appeared to be discrete phase-space gauge transformations between various
fixed gauges of a mysterious gauge symmetry in M-theory \cite{BarsYank-Udual},
and $\left(  ii\right)  $ there was a hint of an extra time dimension in
M-theory because the 11-dimensional extended supersymmetry of M-theory is
really a 12-dimensional SO$\left(  10,2\right)  $ covariant supersymmetry
written in the disguise of 11-dimensions \cite{IB-2TinMtheory}. Exploration of
these notions \cite{S-theory} raised the question of whether the unknown
M-theory might be a two-time theory with a global supersymmetry OSp$\left(
1|64\right)  $ whose BPS sectors that explained the five dual corners of
M-theory \cite{S-theory} could naturally arise from the constraints of an
underlying gauge symmetry? So what could the underlying gauge symmetry be? and
how could a theory with two timelike dimensions be unitary?

A ghost free unitary theory in a target space with two timelike dimensions
could not be viable without the presence of a new type of more powerful gauge
symmetry that could eliminate the problems of causality and ghosts from both
timelike dimensions. After figuring out that such a gauge symmetry does not
exist in position space, but it does exist in phase space \cite{2T-BDA}, it
became evident that the same phase space framework could also provide a
natural connection to dualities. Starting in 1998, 2T-physics was developed in
phase space for particles in the worldline formalism with a target space in
$d+2$ dimensions with two times, progressively including spin \cite{2Tspin1}%
-\cite{2Tspin4}, background fields \cite{2Tspin3GravGauge}\cite{2001HighSpin},
supersymmetry \cite{2TsusyParticle}, and twistors \cite{twistors}%
\cite{2Tspin4} (for a recent overview see \cite{2TphaseSpace}). That M-theory
could be formulated naturally in $11+2$ dimensions, with an OSp$\left(
1|64\right)  $ global supersymmetry and a gauge symmetry in the phase space of
branes, was illustrated with a toy M-model \cite{Mtheory}. 2T-physics was also
extended to the framework of field theory \cite{2Tspin2Fields}%
\cite{2Tspin3GravGauge}, including the standard model in $4+2$ dimensions
\cite{2Tsm}, gravity in $d+2$ dimensions \cite{2Tgravity}, SUSY field theory
with $N=1,2,4$ supersymmetries in $4+2$ dimensions \cite{2Tsusy}, SUSY
Yang-Mills in $10+2$ dimensions in 2010 \cite{2T-SYM-12D}, and finally
supergravity \cite{ibsuper}. It is still under construction for strings \&
branes \cite{stringsBranes}\cite{stringTwistor}\ and M-theory \cite{Mtheory},
and is expected that the most powerful eventual form of 2T-physics will be in
the framework of field theory in phase space as initiated in \cite{NC-U11}. By
now it is evident that, an underlying $4+2$ dimensional phase space, with
appropriate extra gauge symmetry, fits all known physics in $3+1$ dimensions,
from classical and quantum dynamics of particles, to field theory including
the realistic conformal standard model coupled to gravity, all the way to
supergravity. This $\left(  4+2\right)  $-dimensional approach has provided
the useful technical tools for the recent advances in $\left(  3+1\right)
$-dimensional cosmology reported in \cite{inflationBC}%
-\cite{IB-completeJourneys} and \cite{BSTconformalSM}-\cite{BSTsailing}.

The physics content in the 2T-physics formalism in $d+2$ dimensions is the
same as the physics content in the conventional 1T-physics formalism in
$(d-1)+1$ dimensions except that 2T-physics provides a holographic-type
perspective (as described below) with a much larger set of gauge symmetries,
and naturally makes predictions that are not anticipated in 1T-physics. Some
of the predictions take the forms of hidden symmetries and dualities; in this
paper we concentrate mainly on the dualities. The dualities are similar in
spirit to dualities encountered in M-theory or string theory, in the broader
sense of relating theories that look different in conventional 1T formalism,
but in reality contain the same physics information once a map is established
between them. In fact a lot of the new information from 2T-physics, which is
not contained systematically in 1T-physics, can be expressed in the language
of dualities directly in 1T-physics. Developing such dualities is our primary
objective in this paper.

The idea of using an embedding space $X^{M}$ in $4+2$ dimensions, which is
restricted to the cone, $X\cdot X=0,$ in order to realize SO$\left(
4,2\right)  $ conformal symmetry in $3+1$ dimensions, originated with Dirac
\cite{Dirac}. This idea was further developed over the years \cite{salam}%
-\cite{vasiliev}. 2T-physics connects to this notion of conformal symmetry in
one of its duality corners that we discuss in this paper, namely the
\textit{conformal shadow}, which is a gauged fixed version of 2T-physics in
$4+2$ dimensional \textit{flat space-time}. Thus, more recent works, based on
the same conformal symmetry notion in flat $4+2$ dimensions, are automatically
connected to 2T-physics; these include the $4+2$ dimensional formulation of
high-spin theory \cite{2001HighSpin},\cite{vasiliev2}-\cite{taronna2},
computation of conformal correlators in 3+1 dimensions using $4+2$ dimensions
\cite{rivelles}\cite{weinberg}, conformal bootstrap in the embedding formalism
\cite{rychkov}, and new mathematical notions related to conformal symmetry
\cite{waldron}\cite{waldron2}. We emphasize that these growing set of
connections correspond to only one corner of 2T-physics. 2T-physics is much
more than conformal symmetry in $3+1$ dimensions both conceptually and
practically. This is because 2T-physics is a gauge theory in phase space
$\left(  X^{M},P_{M}\right)  ,$ generally in $d+2$ curved space-time and, like
M-theory, has many 1T-physics corners with different physical interpretations
as illustrated with five specific shadows in this paper. When the idea of
\textit{a gauge symmetry in phase space }was introduced in \cite{2T-BDA}
Dirac's idea had faded away; so 2T-physics developed as a much richer theory,
unaware of Dirac's reasoning or motivation for conformal symmetry. That
connection was realized only after the notions of phase space gauge symmetry
had taken root and had already revealed new corners of 1T-physics well beyond
the conformal shadow. We now know that Dirac's idea and modern applications
\cite{vasiliev2}-\cite{waldron2} are automatically part of 2T-physics in the
special case when the Sp$\left(  2,R\right)  $ gauge symmetry generators
$Q_{ij}\left(  X,P\right)  $ take their simplest form shown in
Eq.(\ref{simple}), and only when the conformal shadow (or gauge) is chosen to
connect to 1T-physics. This suggests that the broader phase space properties
of 2T-physics, such as the multi-shadows and dualities discussed in this
paper, that continue to elude the practioners of the $X\cdot X=0$ constraint
even in modern times, can be used to obtain further physical consequences in
those settings. Also, 2T-physics is a general theory that goes well beyond the
\textit{flat} $4+2$ dimensional space-time constraint $X\cdot X=0$: it should
be noted that the generalized Sp$\left(  2,R\right)  $ generators
$Q_{ij}\left(  X,P\right)  $ in curved phase space with background fields
\cite{2Tspin3GravGauge}\cite{2001HighSpin}\cite{NC-U11}\cite{2TphaseSpace} and
interactions in field theory, including the standard model \cite{2Tsm} and
gravity \cite{2Tgravity}, lead to far reacher applications of 2T-physics.

In this paper we will extend previous results on dualities in 1T-physics
predicted by 2T-physics \cite{2TphaseSpace}. These take the form of explicit
canonical transformations among relativistic and non-relativistic 1T-physics
systems in $d$-dimensions, $\tilde{x}^{\mu}=\mathcal{X}^{\mu}\left(
x,p\right)  $ and $\tilde{p}_{\mu}=\mathcal{P}_{\mu}\left(  x,p\right)  ,$
that were not obtained before in classical mechanics with 1-time. These
include canonical transformations among some newly constructed solvable 1T
systems, such as a relativistic particle in an arbitrary potential, and
previously studied simpler systems, such as the relativistic massless
particle, relativistic massive particle, non-relativistic massive particle,
H-atom, and several others. All these cases are further generalized in this
paper by including arbitrary interactions of a particle with classical
background fields (electromagnetic, gravitational, high-spin). It is shown
that these more general systems are mapped from one dual system to another by
the same duality transformations that are independent of the backgrounds. So
the dual systems considered here cover a broad spectrum of interacting
1T-physics models. In principle, these classical canonical transformations
have counterparts in the quantum version of the same systems and can also be
extended to field theory, as has already been demonstrated with simpler
examples in the past \cite{dualitiesFields}.

One of our aims is to concentrate on the practical aspects of these canonical
transformations and to use them for developing new computational methods
within the traditional framework of 1T-physics. Indeed, our duality methods
are useful to perform computations in 1T-physics that would be hard or
impossible otherwise. The idea is to solve complicated systems by solving much
simpler dual systems. As an illustration, we will solve exactly the classical
mechanics of a relativistic particle in $d$-dimensions, which is constrained
to satisfy $p^{2}+V\left(  x^{2}\right)  =0$ for any potential $V\left(
x^{2}\right)  ,$ such as any power law $V\left(  x^{2}\right)  =c\left(
x^{2}\right)  ^{b},$ that we believe has not been solved before, and cannot
imagine how to solve without our dualities.

These dualities are predicted in the context of gauge symmetries in phase
space that generalize the notion of general coordinate invariance in position
space. The examples discussed here are only some representatives of a much
larger group of dualities that belong together in a unique symmetric theory in
2T-physics as reviewed in section \ref{2Tphys}. Each one of these 1T-systems
in $d$ dimensions captures holographically all of the gauge invariant
information in the 2T theory in $d+2$ dimensions. We call such 1T-systems
\textquotedblleft shadows\textquotedblright\ at $d$ dimensional boundaries of
the bulk in $d+2$ dimensions. Since each shadow contains all the physical
information, the parent theory in the bulk predicts that all shadows must be
holographic duals of each other.

Before we discuss specific dualities or the underlying theory, it is useful to
outline some concepts that give a sense of direction for why we are interested
in examining these dualities. Our canonical transformations in $d$-dimensions,
$\tilde{x}^{\mu}=\mathcal{X}^{\mu}\left(  x,p\right)  $ and $\tilde{p}_{\mu
}=\mathcal{P}_{\mu}\left(  x,p\right)  ,$ include the time coordinate and its
canonical conjugate Hamiltonian. Since time and Hamiltonian transform, it is
not surprising that we will establish relations among dynamical systems that
\textit{\`{a} priori} are considered to be different 1T-physics dynamical
systems with different Hamiltonians. We conceptualize a given phase space
$\left(  x^{\mu},p_{\mu}\right)  $ as the coordinates of a chosen phase space
frame for an observer that rides along with a particle on a worldline whose
time development $\left(  x^{\mu}\left(  \tau\right)  ,p_{\mu}\left(
\tau\right)  \right)  $ is determined by a phase space constraint $Q\left(
x,p\right)  =0$. An example of such a frame is the massless relativistic
particle that satisfies the constraint $p^{2}=0$. This observer is set up to
describe all physical phenomena in the universe (not only the motion of this
particle) from the point of view of this frame. A different phase space
$\left(  \tilde{x}^{\mu},\tilde{p}_{\mu}\right)  $ with a different constraint
$\tilde{Q}\left(  \tilde{x},\tilde{p}\right)  =0,$ such as the constrained
relativistic harmonic oscillator, $\left(  \tilde{p}^{2}+\omega^{2}\tilde
{x}^{2}\right)  =0,$ represents the frame of a different observer that also
examines all phenomena from this other perspective. The canonical
transformation, $\tilde{x}^{\mu}=\mathcal{X}^{\mu}\left(  x,p\right)  $ and
$\tilde{p}_{\mu}=\mathcal{P}_{\mu}\left(  x,p\right)  ,$ that maps the 1T
dynamics $Q\left(  x,p\right)  =0$ to the 1T dynamics $\tilde{Q}\left(
\tilde{x},\tilde{p}\right)  =0$ establishes the relations between the frames
and therefore all observations made by the two different observers are also
related to each other. The reader is invited to think of this setup as the
analog of Einstein's observers in different frames that are related to each
other by canonical transformations in phase space which generalize Einstein's
special or general coordinate transformations. The key in our theory is that,
the worldlines. $\left(  x^{\mu}\left(  \tau\right)  ,p_{\mu}\left(
\tau\right)  \right)  $ and $\left(  \tilde{x}^{\mu}\left(  \tau\right)
,\tilde{p}_{\mu}\left(  \tau\right)  \right)  ,$ that define the frames of the
two observers, are actually two shadows of the \textit{same worldline} in the
bulk in $d+2$ dimensions $\left(  X^{M}\left(  \tau\right)  ,P_{M}\left(
\tau\right)  \right)  .$ The two observers see very different 1T-physics
phenomena from the perspective of their own frames, however in our setup there
is already a predicted relationship between the observers since their
1T-physics equations are really two gauge choices of the same gauge invariant
equations in $d+2$ dimensions. There is a unique set of equations in $d+2$
dimensions supplied by 2T-physics that unify the vastly different 1T equations
of all such observers in $d$ dimensions. This unification is not at all
apparent in the conventional setup of 1T-physics. The unification makes
predictions of real physical phenomena in 1T-physics that can be tested by
studying the dualities that capture the hidden correlations of the various 1T
observers. Our purpose in this paper is to establish a few examples of such
dualities, which are surprising in 1T-physics, and in this way show that there
is much more physics to be learned from 2T-physics predictions that are not
supplied systematically in conventional 1T-physics.

In this paper we will first present our results for a few specific dualities
as canonical transformations, $\tilde{x}^{\mu}=\mathcal{X}^{\mu}\left(
x,p\right)  $ and $\tilde{p}_{\mu}=\mathcal{P}_{\mu}\left(  x,p\right)  ,$
purely in the context of conventional 1T-physics. Afterwards we will show how
they were obtained in the first place as the natural predictions of
2T-physics, and also indicate how a vast extension of such dualities can be
further obtained from the 2T approach.

The canonical transformations $\tilde{x}^{\mu}=\mathcal{X}^{\mu}\left(
x,p\right)  $ and $\tilde{p}_{\mu}=\mathcal{P}_{\mu}\left(  x,p\right)  $
discussed in this paper take a special mathematical form. It is shown that
they involve a 2$\times2$ matrix $M=\left(
\genfrac{}{}{0pt}{}{\alpha}{\gamma}%
\genfrac{}{}{0pt}{}{\beta}{\delta}%
\right)  $, of determinant $1$, that belongs to the group Sp$\left(
2,R\right)  =$SL$\left(  2,R\right)  ,$ with entries $\left(  \alpha
,\beta,\gamma,\delta\right)  $ that are non-linear functions of phase space
$\left(  x^{\mu},p_{\mu}\right)  $ including timelike directions. For example,
when the origin and target systems are both Lorentz covariant systems, the
transformation takes the form,%
\begin{align}
\tilde{x}^{\mu}  &  =x^{\mu}\alpha\left(  x,p\right)  +p^{\mu}\beta\left(
x,p\right)  \equiv\mathcal{X}^{\mu}\left(  x,p\right) \\
\tilde{p}^{\mu}  &  =x^{\mu}\gamma\left(  x,p\right)  +p^{\mu}\delta\left(
x,p\right)  \equiv\mathcal{P}_{\mu}\left(  x,p\right)  ,
\end{align}
where $\alpha,\beta,\gamma,\delta$ are functions of phase space$.$ This means
that under the dualities, $\left(  x^{\mu},p^{\mu}\right)  $ form Sp$\left(
2,R\right)  $ doublets covariantly in every direction $\mu$ of spacetime. When
one or both systems, $\left(  x^{\mu},p_{\mu}\right)  $ or $\left(  \tilde
{x}^{\mu},\tilde{p}_{\mu}\right)  ,$ are non-relativistic, the $\alpha
,\beta,\gamma,\delta$ are not as simple and not Lorentz invariant, but they
still belong to the phase-space-local Sp$\left(  2,R\right)  .$ It should be
emphasized that the set of dualities discussed in this paper (as linear
Sp$\left(  2,R\right)  $ transformations) is just a special case. Our
formalism is covariant under the most general non-linear Sp$\left(
2,R\right)  $ as the underlying gauge symmetry in phase space. Either the
linear or non-linear Sp$\left(  2,R\right)  $ transformations are broader than
the familiar local gauge transformations or general coordinate transformations
since the gauge parameters $\left(  \alpha,\beta,\gamma,\delta\right)  $ are
\textit{local in phase space}, not just in position space\footnote{An
infinitesimal gauge parameter as a function of phase $\varepsilon\left(
x,p\right)  $ packs together the parameters for local gauge transformations
$\varepsilon_{0}\left(  x\right)  $, general coordinate transformations
$\varepsilon_{1}^{\mu}\left(  x\right)  ,$ and much more, as seen in an
expansion in powers of momentum just as in Eq.(\ref{backgrounds-d}),
$\varepsilon\left(  x,p\right)  =\varepsilon_{0}\left(  x\right)
+\varepsilon_{1}^{\mu}\left(  x\right)  \left(  p_{\mu}+A_{\mu}\left(
x\right)  \right)  +\cdots..$ As an example, see the familiar transformation
on gauge fields, gravitational metric and high spin fields, organized as phase
space transformations, in Eqs.(33-37) in \cite{2001HighSpin}.}.

This paper is organized as follows. In section (\ref{1Trevisit}) we review and
clarify the gauge symmetries and constraints of the 1T system consisting of a
spinless particle in interaction with an arbitrary set of background fields in
$d$-dimensions. In section (\ref{canTransf}), we use the notation developed in
section (\ref{1Trevisit}) to present our canonical transformations between
five different 1T-physics systems. These are just examples to illustrate our
ideas which apply to a much larger class of 1T-physics systems connected to
each other by canonical transformations. In section (\ref{2Tphys}) we review
the idea of general gauge symmetry in phase space, apply it to 2T-physics
based on the Sp$\left(  2,R\right)  $ gauge symmetry, and then present five
different gauge choices in section (\ref{5gauges}) in which the gauge fixed
forms yield the five different 1T-physics systems that appear in section
(\ref{canTransf}). In section (\ref{dualities}) we show how to map the five
fixed gauge choices to one another by Sp$\left(  2,R\right)  $ gauge
transformations from one fixed gauge to another fixed gauge, thus obtaining
the 1T-physics canonical transformations described in section (\ref{canTransf}%
). In section (\ref{invariants}) we identify the invariant observables under
duality transformations and discuss special circumstances when there is a
hidden global SO$\left(  d,2\right)  $ symmetry associated with these
invariants. This SO$\left(  d,2\right)  $ is related to conformal symmetry in
one special shadow which we call the conformal shadow, but it is the
equivalent of conformal symmetry in all other shadows, including shadows for
massive particles. In section (\ref{solving5}) we illustrate how to use
dualities to explicitly solve the dynamics of a relativistic spinless particle
with a constraint $p^{2}+V(x^{2})=0$ in an arbitrary potential, a problem that
could not be solved before. Finally in section (\ref{conclude}) we interpret
these results from the point of view of $\left(  d+2\right)  $-dimensions,
comment on generalizing the concepts of dualities, and discuss what this means
for physics and spacetime in $d$-dimensions.

\section{Gauge Symmetry in 1T-Physics Revisited \label{1Trevisit}}

In this section we present all 1T-physics systems for a spinless particle in a
unified form that will be useful to discuss the dualities and canonical
transformations among 1T-physics systems that will be the subject of this
paper. In the following sections we will use this unified framework in
1T-physics to discuss canonical transformations that include spacelike as well
as timelike directions (including a change of Hamiltonian) to map various 1T
dynamical systems to each other.

To insure that our ideas are well understood we will begin with a simple
familiar example. The worldline action of a freely moving relativistic
particle of zero spin and mass $m$ is the familiar expression $S\left(
x\right)  =-m\int_{1}^{2}d\tau\sqrt{-\dot{x}^{2}}$. Here $\dot{x}^{\mu}%
\equiv\partial_{\tau}x^{\mu}$ is the velocity of a particle, whose position
$x^{\mu}\left(  \tau\right)  $ as a function of the worldline parameter
$\tau,$ is a covariant vector in $\left(  d-1\right)  $ space and $1$ time
dimensions. The Euler-Lagrange equations derived from the action are
$\partial_{\tau}p^{\mu}\left(  \tau\right)  =0,$ where $p^{\mu}=m\dot{x}^{\mu
}/\sqrt{-\dot{x}^{2}}$ is the canonical momentum derived from the action. The
particle moves freely since the momentum is a constant of motion - indeed this
is guaranteed by the fact that this Lagrangian is translationally invariant.

As is well known, this action has a local symmetry under $\tau$%
-reparametrizations, namely $x^{\mu}\left(  \tau\right)  \rightarrow x^{\mu
}\left(  \tau\right)  +\delta_{\varepsilon}x^{\mu}$ with $\delta_{\varepsilon
}x^{\mu}\left(  \tau\right)  =\varepsilon\left(  \tau\right)  \dot{x}^{\mu
}\left(  \tau\right)  ,$ is a symmetry of this action $S\left(  x+\delta
_{\varepsilon}x\right)  =S\left(  x\right)  ,$ as long as the end points are
not transformed, $\varepsilon\left(  \tau_{1}\right)  =\varepsilon\left(
\tau_{2}\right)  =0$. This is a transformation that mixes position and
momentum locally on the worldline, since we could write $\delta_{\varepsilon
}x^{\mu}\left(  \tau\right)  =\Lambda\left(  \tau\right)  p^{\mu}\left(
\tau\right)  ,$ with another local parameter $\Lambda\left(  \tau\right)
\equiv\varepsilon\left(  \tau\right)  \sqrt{-\dot{x}^{2}\left(  \tau\right)
}/m.$

This phase space gauge symmetry is crucial to remove the ghost degrees of
freedom in the timelike direction of $x^{\mu}\left(  \tau\right)  .$ As usual,
any gauge symmetry leads to constraints among the degrees of freedom. A
constraint is an equation satisfied by phase space degrees of freedom $\left(
x^{\mu},p_{\mu}\right)  $ such that no time derivatives occur, and hence it is
valid for all times $\tau$. In this case the constraint takes the form
$p^{2}+m^{2}=0,$ which is evidently satisfied by $p^{\mu}=m\dot{x}^{\mu}%
/\sqrt{-\dot{x}^{2}}.$ The physical meaning of the constraint is that this is
a massive relativistic particle at all times.

The same physical content is encoded in another form of the action in the
first order formalism which treats the phase space degrees of freedom $\left(
x^{\mu}\left(  \tau\right)  ,p_{\mu}\left(  \tau\right)  \right)  $ as two
independent vectors, whose equations of motion are derived by extremizing$\;$%
with respect to all degrees of freedom $\left(  x,p,e\right)  $ in the
following Lagrangian
\begin{equation}
L\left(  x,p,e\right)  =\left[  \dot{x}^{\mu}\left(  \tau\right)  p_{\mu
}\left(  \tau\right)  -\frac{1}{2}e\left(  \tau\right)  \left(  p^{2}\left(
\tau\right)  +m^{2}\right)  \right]  . \label{1TmassiveL}%
\end{equation}
Here a new degree of freedom $e\left(  \tau\right)  $ has been added. If first
the $p_{\mu},$ and then the $e,$ degrees of freedom are integrated out in that
order, then this action reduces to $S\left(  x\right)  =-m\int_{1}^{2}%
d\tau\sqrt{-\dot{x}^{2}},$ and hence the two versions have the same content.
However, the first order formalism reveals more clearly the nature of the
gauge symmetry, and leads to a full generalization to cover all possible
physical systems for a single spinless particle, massive or massless and in
interaction with all possible background fields, as seen below.

The phase space gauge symmetry of this first order action is given by%
\begin{equation}
\delta_{\Lambda}x^{\mu}=\Lambda\left(  \tau\right)  p^{\mu}\left(
\tau\right)  ,\;\delta_{\Lambda}p_{\mu}=0,\;\delta_{\Lambda}e=\partial_{\tau
}\Lambda\left(  \tau\right)  .
\end{equation}
Then the action is invariant, $\delta_{\Lambda}S\left(  x,p,e\right)  =0,$
because the Lagrangian transforms to a total derivative $\delta_{\Lambda
}L\left(  x,p,e\right)  =\partial_{\tau}\left(  \frac{1}{2}\Lambda\left(
\tau\right)  \left(  p^{2}\left(  \tau\right)  -m^{2}\right)  \right)  ,$
while $\Lambda\left(  \tau_{1}\right)  =\Lambda\left(  \tau_{2}\right)  =0$.

This is an example of a more general worldline gauge symmetry formalism that
applies to all physical systems as discussed presently. Consider the action
$S=\int_{1}^{2}d\tau L$ with the Lagrangian
\begin{equation}
L\left(  x,p,e\right)  =\dot{x}^{\mu}\left(  \tau\right)  p_{\mu}\left(
\tau\right)  -e\left(  \tau\right)  Q\left(  x\left(  \tau\right)  ,p\left(
\tau\right)  \right)  . \label{unified1T}%
\end{equation}
This general $Q\left(  x,p\right)  $ is to be regarded as a generator of
\textit{local} canonical transformations for any observable $A\left(
x,p\right)  $ by applying the Poisson bracket, $\delta_{\Lambda}%
A=\Lambda\left(  \tau\right)  \left\{  A,Q\right\}  ,$ where $\Lambda\left(
\tau\right)  $ is the local parameter on the worldline\footnote{It is possible
to generalize this first order Lagrangian by including also a Hamiltonian $U$,
$L=\dot{x}\cdot p-eQ\left(  x,p\right)  -U\left(  x,p\right)  ,$ as long as
the Hamiltonian is gauge invariant, meaning a vanishing Poisson bracket
$\left\{  Q,U\right\}  =0.$ The inclusion of $U$ does not change our
discussion and also does not really provide more physical (gauge invariant)
models than those obtainable from all possible expressions for $Q\left(
x,p\right)  .$ For this reason we do not find it useful to discuss $U$ any
further in this paper.}. Furthermore, $e\left(  \tau\right)  $ is to be
regarded as a Maxwell-Yang-Mills type Abelian gauge field in 1-dimension
(analog of the time-component of the gauge field $A_{0}$ in
Maxwell-Yang-Mills). Note that the gauge field $e$ is coupled to the generator
of gauge transformation $Q$ as would be the case in familiar gauge theories.
With this point of view, now define a gauge transformation on the phase space
degrees of freedom $\left(  x^{\mu},p_{\mu}\right)  $ by using Poisson
brackets to compute $\delta_{\Lambda}x^{\mu},\delta_{\Lambda}p_{\mu}$, with
$Q\left(  x,p\right)  $ as the generator, as follows
\begin{equation}
\delta_{\Lambda}x^{\mu}=\Lambda\left(  \tau\right)  \frac{\partial Q}{\partial
p_{\mu}},\;\;\;\delta_{\Lambda}p_{\mu}=-\Lambda\left(  \tau\right)
\frac{\partial Q}{\partial x^{\mu}},\;\;\;\delta_{\Lambda}e=\partial_{\tau
}\Lambda\left(  \tau\right)  .
\end{equation}
Note that $e\left(  \tau\right)  $ does indeed transform like an Abelian gauge
field independent of the \textquotedblleft matter\textquotedblright\ content,
while the specific choice of $Q\left(  x,p\right)  $ determines the dynamics
of the \textquotedblleft matter\textquotedblright\ degrees of freedom $\left(
x^{\mu}\left(  \tau\right)  ,p_{\mu}\left(  \tau\right)  \right)  $ through
the equations of motion. It can be checked that the action is invariant
because the Lagrangian transforms to a total $\tau$-derivative
\begin{equation}
\delta_{\Lambda}L=\frac{d}{d\tau}\left[  \Lambda\left(  \tau\right)  \left(
p\cdot\partial_{p}-1\right)  Q\left(  x\left(  \tau\right)  ,p\left(
\tau\right)  \right)  \right]  .
\end{equation}

The equation of motion for the gauge field $\partial L/\partial e\left(
\tau\right)  =0$ imposes the constraint%
\begin{equation}
Q\left(  x,p\right)  =0.
\end{equation}
This is the analog of Gauss's law that follows from $\partial L/\partial
A_{0}=0$ in Maxwell-Yang-Mills theory. Since $Q\left(  x,p\right)  $ is the
generator of gauge transformations, $Q=0$ identifies the sector of the theory
that has zero gauge charge, that is, the gauge invariant sector. So, the
meaning of this constraint is that only the gauge invariant subspace of phase
space, as identified by the solutions of $Q=0,$ is physical.

In this light, in the simple example where $Q=p^{2}+m^{2},$ the mass-shell
constraint $p^{2}+m^{2}=0$ implies not only that this is a massive particle
for all times, but also that the solutions of the constraint identify the
gauge invariant sub-phase-space for all times.

The first quantization of the general gauge theory with any $Q\left(
x,p\right)  $ can be performed by using covariant quantization, in which
$\left(  x^{\mu},p_{\mu}\right)  $ are quantized as if they are unconstrained
variables. The Hilbert space of this quantum phase space cannot be all
physical because it does not take into account the constraint $Q=0$. However,
in this larger Hilbert space, the physical subspace is found by imposing the
constraint on the quantum states $\hat{Q}|\Phi\rangle=0$, where the quantum
operator $\hat{Q}$ is defined by an appropriate ordering of the quantum
operators $\left(  \hat{x}^{\mu},\hat{p}_{\mu}\right)  $ that appear in
$\hat{Q}\left(  \hat{x},\hat{p}\right)  $. In particular, in position space
$\Phi\left(  x^{\mu}\right)  \equiv\langle x^{\mu}|\Phi\rangle$, where the
momentum is represented as a derivative on the complete basis for quantum
states $\langle x^{\mu}|$, the constraint takes the form of a differential
equation to be satisfied by the physical subset of quantum states $\hat
{Q}\left(  x,-i\hbar\partial\right)  \Phi\left(  x^{\mu}\right)  =0.$ For the
example when $\hat{Q}=\hat{p}^{2}+m^{2},$ this becomes the Klein-Gordon
equation $\left(  -\hbar^{2}\partial_{x}^{2}+m^{2}\right)  \Phi\left(  x^{\mu
}\right)  =0.$ For more complicated cases, the proper definition of the
physical sector in the quantum theory is complete only after a quantum
ordering of phase space operators is specified for $\hat{Q}\left(  \hat
{x},\hat{p}\right)  .$

General examples of physical interest that include electromagnetic,
gravitational and high-spin relativistic background fields are%
\begin{equation}
Q\left(  x,p\right)  =\left[
\begin{array}
[c]{c}%
\phi\left(  x\right)  +\frac{1}{2}g^{\mu\nu}\left(  x\right)  \left(  p_{\mu
}+A_{\mu}\left(  x\right)  \right)  \left(  p_{\nu}+A_{\nu}\left(  x\right)
\right) \\
+\sum_{n\geq3}\phi^{\mu_{1}\cdots\mu_{n}}\left(  x\right)  \left(  p_{\mu_{1}%
}+A_{\mu_{1}}\left(  x\right)  \right)  \cdots\left(  p_{\mu_{n}}+A_{\mu_{n}%
}\left(  x\right)  \right)
\end{array}
\right]  \label{backgrounds-d}%
\end{equation}
Here we have assumed that $Q\left(  x,p\right)  $ has a Taylor expansion in
powers of $p_{\mu},$ which is a common assumption for many physical systems.
If this assumption is not valid for some reason, then we can just as well
treat $Q\left(  x,p\right)  $ without an expansion. In any case, when the
expansion is valid, $\phi\left(  x\right)  ,$ $A_{\mu}\left(  x\right)  ,$
$h^{\mu\nu}\left(  x\right)  ,$ $\phi^{\mu_{1}\cdots\mu_{n}}\left(  x\right)
$ are the background fields (where $g^{\mu\nu}\left(  x\right)  =\eta^{\mu\nu
}+h^{\mu\nu}\left(  x\right)  ,$ with $\eta^{\mu\nu}$ the flat metric). Taken
as the generator of gauge transformations, the vanishing of this generalized
$Q\left(  x,p\right)  $ defines the gauge invariant sector at the classical
level. The quantum version (defined after an ordering of quantum operators
$\hat{x},\hat{p}$) is a differential operator acting on the gauge invariant
physical space, $Q\left(  x,-i\hbar\partial\right)  \Phi\left(  x\right)  =0,$
as indicated above. A good first rule for correct quantum ordering is to
replace the operators $\hat{p}_{\mu}$ by generally covariant derivatives,
$\hat{p}_{\mu}\rightarrow-i\hbar\nabla_{\mu},$ which commute with the
background metric $g_{\mu\nu}\left(  x\right)  .$ Clearly, beyond this,
quantum ordering is hard to settle uniquely in the general case without
additional guidance from symmetries of the system $Q\left(  x,p\right)  $, or
a more complete theory such as field theory. In this paper we do not tackle
the quantum issues any further since we will only discuss the purely classical
limit here, but instructive examples are treated in \cite{HatomETC}%
\cite{ADSetc}.

It must be noted that not only relativistic mechanics, but also all
non-relativistic mechanics may be presented in this formalism by taking any
$Q\left(  t,h,\mathbf{r,p}\right)  ,$ where space and time are considered on
the same footing, just as in relativity. Consider the usual non-relativistic
$\left(  d-1\right)  $ dimensional phase space vectors $\mathbf{r}\left(
\tau\right)  $ and $\mathbf{p}\left(  \tau\right)  ,$ plus the time degree of
freedom as a dynamical variable $t\left(  \tau\right)  $ as well as its
conjugate variable $h\left(  \tau\right)  ,$ with their Poisson brackets
$\left\{  r^{i},p^{j}\right\}  =\delta^{ij}$ and $\left\{  t,h\right\}  =-1$.
Then take the Lagrangian comparable to Eq.(\ref{unified1T})%
\begin{equation}
L=\mathbf{\dot{r}}\left(  \tau\right)  \cdot\mathbf{p}\left(  \tau\right)
-\dot{t}\left(  \tau\right)  h\left(  \tau\right)  -e\left(  \tau\right)
~Q\left(  t\left(  \tau\right)  ,h\left(  \tau\right)  ,\mathbf{r}\left(
\tau\right)  \mathbf{,p}\left(  \tau\right)  \right)  . \label{L-generalNR}%
\end{equation}
As already argued above, for any choice of $Q\left(  t,h,\mathbf{r,p}\right)
$ there is a gauge symmetry. Now consider the special case of $Q\left(
t,h,\mathbf{r,p}\right)  $ given by
\begin{equation}
Q\left(  t,h,\mathbf{r,p}\right)  =\left(  H\left(  \mathbf{r,p}\right)
-h\right)  \label{NRconstraint}%
\end{equation}
which is independent of $t$ and where $H\left(  \mathbf{r,p}\right)  $ is any
function of the phase space in $\left(  d-1\right)  $ dimensions. To make
contact with usual non-relativistic physics we may choose the gauge $t\left(
\tau\right)  =\tau$ and then solve the constraint in Eq.(\ref{NRconstraint})
for the canonical conjugate to $t$ in the form $h=H\left(  \mathbf{r,p}%
\right)  .$ Inserting this back in the action, and using $\dot{t}=1$ and
$h=H\left(  \mathbf{r,p}\right)  ,$ results in the familiar non-relativistic
formulation of the system for any Hamiltonian $H\left(  \mathbf{r,p}\right)  $%
\begin{equation}
L=\mathbf{\dot{r}}\left(  \tau\right)  \cdot\mathbf{p}\left(  \tau\right)
-H\left(  \mathbf{r,p}\right)  . \label{L-NRgaugefixed}%
\end{equation}
This shows that, like relativistic systems, non-relativistic systems,
including more complicated versions of $Q\left(  t,h,\mathbf{r,p}\right)  ,$
may also be regarded as gauge symmetric theories, with a dynamical timelike
dimension $t\left(  \tau\right)  $ and an appropriate constraint that can be
used to determine $h\left(  \tau\right)  ,$ as described in the unified
1T-Physics formalism of Eq.(\ref{unified1T}).

To discuss examples for the non-relativistic (cases $3,4$ below$)$ and
relativistic (cases $1,2,5$ below) systems in a unified form, we make up the
notation, $t=x^{0},$ $h=p^{0}=-p_{0},$ even though Lorentz
covariance/invariance is not implied in the rest of the expressions, such as
$Q\left(  x,p\right)  $, for the non-relativistic cases.

\section{The Canonical Transformations \label{canTransf}}

In the remainder of this paper, we will use the approach of the previous
section to discuss canonical transformations among a few 1T systems that are
the illustrative examples of interest in this paper. These will include the
following cases%

\begin{equation}%
\begin{tabular}
[c]{|l|l|l|}\hline
Case & \ \ \ \ \ \ \ \ \ Name & $Q\left(  x,p\right)  $ $%
\genfrac{}{}{0pt}{}{\text{background fields are}}{\text{represented by
\textquotedblleft}\cdots\text{\textquotedblright\ ~}}%
$\\\hline
\multicolumn{1}{|c|}{1} & massless relativistic & $p_{1}^{2}+\cdots$\\\hline
\multicolumn{1}{|c|}{2} & massive relativistic & $p_{2}^{2}+m_{2}^{2}+\cdots
$\\\hline
\multicolumn{1}{|c|}{3} & massive non-relativistic \  & $\mathbf{p}_{3}%
^{2}-2m_{3}h_{3}+\cdots$\\\hline
\multicolumn{1}{|c|}{4} & H-atom & $\mathbf{p}_{4}^{2}-2m_{4}\frac{\alpha
}{\left\vert \mathbf{r}_{4}\right\vert }-2m_{4}h_{4}+\cdots$\\\hline
\multicolumn{1}{|c|}{5} & relativistic potential & $p_{5}^{2}+V\left(
x_{5}^{2}\right)  +\cdots$\\\hline
\end{tabular}
\medskip\label{CaseList}%
\end{equation}
We have labelled the phase space for each case $\left(  x_{i}^{\mu},p_{i\mu
}\right)  $ with the corresponding case number $i=1,2,3,4,5.$ Bold characters
such as $\mathbf{r,p}$ in cases 3 and 4 imply vectors in $\left(  d-1\right)
$ space dimensions, and in those cases $h$ is the canonical conjugate to $t;$
otherwise $x,p$ imply relativistic vectors as in cases 1,2,5, and in those
cases $p_{0}$ is the canonical conjugate to the timelike coordinate $x^{0}.$
The choice of $Q\left(  x,p\right)  ,$ including background fields as in
Eq.(\ref{backgrounds-d}), is what defines the 1T-physics dynamics in each
case. In the table we indicated the form of $Q\left(  x,p\right)  $ in the
limit when all background fields vanish. It is understood that backgrounds
represented by \textquotedblleft$\cdots$\textquotedblright\ are to be included
as follows.

The canonical transformations discussed below are \textit{independent of any
set of background fields}. They apply equally well when background fields
vanish or when they are included according to the following prescription:
first generalize only one of the systems in the table above (say case 1) with
any set of background fields as in Eq.(\ref{backgrounds-d}), and then apply
the background-independent canonical transformations below to generate the
background fields in all the other dual systems. This is the 1T prescription
that emerges from the unified gauge invariant 2T theory including all
backgrounds in $d+2$ dimensions, as discussed in section (\ref{2Tphys}).

We found that for each pair $i,j$ the corresponding systems are related by
non-linear canonical transformations $\left(  j\leftarrow i\right)  $ of the
form
\begin{equation}
x_{j}^{\mu}=\mathcal{X}_{j}^{\mu}\left(  x_{i},p_{i}\right)  ,\;p_{j\mu
}=\mathcal{P}_{j\mu}\left(  x_{i},p_{i}\right)  , \label{qQjpKj}%
\end{equation}
that satisfy the Poisson brackets $\left\{  \mathcal{X}_{j}^{\mu}\left(
x_{i},p_{i}\right)  ,\mathcal{X}_{j}^{\nu}\left(  x_{i},p_{i}\right)
\right\}  =0=\left\{  \mathcal{P}_{j\mu}\left(  x_{i},p_{i}\right)
,\mathcal{P}_{j\nu}\left(  x_{i},p_{i}\right)  \right\}  $ and $\left\{
\mathcal{X}_{j}^{\mu}\left(  x_{i},p_{i}\right)  ,\mathcal{P}_{j\nu}\left(
x_{i},p_{i}\right)  \right\}  =\delta_{\nu}^{\mu}$, where the brackets are
evaluated in the phase space $\left(  x_{i},p_{i}\right)  $ by taking
derivatives $\left\{  A,B\right\}  =\left(  \partial_{x_{i}^{\mu}}A\right)
\left(  \partial_{p_{i\mu}}B\right)  -\left(  \partial_{x_{i}^{\mu}}B\right)
\left(  \partial_{p_{i\mu}}A\right)  .$ To illustrate, in this section we
exhibit one example, namely the cases $\left(  1\leftarrow2\right)  $ and
$\left(  2\leftarrow1\right)  ,$ as the following $2\times2$ matrix form that
gives explicitly the functions $\mathcal{X}_{j}^{\mu}\left(  x_{i}%
,p_{i}\right)  ,\mathcal{P}_{j}^{\mu}\left(  x_{i},p_{i}\right)  $ as well as
the inverse map
\begin{equation}%
\begin{tabular}
[c]{|c|}\hline
massless relativistic (1) $\leftrightarrow~$massive relativistic (2)\\\hline
\multicolumn{1}{|l|}{$\overset{}{%
\begin{array}
[c]{c}%
\left(
\begin{array}
[c]{c}%
x_{1}^{\mu}\\
p_{1}^{\mu}%
\end{array}
\right)  =\left(
\begin{array}
[c]{cc}%
\left(  \frac{1}{2}+\frac{\left\vert x_{2}\cdot p_{2}\right\vert }%
{2\sqrt{\left(  x_{2}\cdot p_{2}\right)  ^{2}+m_{2}^{2}x_{2}^{2}}}\right)
^{-1} & 0\\
\frac{m_{2}^{2}~\text{sign}\left(  x_{2}\cdot p_{2}\right)  }{2\sqrt{\left(
x_{2}\cdot p_{2}\right)  ^{2}+m_{2}^{2}x_{2}^{2}}} & \frac{1}{2}%
+\frac{\left\vert x_{2}\cdot p_{2}\right\vert }{2\sqrt{\left(  x_{2}\cdot
p_{2}\right)  ^{2}+m_{2}^{2}x_{2}^{2}}}%
\end{array}
\right)  \left(
\begin{array}
[c]{c}%
x_{2}^{\mu}\\
p_{2}^{\mu}%
\end{array}
\right)  \medskip\\
\left(
\begin{array}
[c]{c}%
x_{2}^{\mu}\\
p_{2}^{\mu}%
\end{array}
\right)  =\left(
\begin{array}
[c]{cc}%
\left(  1+\frac{m_{2}^{2}x_{1}^{2}}{4\left(  x_{1}\cdot p_{1}\right)  ^{2}%
}\right)  ^{-1} & 0\\
-\frac{m_{2}^{2}}{2\left(  x_{1}\cdot p_{1}\right)  } & \left(  1+\frac
{m_{2}^{2}x_{1}^{2}}{4\left(  x_{1}\cdot p_{1}\right)  ^{2}}\right)
\end{array}
\right)  \left(
\begin{array}
[c]{c}%
x_{1}^{\mu}\\
p_{1}^{\mu}%
\end{array}
\right)
\end{array}
}$}\\\hline
\end{tabular}
\ \ \ \label{nzToZ}%
\end{equation}
For all cases, the explicit $\left(  \mathcal{X}_{j}^{\mu}\left(  x_{i}%
,p_{i}\right)  ,\mathcal{P}_{j\mu}\left(  x_{i},p_{i}\right)  \right)  $ are
given at the equation numbers specified in the following table.
\begin{equation}%
\begin{tabular}
[c]{||c|c|c|c|c|c||}\hline\hline
target%
$\backslash$%
origin & 1 & 2 & 3 & 4 & 5\\\hline
1 &  & Eq.(\ref{massivemassless1}) & Eq.(\ref{MassiveNRToMassless1}) &
Eq.(\ref{41}) & Eq.(\ref{5to1})\\\hline
2 & Eq.(\ref{maslessmassive1}) &  & Eq.(\ref{Mtom-transform1}) & $\ \left(
2\leftarrow1\leftarrow4\right)  $ \  & $\ \left(  2\leftarrow1\leftarrow
5\right)  $ \ \\\hline
3 & Eq.(\ref{masslessToMassiveNR1}) & Eq.(\ref{mToM-transform1}) &  &
$\ \left(  3\leftarrow1\leftarrow4\right)  $ \  & $\ \left(  3\leftarrow
1\leftarrow5\right)  $ \ \\\hline
4 & Eq.(\ref{1to4}) & $\ \left(  4\leftarrow1\leftarrow2\right)  $ \  &
$\ \left(  4\leftarrow1\leftarrow3\right)  $ \  &  & $\ \left(  4\leftarrow
1\leftarrow5\right)  $ \ \\\hline
5 & Eq.(\ref{1to5}) & $\ \left(  5\leftarrow1\leftarrow2\right)  $ \  &
$\ \left(  5\leftarrow1\leftarrow3\right)  $ \  & $\ \left(  5\leftarrow
1\leftarrow4\right)  $ \  & \\\hline\hline
\end{tabular}
\ \ \ \ \ \ \ \ \label{canonList}%
\end{equation}
As an example, the contents of Eq.(\ref{nzToZ}) are indicated at the $\left(
12\right)  $ and $\left(  21\right)  $ entries of this table. The expressions
for $\mathcal{X}_{j}^{\mu}\left(  x_{i},p_{i}\right)  ,\mathcal{P}_{j\mu
}\left(  x_{i},p_{i}\right)  $ are used directly in 1T-physics as canonical
transformations, but these results were obtained as predictions from
2T-physics. The notation $\left(  j\leftarrow1\leftarrow i\right)  $ means the
composition of two transformations $\left(  1\leftarrow i\right)  $ followed
by $\left(  j\leftarrow1\right)  ,$ which gives the transformation $\left(
j\leftarrow i\right)  $. We used this notation for cases $\left(  j\leftarrow
i\right)  $ in which the direct transformation $\left(  \mathcal{X}_{j}^{\mu
}\left(  x_{i},p_{i}\right)  ,\mathcal{P}_{j\mu}\left(  x_{i},p_{i}\right)
\right)  $ looks algebraically too involved to be transparent to the reader,
and hence we opted for the more transparent notation $\left(  j\leftarrow
1\leftarrow i\right)  $ even though the direct transformation $\left(
j\leftarrow i\right)  $ is certainly available explicitly. The derivation of
these transformations using 2T-physics techniques is given in
Sec.(\ref{explicit}).

In this section we describe some of the general properties of these dualities
for all the cases. By definition of momentum as $p_{\mu}=\partial
L/\partial\dot{x}^{\mu},$ which is in agreement with the only term that
contains velocity in the first order Lagrangian (\ref{unified1T}), $L=\dot
{x}\cdot p+\cdots,$ the Poisson brackets must be $\left\{  x_{i}^{\mu}%
,p_{i\nu}\right\}  =\delta_{\nu}^{\mu}$ for each case $i$. The claim that we
found a canonical map $\left(  i\leftrightarrow j\right)  $ between cases $j$
and $i$ implies that our maps satisfy the following defining property that the
first term in the Lagrangian maintains the same form up to a total time
derivative
\begin{equation}
\dot{x}_{j}\cdot p_{j}=\frac{d}{d\tau}\mathcal{X}_{j}\left(  x_{i}%
,p_{i}\right)  \cdot\mathcal{P}_{j}\left(  x_{i},p_{i}\right)  =\dot{x}%
_{i}\cdot p_{i}+\frac{d}{d\tau}\Lambda_{ji}\left(  x_{i},p_{i}\right)  .
\label{p1-derivatives}%
\end{equation}
The total derivative may be dropped because it does not contribute to the
action or to the equations of motion. This is verified for each duality
$\left(  i\leftarrow j\right)  $ and the $\Lambda_{ji}\left(  x_{i}%
,p_{i}\right)  $ is computed in section (\ref{explicit}). Consequently our
canonical maps have to satisfy the Poisson bracket property (no sum on $i$ or
$j)$
\begin{equation}
\left\{  x_{j}^{\mu},p_{j\nu}\right\}  =\frac{\partial\mathcal{X}_{j}^{\mu
}\left(  x_{i},p_{i}\right)  }{\partial x_{i}^{\lambda}}\frac{\partial
\mathcal{P}_{j\nu}\left(  x_{i},p_{i}\right)  }{\partial p_{i\lambda}}%
-\frac{\partial\mathcal{P}_{j\nu}\left(  x_{i},p_{i}\right)  }{\partial
x_{i}^{\lambda}}\frac{\partial\mathcal{X}_{j}^{\mu}\left(  x_{i},p_{i}\right)
}{\partial p_{i\lambda}}=\delta_{\nu}^{\mu}. \label{p2-Poisson}%
\end{equation}
We have checked that this is indeed true, but have not included the tedious
algebra in this paper. This also guarantees that the Poisson brackets for any
observables $\left\{  A,B\right\}  $ give the same result if evaluated in
terms of any of the phase spaces listed in table (\ref{CaseList}).

Our duality maps satisfy the Poisson bracket (\ref{p2-Poisson}) or the
canonical property (\ref{p1-derivatives}) \textit{off-shell}, meaning that
they hold for the bigger phase space (including physical and unphysical
sectors of phase space) before any equation of motion is used or any
constraint $Q\left(  x,p\right)  $ is imposed. That is, they are properties of
just the duality transformations among the phase spaces and they are satisfied
independently of any specific dynamics or physical model. This means that any
set of background fields may be introduced as outlined above without changing
the duality transformations. The duality transformations may be thought of as
transformations between observers which are set up to describe physics in
their own phase-space frames, with their own definition of 1T phase space. The
canonical transformations connect the frames of such observers to one another
in a way that is analogous to general coordinate transformations connecting
observers in different frames. In the present case we are considering
transformations that connect observers that are local in phase space rather
than only in coordinate subspace.

In addition to the model independent properties (\ref{p1-derivatives}%
,\ref{p2-Poisson}), these canonical transformations have the following
remarkable property. The 5 quantities $Q\left(  x,p\right)  $ listed in
(\ref{CaseList}) transform into each other under the dualities. So, up to
overall factors these expressions are proportional to each other
\begin{equation}
p_{1}^{2}\sim\left(  p_{2}^{2}+m_{2}^{2}\right)  \sim\left(  \mathbf{p}%
_{3}^{2}-2m_{3}h_{3}\right)  \sim\left(  \mathbf{p}_{4}^{2}-2m_{4}\frac
{\alpha}{\left\vert \mathbf{r}_{4}\right\vert }-2m_{4}h_{4}\right)
\sim\left(  p_{5}^{2}+V\left(  x_{5}^{2}\right)  \right)  .
\label{dualConstraints}%
\end{equation}
The proportionality factors are given precisely by multiplying each constraint
$Q\left(  x,p\right)  $ by $\left(  X^{+^{\prime}}\right)  ^{2}$ in the same
Sp$\left(  2,R\right)  $ gauge, such as
\begin{equation}
\left(  X_{1}^{+^{\prime}}\right)  ^{2}\left(  p_{1}^{2}+\cdots\right)
=\left(  X_{2}^{+^{\prime}}\right)  ^{2}\left(  p_{2}^{2}+m_{2}^{2}%
+\cdots\right)  ,\text{ etc.} \label{dualConstraints2}%
\end{equation}
where the gauge fixed $X_{i}^{+^{\prime}}\left(  x_{i},p_{i}\right)
,~i=1,\cdots,5,$ are given for each gauge in section \ref{5gauges}. Thus, when
a constraint holds in one of the frames, e.g. $Q_{1}\left(  x_{1}%
,p_{1}\right)  =0$, it holds automatically also in all the dual frames,
including the backgrounds. Although we are considering only 5 explicit cases
in this paper, there are an infinite number of such cases (including their
generalizations with background fields represented by the ellipsis
\textquotedblleft$\cdots$\textquotedblright\ ). That is, there are an infinite
number of observers defined by their own frames in phase space, which are
related to each other by canonical transformations, as we will illustrate in
section \ref{explicit}. The 1T-physics dynamics in each frame is captured by
the expression of $Q\left(  x,p\right)  $ as discussed in the previous
section. The relations among these $Q\left(  x,p\right)  $ as in
(\ref{dualConstraints}) allows us to give physical meaning to observations (in
the sense of 1T-physics) and to the dualities among them.

For example, for simplicity we consider the free massless relativistic
particle, with the constraint $Q=p_{1}^{2}=0$ (no background fields), then via
our dualities all the expressions in Eq.(\ref{dualConstraints}) must vanish.
This means that, while observer 1 interprets this system as the free massless
relativistic particle $p_{1}^{2}=0,$ observer 2 interprets it as the free
massive relativistic particle $p_{2}^{2}+m_{2}^{2}=0,$ observer 3 sees it as
the free massive non-relativistic particle with Hamiltonian $h_{3}%
=\frac{\mathbf{p}_{3}^{2}}{2m_{3}}$, observer 4 thinks it is a planetary-type
or H-atom type interacting system with Hamiltonian $h_{4}=\frac{\mathbf{p}%
_{4}^{2}}{2m_{4}}-\frac{\alpha}{\left\vert \mathbf{r}_{4}\right\vert },$ and
observer 5 believes it is the relativistic particle in an arbitrary Lorentz
invariant potential that satisfies the constraint $p_{5}^{2}+V\left(
x_{5}^{2}\right)  =0$.

In 1T-physics, the dynamics of these systems are considered to be independent
with no particular relations among them. However, we will show in section
(\ref{invariants}) that there are duality invariant quantities that do not
transform, and are exactly equal to each other in all these systems. Hence
there are an infinite number of relations among them which are instant
predictions that can be verified by experiment or computation. The duality
invariants contain all the physical information about the whole collection of
these systems. For example, the initial conditions for solving the equations
of motion in any one of these systems can be expressed in terms of the duality
invariants. If the equations of motion are solved in one system (which is easy
for the free cases 1,2,3) then they are automatically solved in the difficult
systems, such as case 4 and especially 5, by using the duality transformations
as well as the duality invariants to relate the initial conditions. Such
hidden information is not available in 1T-physics, but it is a property of
Nature which can be verified by physicists in frames related to each other by
our transformations. The frame of such observers can in principle be created
with proper conditions in a laboratory and the predictions can be verified experimentally.

It is now clear that, based on the model independent properties of our
transformations, we can construct large classes of physical models that are
dual to each other by including background fields as in
Eq.(\ref{backgrounds-d}). For each case $i=1,2,\cdots$ one may introduce
background fields. If the backgrounds are related to each other by the
background independent canonical transformations in Eq.(\ref{canonList}) then
the models with such backgrounds continue to be duals of each other. For
example, if the relativistic massless particle in case \#1 is taken with a
background electromagnetic field as described by $Q_{1}\left(  x_{1}%
,p_{1}\right)  =\frac{1}{2}\left(  p_{1}+A(x_{1})\right)  ^{2}$, what are the
set of background fields in the other dual cases? This is computed by applying
the canonical transformation $\left(  i\leftarrow1\right)  $ to obtain the
form for $Q_{i}\left(  x_{i},p_{i}\right)  $ and then expand it in powers of
$p_{i}$ as in Eq.(\ref{backgrounds-d}) to read off the dual version of the
backgrounds. This is sufficient to see that the type of duality we have been
discussing is the norm rather than the exception. There is a huge amount of
testable physical predictions that can be made in this way by first compiling
a list of canonical transformations without background fields, as in the
illustrative examples of Eq.(\ref{CaseList}). This list is in principle
infinitely long. The transformations among the members of the list can all be
derived from the gauge invariant form of the theory in the framework of
2T-physics as will be discussed in section (\ref{explicit}). Hence all the
corresponding physical predictions are natural consequences of 2T-physics.

\section{Sp$\left(  2,R\right)  $ Gauge Symmetry in 2T-Physics \label{2Tphys}}

The notion of gauge symmetry in phase space, based on gauging Sp$\left(
2,R\right)  $ that led to 2T-physics, appears at first sight to be
generalizable. This generalization is reviewed in \cite{2TphaseSpace} where it
is shown that the formulation of phase space gauge symmetry for any Lie group
would start by constructing a set of Lie algebra generators $Q_{a}\left(
X,P\right)  ,$ $a=1,2,\cdots,N,$ that close under Poisson brackets in phase
space. The closure of the Lie algebra is required for the consistency of first
class constraints $Q_{a}\left(  X,P\right)  =0$ for physical states which
follows from gauge invariance.

The case of a single \textit{non-compact} generator $Q\left(  X,P\right)  $
leads to the formulation of all 1T-physics as shown in section
(\ref{1Trevisit}). The case of the simplest non-Abelian \textit{non-compact}
group Sp$\left(  2,R\right)  =$SL$\left(  2,R\right)  $ with 3 generators
leads uniquely to all 2T-physics without any ghosts and consistent with
causality. One may be tempted to speculate that larger non-compact Lie groups
may lead to reasonable unitary formulations of physics with more timelike
dimensions as formulated in \cite{2TphaseSpace}. However, with the phase space
degrees of freedom of \textit{a single particle} such attempts have repeatedly
failed because we could not find expressions for $Q_{a}\left(  X,P\right)  $
that yielded non-trivial and ghost-free solutions of the constraints
$Q_{a}\left(  X,P\right)  =0$, except for the cases of one or three
generators. The failure of the attempts may suggest the possibility of a
theorem that generalizations with larger non-compact groups
\cite{2TphaseSpace} must always fail for a \textit{single} spinless particle.
A brief review of the Sp$\left(  2,R\right)  $ case follows.

In the first order formalism, in which position $X^{M}\left(  \tau\right)  $
and momentum $P_{M}\left(  \tau\right)  $ are treated on an equal footing, we
require our theory to have an Sp$\left(  2,R\right)  $ gauge symmetry, which
is a subset of canonical transformations that mix $X$ and $P$ locally on the
worldline. This gauged subset of canonical transformations is generated by 3
generators written in the form of a symmetric $2\times2$ tensor $Q_{ij}.$ The
indices $i,j$ correspond to doublet indices under Sp$\left(  2,R\right)  ,$
$i,j=1,2,$ while the symmetric tensor $Q_{ij}$ is the triplet that corresponds
to the adjoint representation. These generators are constructed from the phase
space degrees of freedom $Q_{ij}\left(  X,P\right)  .$ The Sp$\left(
2,R\right)  $ Lie algebra is
\begin{equation}
\left\{  Q_{12},Q_{11}\right\}  =-2Q_{11},\;\left\{  Q_{12},Q_{22}\right\}
=2Q_{22},\;\left\{  Q_{11},Q_{22}\right\}  =4Q_{12}, \label{sp2rLie}%
\end{equation}
where the Poisson brackets $\left\{  Q_{ij},Q_{kl}\right\}  $ are computed in
terms of the $\left(  X^{M},P_{M}\right)  $ phase space. So, to proceed one
must find expressions for the $Q_{ij}\left(  X,P\right)  $ that satisfy this
Lie algebra. There are an infinite number of such phase space structures for
Sp$\left(  2,R\right)  $, which have been classified in \cite{2001HighSpin}.
Assuming some such expression for $Q_{ij}\left(  X,P\right)  $, we proceed as follows.

This Sp$\left(  2,R\right)  $, which algebraically is the same as SO$\left(
1,2\right)  ,$ is equivalent to a \textit{local conformal symmetry} SO$\left(
1,2\right)  $ on the worldline (i.e. SO$\left(  d,2\right)  $ with $d=1$) as
seen in a second order formalism where $P_{M}$ is integrated out
\cite{2T-BDA}. As a guide to readers familiar with string theory, it may be
useful to mention that this gauge symmetry may be regarded as being analogous
to the local conformal symmetry on the worldsheet generated by the Virasoro
algebra in string theory. Recall that, like here, the Virasoro algebra is also
constructed from the phase space degrees of freedom of the string (harmonic
oscillators). As a further guide, it may also be useful to mention that
background fields, that are restricted in string theory by equations that come
from imposing local conformal symmetry on the worldsheet (closure of Virasoro
algebra), also appear with analogous restrictions in the Sp$\left(
2,R\right)  $ gauge theory on the worldline, as seen below.

To implement the gauge symmetry generated by $Q_{ij}\left(  X,P\right)  $, we
introduce the gauge field $A^{ij}\left(  \tau\right)  $ in the adjoint
representation of Sp$\left(  2,R\right)  $ and then write the gauge invariant
action on the worldline in the first order formalism as follows\footnote{To
continue the analogies to string theory, we mention that string theory, which
is usually presented in the second order formulation, could also be
re-organized in the first order formalism as here. The second order
formulation of the Sp$\left(  2,R\right)  $ theory could be pursued, but this
would be very messy for the general case with all possible background fields,
and hence we prefer the first order formalism.}
\begin{align}
L  &  =\dot{X}^{M}\left(  \tau\right)  P_{M}\left(  \tau\right)  -\frac{1}%
{2}A^{ij}\left(  \tau\right)  Q_{ij}\left(  X\left(  \tau\right)  ,P\left(
\tau\right)  \right)  -\mathcal{H}\left(  X\left(  \tau\right)  ,P\left(
\tau\right)  \right)  ,\label{2TAction1}\\
\text{where}  &  \text{: }\mathcal{H}(X,P)\text{ is anything invariant under
Sp}\left(  2,R\right)  ,\text{ i.e., }\left\{  Q_{ij},\mathcal{H}\right\}
=0.\nonumber
\end{align}
If the gauge generators satisfy the algebra given in Eq.(\ref{sp2rLie}), the
action is invariant under the following infinitesimal transformation with
local parameters $\omega^{ij}\left(  \tau\right)  $,%

\begin{align}
\delta_{\omega}X^{M}  &  =\frac{1}{2}\omega^{ij}\left\{  X^{M},Q_{ij}\right\}
=\frac{1}{2}\omega^{ij}\frac{\partial Q_{ij}\left(  X,P\right)  }{\partial
P_{M}},\label{gtrans1}\\
\delta_{\omega}P_{M}  &  =\frac{1}{2}\omega^{ij}\left\{  P_{M},Q_{ij}\right\}
=-\frac{1}{2}\omega^{ij}\frac{\partial Q_{ij}\left(  X,P\right)  }{\partial
X^{M}},\label{gtrans2}\\
\delta_{\omega}A^{ij}  &  =\frac{d}{d\tau}\left(  \omega^{ij}\right)
+\omega^{ik}\varepsilon_{kl}A^{lj}+\omega^{jk}\varepsilon_{kl}A^{li}%
,\label{gtrans3}\\
\delta_{\omega}\mathcal{H}  &  =\frac{1}{2}\omega^{ij}\left\{  \mathcal{H}%
,Q_{ij}\right\}  =0. \label{gtrans4}%
\end{align}
These lead to $\delta_{\omega}Q_{kl}=\frac{1}{2}\omega^{ij}\left\{
Q_{kl},Q_{ij}\right\}  $, where the right hand side is given by
Eq.(\ref{sp2rLie}). Then it is easy to verify that the Lagrangian transforms
into a total derivative
\begin{equation}
\delta_{\omega}L=\frac{d}{d\tau}\left(  \frac{1}{2}\omega^{ij}\left(
\tau\right)  P_{M}\frac{\partial Q_{ij}}{\partial P_{M}}-\frac{1}{2}%
\omega^{ij}\left(  \tau\right)  Q_{ij}\right)  ,
\end{equation}
and therefore the action $S=\int_{\tau_{1}}^{\tau_{2}}d\tau L\left(
\tau\right)  $ is invariant, $\delta_{\omega}S=0,$ provided $\omega
^{ij}\left(  \tau\right)  $ vanishes at the end points $\tau_{1},\tau_{2}.$
This is the Sp$(2,R)$ gauge symmetry that underlies all 2T-physics.

An example of $Q_{ij}\left(  X,P\right)  $ that satisfy the Sp$\left(
2,R\right)  $ Lie algebra under Poisson brackets is
\begin{equation}
\text{example:\ }Q_{11}=X\cdot X,\;Q_{12}=X\cdot P,\;Q_{22}=P\cdot P,
\label{simple}%
\end{equation}
where the dot products are constructed with a flat metric $\eta_{MN}$ of any
signature. But only for $d+2$ dimensions with a signature with two times there
are non-trivial solutions to the constraints $Q_{ij}=0$; this means there is a
non-trivial gauge invariant physical sub-phase-space only when the formalism
admits two times or more. Only two times can be admitted because for more
timelike dimensions there would be ghosts and the theory would fail to be
unitary. Furthermore, with less than two times all solutions of $Q_{ij}=0$ are
either identically zero phase space (0 times) or physically trivial phase
space (1 time, with $X$ and $P$ parallel, so no angular momentum). Hence, only
two times, no less and no more, are possible when we demand the Sp$\left(
2,R\right)  $ gauge symmetry. In the simple case of Eq.(\ref{simple}) the
infinitesimal transformations $\delta_{\omega}X^{M},\delta_{\omega}P_{M}$
above are linear in $\left(  X,P\right)  ,$ and therefore in that case
$\left(  X,P\right)  $ behaves like the doublet of Sp$\left(  2,R\right)  $
under the local transformation. Hence, if the $Q_{ij}\left(  X,P\right)  $
have the quadratic form (\ref{simple}), then the finite Sp$\left(  2,R\right)
$ transformation takes the linear form with a matrix of determinant 1 as follows%

\begin{equation}%
\begin{pmatrix}
X^{\prime M}\\
P^{\prime M}%
\end{pmatrix}
=%
\begin{pmatrix}
\alpha\left(  \tau\right)  & \beta\left(  \tau\right) \\
\gamma\left(  \tau\right)  & \frac{1+\beta\left(  \tau\right)  \gamma\left(
\tau\right)  }{\alpha\left(  \tau\right)  }%
\end{pmatrix}%
\begin{pmatrix}
X^{M}\\
P^{M}%
\end{pmatrix}
. \label{2TFiniteLocalTransf}%
\end{equation}
More general examples of $Q_{ij}\left(  X,P\right)  $ involve all possible
background fields as in Eq.(\ref{backgrounds-d}). So, when there are
background fields, the local infinitesimal transformations $\delta_{\omega
}X^{M},\delta_{\omega}P_{M}$ in (\ref{gtrans1},\ref{gtrans2}) are non-linear
and cannot be written in this linear matrix form. Nevertheless, the
transformation of the gauge field $A^{ij}$ is necessarily of the Yang-Mills
form, and for finite transformations it can always be written in terms of the
matrix with one lower index, $A_{i}^{~j}\equiv\varepsilon_{ik}A^{kj}$ where
$\varepsilon_{ij}$ is the Sp$\left(  2,R\right)  $ metric, as follows%
\[
\left(
\begin{array}
[c]{cc}%
A^{12} & A^{22}\\
-A^{11} & -A^{12}%
\end{array}
\right)  ^{\prime}=%
\begin{pmatrix}
\alpha & \beta\\
\gamma & \frac{1+\beta\gamma}{\alpha}%
\end{pmatrix}
\left(  \left(
\begin{array}
[c]{cc}%
A^{12} & A^{22}\\
-A^{11} & -A^{12}%
\end{array}
\right)  -\partial_{\tau}\right)
\begin{pmatrix}
\alpha & \beta\\
\gamma & \frac{1+\beta\gamma}{\alpha}%
\end{pmatrix}
^{-1}%
\]
which gives%

\begin{equation}%
\begin{array}
[c]{l}%
A^{\prime11}=\left(
\begin{array}
[c]{c}%
\gamma\left(  1+\beta\gamma\right)  \partial_{\tau}\alpha^{-1}+\gamma
^{2}\alpha^{-1}\partial_{\tau}\beta-\alpha^{-1}\partial_{\tau}\gamma\\
+\left(  \frac{1+\beta\gamma}{\alpha}\right)  ^{2}A^{11}-2\frac{\gamma}%
{\alpha}\left(  1+\beta\gamma\right)  A^{12}+\gamma^{2}A^{22}%
\end{array}
\right)  ,\\
A^{\prime12}=\left(
\begin{array}
[c]{c}%
\frac{1}{\alpha}\left(  1+\gamma\beta\right)  \partial_{\tau}\alpha
-\beta\partial_{\tau}\gamma\\
-\frac{\beta}{\alpha}\left(  1+\beta\gamma\right)  A^{11}+\left(
1+2\beta\gamma\right)  A^{12}-\gamma\alpha A^{22}%
\end{array}
\right)  ,\\
A^{\prime22}=\left(
\begin{array}
[c]{c}%
\alpha\partial_{\tau}\beta-\beta\partial_{\tau}\alpha\\
+\beta^{2}A^{11}-2\beta\alpha A^{12}+\alpha^{2}A^{22}%
\end{array}
\right)  .
\end{array}
\label{A22Transf}%
\end{equation}

For the more general case with background fields, as in \cite{2001HighSpin}
one may argue that, up to canonical transformations of $X^{M},P_{M}$, the
generators $Q_{11}\left(  X,P\right)  $ and $Q_{12}\left(  X,P\right)  $ may
be simplified to the following forms\footnote{A generally covariant form that
avoids the appearance of explicit $X^{M}$ is given in \cite{2Tspin3GravGauge}%
\cite{2001HighSpin}\cite{2Tgravity} as follows: $Q_{11}=W\left(  X\right)  $
and $Q_{12}=V^{M}\left(  X\right)  P_{M},$ where $W\left(  X\right)
,V^{M}\left(  X\right)  $ are background fields like the others, and instead
of $h_{2}^{MN}\left(  X\right)  +\eta^{MN}$ in $Q_{22}$ we simply write the
general metric $g^{MN}\left(  X\right)  .$ Then closure for Sp$\left(
2,R\right)  $ restricts these background fields to obey some homothety
conditions as given in \cite{2Tspin3GravGauge}\cite{2001HighSpin}%
\cite{2Tgravity}. The simplified version, with the explicit $X^{M}$ used in
this paper, is a choice of coordinates under general coordinate
transformations that is equivalent to the general version, while maitaning
covariance with respect to the SO$\left(  d,2\right)  $ global transformations
as a subset of general coordinate transformations. The simplified version
satisfies the homothety conditions automatically. \label{specialXPbasis}}
\begin{equation}
Q_{11}\left(  X,P\right)  =X^{M}X^{N}\eta_{MN},\;\;Q_{12}\left(  X,P\right)
=X^{M}P_{M}. \label{Q11simplified}%
\end{equation}
while the most general form of $Q_{22}\left(  X,P\right)  $ that satisfies the
Sp$\left(  2,R\right)  $ Lie algebra in Eq.(\ref{sp2rLie}) may be
parameterized in a power expansion of momentum (when this is permitted) and
contains background fields as functions of $X$ as follows \cite{2001HighSpin}
\begin{align}
Q_{22}(X,P)  &  =h_{0}\left(  X\right)  +\left(  \eta^{M_{1}M_{2}}%
+h_{2}^{M_{1}M_{2}}\left(  X\right)  \right)  \left(  P_{M_{1}}+A_{M_{1}%
}\left(  X\right)  \right)  \left(  P_{M_{2}}+A_{M_{2}}\left(  X\right)
\right) \nonumber\\
&  +\sum_{n\geq3}h_{n}^{M_{1}M_{2}\cdots M_{n}}\left(  X\right)  \left(
P_{M_{1}}+A_{M_{1}}\left(  X\right)  \right)  \left(  P_{M_{2}}+A_{M_{2}%
}\left(  X\right)  \right)  \cdots\left(  P_{M_{n}}+A_{M_{n}}\left(  X\right)
\right)  . \label{Q22simplified}%
\end{align}
The background fields are%
\begin{equation}
h_{0}\left(  X\right)  ,~A_{M}\left(  X\right)  \;\text{and }h_{n}^{M_{1}%
M_{2}\cdots M_{n}}\left(  X\right)  ,~\text{with }n=2,3,\cdots.
\end{equation}
When all of these vanish we obtain the simple case $Q_{22}(X,P)=P^{2}$ in
Eq.(\ref{simple}). The vector $A_{M}\left(  X\right)  $ is a $U\left(
1\right)  $ gauge field coupled covariantly to momentum $\left(  P_{M}%
+A_{M}\left(  X\right)  \right)  .$ The 2-tensor $g^{M_{1}M_{2}}\left(
X\right)  =\eta^{M_{1}M_{2}}+h_{2}^{M_{1}M_{2}}\left(  X\right)  $ is a
general metric in curved space. $h_{0}\left(  X\right)  $ is a scalar field,
while the $h_{n}^{M_{1}M_{2}\cdots M_{n}}\left(  X\right)  ,$ which are
symmetric traceless tensors with $n\geq3$ indices, are higher spin fields with
spin $n.$ There is no independent vector $h_{1}^{M}\left(  X\right)  $
associated with the first power of $P_{M},$ because in a rearrangement in
powers of $P$ rather than $\left(  P+A\right)  ,$ the vector $h_{1}^{M}\left(
X\right)  $ emerges as a combination of the vector $A_{M}\left(  X\right)  $
and the other $h_{n}^{M_{1}M_{2}\cdots M_{n}},$ i.e. $h_{1}^{M}=2\left(
\eta^{M_{1}M_{2}}+h_{2}^{M_{1}M_{2}}\right)  A_{M_{2}}+\cdots.$

For the Sp$\left(  2,R\right)  $ Lie algebra to close properly as in
Eq.(\ref{sp2rLie}) it is necessary to impose restrictions on the background
fields. The closure requires that the two form, $F_{MN}\equiv\partial_{M}%
A_{N}-\partial_{N}A_{M},$ and all the high spin fields be transverse to the
vector $X^{M}$ \cite{2001HighSpin}
\begin{equation}
X^{M}F_{MN}=0,\;\text{and }\eta_{MM_{1}}X^{M}h_{n}^{M_{1}M_{2}\cdots M_{n}%
}=0,\;n=2,3,\cdots\label{cond-transverse}%
\end{equation}
and that all other backgrounds are homogeneous fields with definite scaling
dimensions for $n=0,2,3,\cdots$ \cite{2001HighSpin}
\begin{equation}
\left(  X^{M}\partial_{M}-\left(  n-2\right)  \right)  h_{n}^{M_{1}M_{2}\cdots
M_{n}}\left(  X\right)  =0,\;\text{or }h_{n}^{M_{1}M_{2}\cdots M_{n}}\left(
\lambda X\right)  =\lambda^{n-2}h_{n}^{M_{1}M_{2}\cdots M_{n}}\left(
X\right)  . \label{cond-homog}%
\end{equation}
The Sp$\left(  2,R\right)  $ algebra among the $Q_{ij}\left(  X,P\right)
~$closes only if the background fields satisfy the transversality and
homogeneity conditions in Eqs.(\ref{cond-transverse},\ref{cond-homog}). Hence,
to define the model with an Sp$\left(  2,R\right)  $ gauge symmetry, it is
necessary to impose these as \textit{\`{a} priori} conditions on the
background fields.

For the reader familiar with string theory, these Sp$\left(  2,R\right)  $
conditions on the backgrounds in the worldline formalism are analogous to the
conditions on backgrounds that emerge from conformal symmetry on the
worldsheet (closure of the Virasoro algebra).

It is useful to work in a fixed axial-type gauge for the U$\left(  1\right)  $
background gauge field, $X\cdot A=0,$ which makes it a transverse vector, just
like all other tensors as in Eq.(\ref{cond-transverse}). In that case the
constraint $X^{M}F_{MN}=0$ simplifies to the following homogeneity condition
on $A_{M}$ \cite{2001HighSpin}$,$ which is also similar to all other tensors
as in Eq.(\ref{cond-homog})
\begin{equation}
X\cdot A=0,\;\left(  X^{M}\partial_{M}+1\right)  A_{M}=0,\text{ or }%
A_{M}\left(  \lambda X\right)  =\lambda^{-1}A_{M}\left(  X\right)  .
\label{cond-homog-A}%
\end{equation}
The generalization of these equations to spinning systems was given in
\cite{2Tspin1}-\cite{2Tspin3GravGauge} but we will not discuss this here since
in this paper we are concentrating only on spinless particles.

It may be of interest to emphasize that the constraints on 6-dimensional
fields found by trial and error by Weinberg \cite{weinberg} in order to have
6-dimensional correlelators consistent with conformal symmetry in 3+1
dimensions, are identical to the Sp$\left(  2,R\right)  $ gauge symmetry
conditions on fields that were already derived in \cite{2Tspin1}%
-\cite{2Tspin3GravGauge},\cite{2001HighSpin},\cite{NC-U11} as given above. So
these constraints on fields, which were also naturally incorporated in the 2T
standard model \cite{2Tsm} and 2T gravity \cite{2Tgravity}, including fermions
and gauge bosons, follow directly from a fundamental gauge symmetry Sp$\left(
2,R\right)  $ in phase space, and their underlying role is to insure a unitary
and causal theory with two times in $d+2$ dimensions.

As explained in footnote (\ref{specialXPbasis}), we made a special choice of
basis of phase space $\left(  X^{M},P_{M}\right)  $ such that the expression
for $Q_{11}=X^{M}X^{N}\eta_{MN}$ introduced the flat metric $\eta_{MN}$ which
is invariant under SO$\left(  d,2\right)  .$ Using this flat metric we may
raise or lower indices, such as $P^{M}\equiv\eta^{MN}P_{N}$ or $X_{M}%
\equiv\eta_{MN}X^{N},$ which should not be confused with raising or lowering
indices with the full metric $g^{M_{1}M_{2}}\left(  X\right)  =\eta
^{M_{1}M_{2}}+h_{2}^{M_{1}M_{2}}\left(  X\right)  .$ With this definition of
$P^{M}$ we define the generators of SO$\left(  d,2\right)  $ transformations
\begin{equation}
L^{MN}=X^{M}P^{N}-X^{N}P^{M}. \label{LMN}%
\end{equation}
Under Poisson brackets these commute with all dot products $\left(  X^{M}%
X^{N}\eta_{MN}\right)  $, $\left(  X^{M}P_{N}\right)  $, $\left(  P_{M}%
P_{N}\eta^{MN}\right)  .$ In particular they commute with the two Sp$\left(
2,R\right)  $ generators $Q_{11}=$ $X^{M}X^{N}\eta_{MN}$ and $Q_{12}%
=X^{M}P_{M}$%
\begin{equation}
\left\{  Q_{11},L^{MN}\right\}  =0,\;\left\{  Q_{12},L^{MN}\right\}  =0.
\label{LMNinvar}%
\end{equation}
This means that $Q_{11},Q_{12}$ are invariant under global SO$\left(
d,2\right)  $ transformations, but it also means that the $L^{MN}$ are
\textit{gauge invariant} under the subgroup of Sp$\left(  2,R\right)  $
transformations generated by $Q_{11},Q_{12}.$ Since these two generators are
quadratic, the 2-parameter gauge transformation they induce on $\left(
X^{M},P^{M}\right)  $ is linear just as Eq.(\ref{2TFiniteLocalTransf}), with
the parameter $\beta=0.$ This subgroup of gauge transformations will play an
important role in the dualities we will discuss in this paper. The fact that
$L^{MN}$ are gauge invariant under this subgroup of Sp$\left(  2,R\right)  $
predicts that these $L^{MN}$\textit{ are invariants under the dualities} as
discussed in section (\ref{invariants}).

In the presence of background fields denoted by \textquotedblleft$\cdots
$\textquotedblright\ the third Sp$\left(  2,R\right)  $ generator,
$Q_{22}=\left(  P^{2}+\cdots\right)  ,$ does not commute with $L^{MN}$ except
for its first term $\left\{  P^{2},L^{MN}\right\}  =0$, but when the
background fields vanish then $Q_{22}$ becomes SO$\left(  d,2\right)  $
invariant while $L^{MN}$ becomes gauge invariant under the full Sp$\left(
2,R\right)  .$

\subsection{Five Gauges and Five Shadows \label{5gauges}}

In this section, we give five different gauge fixed configurations of $\left(
X^{M},P_{M}\right)  $ such that, when inserted in the 2T action
(\ref{2TAction1}), result in five shadows in two less dimensions and
interpreted as five different 1T-physics systems. Each 1T shadow is expressed
by 1T Lagrangians $L_{i}$ $,$ $i=1,\cdots,5,$ as in Eq.(\ref{unified1T}), but
with five different constraints $Q_{i}\left(  x_{i},p_{i}\right)  $ as in
(\ref{backgrounds-d}), and parametrized in terms of five canonical sets of
degrees of freedom $\left(  x_{i}\left(  \tau\right)  ,p_{i}\left(
\tau\right)  \right)  ,$ as listed in Eq.(\ref{CaseList}). It should be
mentioned that the emerging 1T Lagrangians $L_{i}$ are defined up to a total
derivative $L_{i}\rightarrow L_{i}+\frac{d\Lambda_{i}}{d\tau}.$ The total
derivative could be dropped since it does not contribute to the action or the
equations of motion, but here we will give the $\Lambda_{i}\left(  x,p\right)
$ that emerge directly from the gauge fixing, so that the interested reader
can verify the result.

It should be emphasized that for these five shadows the parent 2T theory in
general contains any set of background fields, since $Q_{22}\left(
X,P\right)  =P^{2}+\left(  \text{backgrounds}\right)  ,$ but for simplicity we
will not explicitly write down specific backgrounds. Also, we will discuss
only the case of the 2T-system in Eq.(\ref{2TAction1}) in which $\mathcal{H}%
=0$ because this is sufficient to illustrate our methods, while the addition
of a non-trivial $\mathcal{H}$ does not change the essential part of the discussion.

We now give a list of five configurations for $X^{M},P^{M}$ (where $P^{M}%
=\eta^{MN}P_{N}$, using the $\eta^{MN}$ already introduced in
(\ref{Q11simplified},\ref{Q22simplified})), for which two gauges have been
fixed and the two constraints \thinspace$X^{2}=0$ and $X\cdot P$ have been
solved explicitly. So each configuration is parametrized in terms of the
remaining 1T degrees of freedom $\left(  x_{i}^{\mu},p_{i_{\mu}}\right)  $ in
two less dimensions. In each gauge the resulting $Q_{i}\left(  x_{i}%
,p_{i}\right)  $ and $\Lambda_{i}\left(  x_{i},p_{i}\right)  $ are computed.
The algebra to get these results is straightforward. We will illustrate this
in detail for the simplest case \#1 and most complicated case \#5, while cases
2,3,4 are sketched with sufficient detail but leaving a small exercise for the reader.

\subsubsection{Shadow 1, Massless Relativistic:}

The lightcone basis in the extra dimensions $\pm^{\prime}$ is defined as,
$X^{\pm^{\prime}}=\frac{1}{\sqrt{2}}\left(  X^{0^{\prime}}\pm X^{1^{\prime}%
}\right)  ,$ and similarly for the momenta. The two gauge choices are
$X_{1}^{+^{\prime}}\left(  \tau\right)  =1$ and $P_{1}^{+^{\prime}}\left(
\tau\right)  =0$ for all $\tau.$ The components $X_{1}^{-^{\prime}}\left(
\tau\right)  =\frac{1}{2}x_{1}^{2}$ and $P_{1}^{-^{\prime}}\left(
\tau\right)  =x_{1}\cdot p_{1}$ are computed to satisfy the constraints,
$X\cdot X=0=-2X^{+^{\prime}}X^{-^{\prime}}+X^{\mu}X_{\mu},$ and similarly for
$0=X\cdot P.$ The gauge fixed configuration of $\left(  X^{M},P^{M}\right)  $
is then
\begin{equation}%
\begin{tabular}
[c]{|l|l|l|l|}\hline
$M=$ & $+^{\prime}$ & $~~~-^{\prime}$ & $\mu$\\\hline
$X_{1}^{M}=$ & $1~$ & $~\frac{1}{2}x_{1}^{2}~~$ & $x_{1}^{\mu}$\\\hline
$P_{1}^{M}=$ & $0~$ & $~x_{1}\cdot p_{1}~~$ & $p_{1}^{\mu}$\\\hline
\end{tabular}
\ \ \ \ \ ,\;\;%
\begin{tabular}
[c]{|l|}\hline
$P_{1}^{2}=p_{1}^{2}~~$\\\hline
$\Lambda_{1}=0~~$\\\hline
\end{tabular}
\ ~ \label{g1}%
\end{equation}
Now, to obtain the gauge fixed form of the action (\ref{2TAction1}) up to a
total $\tau$ derivative, we compute $\dot{X}_{1}^{M}=\left(  0,~\dot{x}%
_{1}\cdot x_{1},\;\dot{x}_{1}^{\mu}\right)  $ which gives, $\dot{X}_{1}\cdot
P_{1}=\dot{x}_{1}\cdot p_{1}+d\Lambda_{1}/d\tau.$ We see that $\Lambda_{1}=0$
since we find no extra total time derivative$.$ We also compute the third
constraint given in (\ref{Q22simplified}), $Q_{22}=P_{1}^{2}+\cdots,$ which
becomes $Q_{22}=p_{1}^{2}+\cdots$, where \textquotedblleft$\cdots
$\textquotedblright\ stand for background fields consistent with the
constraints (\ref{Q22simplified}-\ref{cond-homog}). Inserting these in the 2T
Lagrangian (\ref{2TAction1}) we obtain the 1T shadow Lagrangian
\begin{equation}
L_{1}=\dot{x}_{1}\cdot p_{1}-\frac{1}{2}A_{1}^{22}\left(  \tau\right)  \left(
p_{1}^{2}+\cdots\right)  . \label{g1L}%
\end{equation}
After imposing the transversality and homogeneity constraints in
(\ref{cond-transverse},\ref{cond-homog}) on the background fields in $d+2$
dimensions, we find that the surviving background fields denoted by
\textquotedblleft$\cdots$\textquotedblright\ are precisely the background
fields in $d$ dimensions displayed in Eq.(\ref{backgrounds-d}) and
\cite{2001HighSpin}. The emergent shadow in $d$ dimensions is evidently the
Lagrangian for the interacting 1T massless relativistic particle as discussed
in Eqs.(\ref{unified1T},\ref{backgrounds-d}).

We call this gauge the conformal shadow. This is the shadow in which linear
SO$\left(  d,2\right)  $ transformations on $\left(  X^{M},P_{M}\right)  ,$
that leave the flat metric $\eta_{MN}$ invariant, become the familiar
non-linear conformal transformations in phase space in $\left(  d-1\right)
+1$ dimensions. To see this, the reader is invited to evaluate the SO$\left(
d,2\right)  $ generators $L^{MN}=X^{M}P^{N}-X^{N}P^{M}$ for the gauged fixed
configuration $\left(  X_{1}^{M},P_{1}^{N}\right)  $ of Eq.(\ref{g1}) and
verify that these $L^{MN}$ take the form of the familiar SO$\left(
d,2\right)  $ conformal transformations in $d$ dimensions. That we should
expect such a hidden symmetry in Eq.(\ref{g1L}) when all background fields
vanish is predicted from the fully covariant parent 2T theory (\ref{2TAction1}%
) before gauges are fixed.

\subsubsection{Shadow 2, Massive relativistic:}

We will be brief because the procedure is the same and the result was given
before (see references in \cite{2TphaseSpace}). The gauge fixed configuration
that also satisfies $X_{2}^{2}=0=X_{2}\cdot P_{2}$ is%
\begin{equation}%
\begin{tabular}
[c]{|l|l|l|l|}\hline
$M=$ & $~~\;\;+^{\prime}$ & $~~~~\;-^{\prime}$ & $\;\mu$\\\hline
$X_{2}^{M}=$ & $\;\;\;\frac{1+a}{2a}$ & $\;\;\;\frac{x_{2}^{2}a}{1+a}~~~$ &
$x_{2}^{\mu}~$\\\hline
$P_{2}^{M}=$ & $\frac{-m_{2}^{2}}{2\left(  x_{2}\cdot p_{2}\right)  a}~$ &
$~\left(  x_{2}\cdot p_{2}\right)  a~$ & $\;p_{2}^{\mu}~$\\\hline
\end{tabular}
\ \ \ \ \ \ \ \ \ ,\;%
\begin{tabular}
[c]{|l|}\hline
$\;a\equiv\sqrt{1+\frac{m_{2}^{2}x_{2}^{2}}{\left(  x_{2}\cdot p_{2}\right)
^{2}}}$\\\hline
$P_{2}^{2}=p_{2}^{2}+m_{2}^{2}$\\\hline
$\Lambda_{2}=\left(  x_{2}\cdot p_{2}\right)  \left(  a-1\right)  ~$\\\hline
\end{tabular}
\ \ \ \ \ \label{g2}%
\end{equation}
The steps leading from the 2T Lagrangian to the 1T shadow are parallel to
those in case 1. We find $\dot{X}_{2}\cdot P_{2}=\dot{x}_{2}\cdot p_{2}+$
$d\Lambda_{2}/d\tau$ with the $\Lambda_{2}$ given in (\ref{g2}), and
$P_{2}^{2}=p_{2}^{2}+m_{2}^{2}.$ Inserting these in the 2T Lagrangian
(\ref{2TAction1}) we obtain the 1T shadow action
\begin{equation}
L_{2}=\dot{x}_{2}\cdot p_{2}-\frac{1}{2}A_{2}^{22}\left(  \tau\right)  \left(
p_{2}^{2}+m_{2}^{2}+\cdots\right)  . \label{g2L}%
\end{equation}
in which we have dropped the total derivative $d\Lambda_{2}/d\tau$. Here the
remaining constraint is the same $Q_{22}$ in (\ref{Q22simplified}), but now
written in gauge 2, $0=Q_{22}=P_{2}^{2}+\cdots=\left(  p_{2}^{2}+m_{2}%
^{2}+\cdots\right)  \equiv Q_{2}\left(  x_{2},p_{2}\right)  ,$ as listed in
(\ref{CaseList}). The background fields in $L_{2}$ are inherited from those in
$d+2$ dimensions by specializing to the gauge 2. This is evidently the
Lagrangian for the 1T massive relativistic particle, with mass $m_{2},$ and
generally interacting with background fields. The mass can now be viewed as a
modulus in the embedding of the $d$-dimensional phase space $\left(
x_{2}^{\mu},p_{2\mu}\right)  $ in the $\left(  d+2\right)  $-dimensional phase
space $\left(  X^{M},P_{M}\right)  .$ So, it is a property of the 1T observer
as he/she parametrizes from this perspective the phenomena that occur in
$\left(  d+2\right)  $-dimensional phase space.

We should expect a relationship between the background fields in shadow \#1
and shadow \#2 since they are both derived from those in $d+2$ dimensions. As
we have summarized in the paragraph just before Eq.(\ref{qQjpKj}), this
relationship is given by the background independent duality transformation
between shadows 1\&2 which takes the form of canonical transformations
displayed in Eq.(\ref{nzToZ}).

When all backgrounds vanish, the massive particle system described by
(\ref{g2L}) has a hidden SO$\left(  d,2\right)  $ symmetry given by the
conserved generators, $L^{MN}=X_{2}^{M}P_{2}^{N}-X_{2}^{N}P_{2}^{M}$, as
demonstrated in \cite{2TphaseSpace}. That we should expect such a hidden
symmetry in Eq.(\ref{g2L}) when all backgrounds vanish is evident from the
fully covariant parent theory (\ref{2TAction1}) before gauges are fixed.

\subsubsection{Shadow 3, Massive Non-relativistic:}

The gauge fixed configuration that also satisfies $X_{3}^{2}=0=X_{3}\cdot
P_{3}$ is (here we use the parameters $t_{3}$ for the time-like coordinate and
$h_{3}$ for its canonical conjugate since these are more intuitive symbols in
non-relativistic physics)
\begin{equation}%
\begin{tabular}
[c]{|l|l|l|l|l|}\hline
$M=$ & $~+^{\prime}$ & $~-^{\prime}$ & $~0$ & $~i$\\\hline
$X_{3}^{M}=$ & $~t_{3}$ & $~u$ & $~s~$ & $~\mathbf{x}_{3}^{i}$\\\hline
$P_{3}^{M}=$ & $~m_{3}$ & $~h_{3}$ & $~0$ & $~\mathbf{p}_{3}^{i}$\\\hline
\end{tabular}
\ \ \ \ \ \ \ \ \ \ ,\;%
\begin{tabular}
[c]{|l|}\hline
$~u\equiv\frac{1}{m_{3}}\left(  \mathbf{x}_{3}\cdot\mathbf{p}_{3}-t_{3}%
h_{3}\right)  $\\\hline
$~s^{2}\equiv\mathbf{x}_{3}^{2}-\frac{2t_{3}}{m_{3}}\mathbf{x}_{3}%
\cdot\mathbf{p}_{3}+\frac{2t_{3}^{2}}{m_{3}}h_{3}$\\\hline
$P_{3}^{2}=\mathbf{p}_{3}^{2}-2m_{3}h_{3}$\\\hline
$\Lambda_{3}=-m_{3}u$\\\hline
\end{tabular}
\ \ \ \ \ \ \label{g3}%
\end{equation}
The steps leading from the 2T Lagrangian to the 1T shadow are parallel to
those in cases 1\&2. We find $\dot{X}_{3}\cdot P_{3}=\mathbf{\dot{x}}_{3}%
\cdot\mathbf{p}_{3}-\dot{t}_{3}h_{3}+$ $d\Lambda_{3}/d\tau,$ and $P_{3}%
^{2}=-2m_{3}h_{3}+\mathbf{p}_{3}^{2}.$ Inserting these in the 2T Lagrangian
(\ref{2TAction1}) we obtain the 1T shadow \#3 action
\begin{equation}
L_{3}=\mathbf{\dot{x}}_{3}\cdot\mathbf{p}_{3}-\dot{t}_{3}h_{3}-\frac{1}%
{2}A_{3}^{22}\left(  \tau\right)  \left(  \mathbf{p}_{3}^{2}-2m_{3}%
h_{3}+\cdots\right)  . \label{g3L}%
\end{equation}
in which we dropped the total derivative $d\Lambda_{3}/d\tau$. The remaining
constraint is the same $Q_{22}$ now written in gauge 3, $0=Q_{22}=P_{3}%
^{2}+\cdots=\left(  \mathbf{p}_{3}^{2}-2m_{3}h_{3}+\cdots\right)  \equiv
Q_{3}\left(  x_{3},p_{3}\right)  $ as listed in (\ref{CaseList}). This is
evidently the Lagrangian for the massive non-relativistic particle, with mass
$m_{3},$ as discussed in Eqs. (\ref{L-generalNR}-\ref{L-NRgaugefixed}). The
mass $m_{3}$ can be viewed as a modulus in the embedding of the $d$%
-dimensional non-relativistic phase space $\left(  \mathbf{x}_{3}%
,\mathbf{p}_{3},t_{3},h_{3}\right)  $ in the $\left(  d+2\right)
$-dimensional phase space $\left(  X^{M},P_{M}\right)  .$ So, it is a property
of the 1T non-relativistic observer as he/she parametrizes from this
perspective the phenomena that occur in $\left(  d+2\right)  $-dimensional
phase space. The background fields represented by \textquotedblleft$\cdots
$\textquotedblright\ are again inherited from those in $d+2$ dimensions, and
therefore are related to the background fields in shadows 1\&2 by the
background independent canonical transformations given in
Eqs.(\ref{MassiveNRToMassless1},\ref{masslessToMassiveNR1}) and
Eqs.(\ref{mToM-transform1},\ref{Mtom-transform1}).

When all backgrounds vanish, the massive particle system described by
(\ref{g3L}) has a hidden SO$\left(  d,2\right)  $ symmetry given by the
conserved generators, $L^{MN}=X_{3}^{M}P_{3}^{N}-X_{3}^{N}P_{3}^{M}$, as
demonstrated in \cite{2TphaseSpace}. That we should expect such a hidden
symmetry in Eq.(\ref{g3L}) when all backgrounds vanish is evident from the
fully covariant parent theory (\ref{2TAction1}) before gauges are fixed.

\subsubsection{Shadow 4, H-atom:}

The gauge fixed configuration that already satisfies $X_{4}^{2}=0=X_{4}\cdot
P_{4}$ is%
\begin{equation}%
\begin{tabular}
[c]{|l|}\hline
$%
\begin{tabular}
[c]{|l|c|c|c|}\hline
$M=$ & $\pm^{\prime}$ & $0$ & $i$\\\hline
$X_{4}^{M}=$ & $~\frac{1}{\sqrt{-4m_{4}h_{4}}}\left\{
\begin{array}
[c]{c}%
\left\vert \mathbf{x}_{4}\right\vert \sqrt{-2m_{4}h_{4}}\sin u\\
+\left(  \mathbf{x}_{4}\cdot\mathbf{p}_{4}\right)  \left(  \cos u\pm1\right)
\end{array}
\right\}  ~,$ & $\left\{
\begin{array}
[c]{c}%
\left\vert \mathbf{x}_{4}\right\vert \cos u\\
-\frac{\mathbf{x}_{4}\mathbf{\cdot p}_{4}}{\sqrt{-2m_{4}h_{4}}}\sin u
\end{array}
\right\}  ~,$ & $\mathbf{x}_{4}^{i}~$\\\hline
$P_{4}^{M}=$ & $\frac{1}{\sqrt{-4m_{4}h_{4}}}\left\{
\begin{array}
[c]{c}%
\pm2m_{4}h_{4}\\
+\frac{m_{4}e_{4}^{2}}{\left\vert \mathbf{x}_{4}\right\vert }\left(  \cos
u\pm1\right)
\end{array}
\right\}  ~,$ & $-\frac{m_{4}e_{4}^{2}}{\left\vert \mathbf{x}_{4}\right\vert
\sqrt{-2m_{4}h_{4}}}\sin u~,$ & $\mathbf{p}_{4}^{i}~$\\\hline
\end{tabular}
$\\\hline
\multicolumn{1}{|c|}{$\text{in this paper it is understood that }\sqrt
{-2m_{4}h_{4}}\text{ means }\pm\sqrt{-2m_{4}h_{4}}.$}\\\hline
\end{tabular}
\ \ \label{g4}%
\end{equation}
with
\begin{equation}%
\begin{tabular}
[c]{|l|}\hline
$u\equiv\frac{\sqrt{-2m_{4}h_{4}}}{m_{4}e_{4}^{2}}(\mathbf{x}_{4}\mathbf{\cdot
p}_{4}-2h_{4}t_{4})$\\\hline
$P_{4}^{2}=\left(  \mathbf{p}_{4}^{2}-2m_{4}\frac{e_{4}^{2}}{\left\vert
\mathbf{x}_{4}\right\vert }-2m_{4}h_{4}\right)  $\\\hline
$\Lambda_{4}=3h_{4}t_{4}-2\left(  \mathbf{x}_{4}\cdot\mathbf{p}_{4}\right)
$\\\hline
\end{tabular}
\ \ \label{lambda4}%
\end{equation}
Here we use the parameters $t_{4}$ for the time-like coordinate and $h_{4}$
for its canonical conjugate, and assumed $h_{4}<0$ for bound
states\footnote{The analytic continuation to the phase space region $h_{4}>0$
for scattering states looks similar, but to insure a real parametrization we
also swap a timelike coordinate with a spacelike coordinate. See Table II in
[Phys.Rev. D76 (2007) 065016, arXiv:0705.2834] for details.}. The remaining
constraint is the same $Q_{22}$ now written in gauge 4, $0=Q_{22}=P_{4}%
^{2}+\cdots=\left(  \mathbf{p}_{4}^{2}-2\frac{m_{4}e_{4}^{2}}{\left\vert
\mathbf{x}_{4}\right\vert }-2m_{4}h_{4}+\cdots\right)  \equiv Q_{4}\left(
x_{4},p_{4}\right)  $ as listed in (\ref{CaseList}). The steps leading from
the 2T Lagrangian to the 1T shadow are parallel to those in case 1. We find
$\dot{X}_{4}\cdot P_{4}=\mathbf{\dot{x}}_{4}\cdot\mathbf{p}_{4}-\dot{t}%
_{4}h_{4}+$ $d\Lambda_{4}/d\tau,$ and $P_{4}^{2}=\left(  \mathbf{p}_{4}%
^{2}-2\frac{m_{4}e_{4}^{2}}{\left\vert \mathbf{x}_{4}\right\vert }-2m_{4}%
h_{4}\right)  .$ Inserting these in the 2T Lagrangian (\ref{2TAction1}) we
obtain the 1T shadow action (in which we drop the total derivative
$d\Lambda_{4}/d\tau$)
\begin{equation}
L_{4}=\mathbf{\dot{x}}_{4}\cdot\mathbf{p}_{4}-\dot{t}_{4}h_{4}-\frac{1}%
{2}A_{4}^{22}\left(  \tau\right)  \left(  \mathbf{p}_{4}^{2}-\frac{2m_{4}%
e_{4}^{2}}{\left\vert \mathbf{x}_{4}\right\vert }-2m_{4}h_{4}+\cdots\right)  .
\label{gL4}%
\end{equation}
To see that this is equivalent to the Lagrangian for a particle in the $1/r$
potential (like the H-atom or planetary motion), we use the remaining gauge
freedom to choose the gauge $t_{4}\left(  \tau\right)  =\tau$ and solve the
constraint $Q_{4}=0$ for the canonical conjugate $h_{4}.$ In the case of no
background fields we get $h_{4}=\frac{\mathbf{p}_{4}^{2}}{2m_{4}}-\frac
{e_{4}^{2}}{\left\vert \mathbf{x}_{4}\right\vert }$ . Then, using $\dot{t}%
_{4}\left(  \tau\right)  =1$ and the solved form for $h_{4},$ the Lagrangian
$L_{4}$ reduces to the familiar form for the particle in the $1/r$ potential
\begin{equation}
L_{4}\rightarrow\mathbf{\dot{x}}_{4}\cdot\mathbf{p}_{4}-\left(  \frac
{\mathbf{p}_{4}^{2}}{2m_{4}}-\frac{e_{4}^{2}}{\left\vert \mathbf{x}%
_{4}\right\vert }\right)  ,\text{ if no backgrounds.} \label{g4L}%
\end{equation}
The mass $m_{4}$ and coupling strength $e_{4}^{2}$ can be viewed as moduli in
the embedding of the $d$-dimensional non-relativistic phase space $\left(
\mathbf{x}_{4},\mathbf{p}_{4},t_{4},h_{4}\right)  $ in the $\left(
d+2\right)  $-dimensional phase space $\left(  X^{M},P_{M}\right)  .$ So,
these are properties of the 1T non-relativistic observer as he/she
parametrizes from this perspective the phenomena that occur in $\left(
d+2\right)  $-dimensional phase space. If there are background fields then
they are inherited from those in $d+2$ dimensions specialized to gauge 4 and
related to those in shadows 1,2,3 by the duality transformations indicated in
Table II.

When all backgrounds vanish, the massive particle system described by
(\ref{g4L}) has a hidden SO$\left(  d,2\right)  $ symmetry given by the
conserved generators, $L^{MN}=X_{4}^{M}P_{4}^{N}-X_{4}^{N}P_{4}^{M}$, as
demonstrated in \cite{2TphaseSpace}. That we should expect such a hidden
symmetry in Eq.(\ref{g4L}) when all backgrounds vanish is evident from the
fully covariant parent theory (\ref{2TAction1}) before gauges are fixed.

\subsubsection{Shadow 5, Relativistic Potential $V\left(  x^{2}\right)  $:}

The discussion of this case will also shed some light on how to proceed to
construct more general shadows. The gauge fixed configuration is%
\begin{equation}%
\begin{tabular}
[c]{|c|c|c|c|}\hline
$M=$ & $+^{\prime}$ & $-^{\prime}$ & $\;\mu$\\\hline
$X_{5}^{M}=$ & $A$ & $\frac{1}{2A}x_{5}^{2}$ & $x_{5}^{\mu}~$\\\hline
$P_{5}^{M}=$ & $~\frac{A}{x_{5}^{2}}\left(  x_{5}\cdot p_{5}-\phi\right)  ~$ &
$~\frac{1}{2A}\left(  x_{5}\cdot p_{5}+\phi\right)  ~$ & $\;p_{5}^{\mu}%
~$\\\hline
\end{tabular}
\ \ \ \ \ \ \ \ \label{g5}%
\end{equation}
The constraints $X_{5}^{2}=0=X_{5}\cdot P_{5}$ are already satisfied for any
$A\left(  x_{5},p_{5}\right)  $ and $\phi\left(  x_{5},p_{5}\right)  .$ The
$A,\phi$ are constructed to obtain the dynamics described by the following
constraint%
\begin{equation}
Q_{5}\left(  x_{5},p_{5}\right)  =\left(  P_{5}^{2}+\cdots\right)  =\left[
p_{5}^{2}+V\left(  x_{5}^{2}\right)  +\cdots\right]  =0,
\end{equation}
where \textquotedblleft$\cdots$\textquotedblright\ \ represents the
contribution of background fields, and $V\left(  x^{2}\right)  $ is any
function of the Lorentz invariant $x_{5}^{2}$. Hence, to obtain
\begin{equation}
P_{5}^{2}=p_{5}^{2}-\frac{1}{x_{5}^{2}}\left(  \left(  x_{5}\cdot
p_{5}\right)  ^{2}-\phi^{2}\right)  =p_{5}^{2}+V\left(  x_{5}^{2}\right)  ,
\end{equation}
we chose $\phi\left(  x_{5},p_{5}\right)  $ such that
\begin{equation}
\phi\left(  x_{5},p_{5}\right)  =\left(  x_{5}\cdot p_{5}\right)
\sqrt{1+\frac{x_{5}^{2}V\left(  x_{5}^{2}\right)  }{\left(  x_{5}\cdot
p_{5}\right)  ^{2}}}. \label{phi2}%
\end{equation}
More general relativistic shadows follow from more general choices of
$\phi\left(  x_{5},p_{5}\right)  .$

To insure that $\left(  x_{5}^{\mu},p_{5\mu}\right)  $ are canonical
conjugates in the emergent $5^{th}$ shadow, rather than being some random
symbols, we must also require
\begin{equation}
\dot{X}_{5}^{M}P_{5M}=\dot{x}_{5}^{\mu}p_{5\mu}+\frac{d\Lambda_{5}}{d\tau},
\end{equation}
where the second term is a total time derivative. Then $\Lambda_{5}\left(
x_{5}\left(  \tau\right)  ,p_{5}\left(  \tau\right)  \right)  $ may be dropped
from the action since it does not contribute to the equations of motion of
$\left(  x_{5}^{\mu},p_{5\mu}\right)  $. Inserting the $\left(  X_{5}%
^{M},P_{5}^{M}\right)  $ of Eq.(\ref{g5}) into this requirement results in a
non-trivial restriction on $X_{5}^{+^{\prime}}\equiv A$ as follows%
\begin{equation}
\frac{1}{2A}\left(  x_{5}\cdot p_{5}+\phi\right)  \frac{dA}{d\tau}+\frac
{A}{x_{5}^{2}}\left(  x_{5}\cdot p_{5}-\phi\right)  \frac{d}{d\tau}\left(
\frac{x_{5}^{2}}{2A}\right)  =\frac{d\Lambda_{5}}{d\tau}.
\end{equation}
The general solution of this equation for the special $\phi\left(  x_{5}%
,p_{5}\right)  $ in Eq.(\ref{phi2}) is given by%
\begin{equation}
A\left(  x_{5},p_{5}\right)  =F\left(  \phi\right)  \sqrt{x_{5}^{2}}%
\exp\left[  -\frac{1}{2}\int^{x_{5}^{2}}\frac{du}{u}\left(  1-\frac{uV\left(
u\right)  }{\phi^{2}}\right)  ^{-1/2}\right]  , \label{A5}%
\end{equation}
and
\begin{equation}
\Lambda_{5}\left(  x_{5},p_{5}\right)  =-\int^{x_{5}^{2}}duV\left(  u\right)
\left(  \phi^{2}-uV\left(  u\right)  \right)  ^{-1/2}+2\int^{\phi}dz~z\frac
{d}{dz}\ln F\left(  z\right)  . \label{lambda5}%
\end{equation}
where $F\left(  \phi\right)  $ is a general function of its argument
$\phi\left(  x_{5},p_{5}\right)  $ given in (\ref{phi2}). Then we see that the
emerging $5^{th}$ shadow Lagrangian which determines the dynamics of the
remaining degrees of freedom $\left(  x_{5}^{\mu},p_{5\mu}\right)  $ as
derived from Eq.(\ref{2TAction1}) is given by%
\begin{equation}
L_{5}=\dot{x}_{5}^{\mu}p_{5\mu}-\frac{1}{2}A_{5}^{22}\left(  p_{5}%
^{2}+V\left(  x_{5}^{2}\right)  +\cdots\right)  . \label{g5L}%
\end{equation}
Note that the dynamics of $\left(  x_{5},p_{5}\right)  $ is independent of the
solution $X_{5}^{+^{\prime}}=A\left(  x_{5},p_{5}\right)  $ given in
Eq.(\ref{A5}), but the expression for $A\left(  x_{5},p_{5}\right)  $ in
(\ref{A5}) is needed to fully determine the embedding of the 5$^{th}$ shadow
in $d+2$ dimensional phase space as given in Eq.(\ref{g5}). Also,
$X_{5}^{+^{\prime}}=A\left(  x_{5},p_{5}\right)  $ is needed to obtain the
duality transformation to the other shadows. Furthermore, $A\left(
x_{5},p_{5}\right)  $ determines also the components $L^{\pm^{\prime}\mu}$ of
the SO$\left(  d,2\right)  $ generators.

When all background fields \textquotedblleft$\cdots$\textquotedblright%
\ vanish, the action in Eq.(\ref{g5L}) has a hidden SO$\left(  d,2\right)  $
symmetry just as in all previous cases discussed above. The generators of this
symmetry are again, $L^{MN}=X_{5}^{M}P_{5}^{N}-X_{5}^{N}P_{5}^{M},$ where we
insert the gauge fixed $\left(  X_{5}^{M},P_{5}^{M}\right)  $ given in
Eq.(\ref{g5}). These symmetry generators are the Noether charges that are
conserved using the equations of motion derived from (\ref{g5L}). In
particular the conserved generator $L^{+^{\prime}-^{\prime}}$ coincides with
$\phi\left(  x_{5},p_{5}\right)  $ given in Eq.(\ref{phi2}), namely
$L^{+^{\prime}-^{\prime}}=X_{5}^{+^{\prime}}P_{5}^{-^{\prime}}-X_{5}%
^{-^{\prime}}P_{5}^{+^{\prime}}=\phi,$ as derived from (\ref{g5}). That we
should expect a hidden SO$\left(  d,2\right)  $ symmetry for the action in
Eq.(\ref{g5L}) when all backgrounds vanish is evident from the fully covariant
parent theory (\ref{2TAction1}) before the gauge is fixed.

As an example, consider $V\left(  x^{2}\right)  =c\left(  x^{2}\right)  ^{b}$
where $c,b$ are arbitrary constants. For this case the integrals in
Eq.(\ref{A5}) can be done explicitly, yielding%
\begin{equation}
A^{1+b}=\left(  F\left(  \phi\right)  \right)  ^{1+b}\frac{\left\vert
x_{5}\cdot p_{5}\right\vert }{\sqrt{c}}\left[  1+\left(  1+\frac{c\left(
x_{5}^{2}\right)  ^{1+b}}{\left(  x_{5}\cdot p_{5}\right)  ^{2}}\right)
^{1/2}\right]  .\text{ } \label{A5bc5}%
\end{equation}
As a check, we compare this result to shadow 2 given in Eq.(\ref{g2}). We find
agreement when we specialize by taking $V\left(  x^{2}\right)  =m_{2}^{2},$ or
$c=m_{2}^{2},\;b=0,$ with $F\left(  \phi\right)  =m_{2}/\left(  2\left\vert
\phi\right\vert \right)  .$

Note that solving for $A\left(  x_{5},p_{5}\right)  $ from the expression
(\ref{A5bc5}) involves branch cuts. Therefore $A$ may need to be re-defined up
to various signs in neighboring patches of phase space $\left(  x_{5}^{\mu
},p_{5}^{\mu}\right)  $ so as to be able to cover continuously the phase space
in $d+2$ dimensions $\left(  X^{M},P^{M}\right)  .$ We leave this issue open
here, but we return to make comments about it in section (\ref{solving5}) in
the context of obtaining solutions of the constrained system (\ref{g5L}) by
using dualities.

This example yields another interesting shadow, namely the relativistic
harmonic oscillator, with the constraint $Q\left(  x,p\right)  =\left(
p^{2}+\omega^{2}x^{2}+\cdots\right)  =0$, when $V\left(  x^{2}\right)
=c\left(  x^{2}\right)  ^{b}$ is taken with the special constants
$c=\omega^{2},b=1$. When the background fields \textquotedblleft$\cdots
$\textquotedblright\ vanish, the physical sector of the constrained
relativistic harmonic oscillator in $\left(  d-1\right)  +1$ dimensions is the
same as the unconstrained non-relativistic harmonic oscillator in $\left(
d-1\right)  $ space dimensions$.$ This was demonstrated in \cite{RelHarmOsc}
by the following canonical transformation for the timelike phase space,%
\begin{equation}
\omega x^{0}\left(  \tau\right)  =\pm\sqrt{2(h\left(  \tau\right)  -E_{0}%
)}\sin\left(  t\left(  \tau\right)  \right)  ,\;p^{0}\left(  \tau\right)
=\pm\sqrt{2(h\left(  \tau\right)  -E_{0})}\cos\left(  t\left(  \tau\right)
\right)  , \label{NRharmonicOsc}%
\end{equation}
where $E_{0}$ is a constant. Choosing the gauge $t\left(  \tau\right)  =\tau,$
and solving the relativistic constraint for $h$ from $\left(  p^{2}+\omega
^{2}x^{2}\right)  =\left(  \mathbf{p}^{2}+\omega^{2}\mathbf{r}^{2}\right)
-2(h\left(  \tau\right)  -E_{0})=0,$ reduces the problem to only the physical
phase space degrees of freedom $\left(  \mathbf{r,p}\right)  $ with the
Hamiltonian $h=\frac{1}{2}\left(  \mathbf{p}^{2}+\mathbf{r}^{2}\right)
+E_{0},$ which describes the non-relativistic oscillator.

The more general case of the constrained system, $p^{2}+c\left(  x^{2}\right)
^{b}=0$ with general $b,c,$ is a rather complicated problem whose solution was
not known until now. But we will show in section (\ref{solving5}) that the
duality methods discussed in this paper will provide the means to solve it
analytically. The same methods apply also to the even more general case
$p^{2}+V\left(  x^{2}\right)  =0,$ with any $V\left(  x^{2}\right)  $, thus
demonstrating the power of our duality methods derived from 2T-physics.

\section{Sp$\left(  2,R\right)  $ gauge transformations and dualities
\label{dualities}}

As was already discussed in the previous section, in the case of 2T-physics,
the 5 gauges that were studied above correspond to different physical systems
in 1T-physics. However, all of them are holographic shadows of the same theory
in 2T-physics, meaning that each shadow contains all the gauge invariant
2T-physics by virtue of being just a gauge choice. So, there must exist
duality relations that map the 1T shadows into each other. The 1T-physics
observed in the respective shadows, although they have different 1T-physics
interpretations, must be related to each other by dualities, and must describe
the same gauge invariant content of the 2T-physics theory from which the
shadows are derived. This is hidden information among 1T-physics systems that
1T-physics does not provide systematically, but is a prediction of the
2T-physics formulation which can be tested and verified directly in 1T-physics
by using our dualities.

These dualities have to be Sp$(2,R)$ gauge transformations acting on the 2T
phase space $\left(  X^{M},P_{M}\right)  $. The parameters of these
transformations are local on the worldline parametrized by $\tau$, but since
the interest is in transforming one fixed gauge $\left(  X_{i}^{M}%
,P_{iM}\right)  $ to another $\left(  X_{j}^{M},P_{jM}\right)  $, the
parameters of the gauge transformation would be written in terms of the $\tau
$-dependent phase space coordinates of the corresponding 1T shadows
themselves. So, these Sp$\left(  2,R\right)  $ gauge transformations must take
the form of canonical transformations among the 1T shadows. In this section we
will illustrate these ideas by considering a special subset of canonical
transformations that connect the five shadows to each other.

We can compute algebraically the gauge transformations that relate the
phase-space degrees of freedom of any two shadows to each other. To do this we
consider the gauge transformations generated by the Sp$\left(  2,R\right)  $
charges of the form given in Eqs.(\ref{Q11simplified}-\ref{cond-homog})
\begin{equation}
Q_{11}=X\cdot X,\;\;Q_{12}=X\cdot P,\;\;Q_{22}=P\cdot P+\cdots
\end{equation}
The gauge transformations (\ref{gtrans1}-\ref{gtrans4}) generated by the first
two charges $Q_{11},Q_{12}$ are linear Sp$(2,R)$ transformations that are
written as a $2\times2$ matrix of the general form (\ref{2TFiniteLocalTransf}%
), but with $\beta=0,$ namely $\left(
\genfrac{}{}{0pt}{}{\alpha}{\mathcal{\gamma}}%
\genfrac{}{}{0pt}{}{0}{1/\alpha}%
\right)  ,$ because the transformation generated by $Q_{22}$ is not included
in the present duality discussion. In fact, the Sp$(2,R)$ transformations
generated by $Q_{22}$ are non-linear in the capital $\left(  X,P\right)  ,$
because $Q_{22}$ generally contains background fields denoted by
\textquotedblleft$\cdots$\textquotedblright\ that we wish to keep as general
as possible in our discussion. By contrast, since $Q_{11},Q_{12}$ are purely
quadratic in phase space these charges induce only linear transformations via
(\ref{gtrans1}-\ref{gtrans4}). Recall that $X\cdot X=X\cdot P=0,$ are already
satisfied explicitly in each shadow. The gauge transformations generated by
$\left(  X\cdot X\right)  ,\left(  X\cdot P\right)  $ close into a subgroup of
Sp$\left(  2,R\right)  $ and hence they act within the physical space that
already satisfies these constraints in each shadow, namely $X\cdot X=X\cdot
P=0$. Furthermore, within this restricted phase space, the subgroup of
transformations generated by $\left(  Q_{11},Q_{12}\right)  $ transform
$Q_{22}$, and the corresponding gauge field $A^{22}$, only by an overall
scaling $Q_{22}\rightarrow\alpha^{-2}Q_{22}$ and $A^{22}\rightarrow\alpha
^{2}A^{22},$ as seen from Eqs.(\ref{A22Transf}) at $\beta=0$. So the remaining
term in the gauge fixed action $A^{22}Q_{22}$ is invariant while compatible
with being in the subspace $X\cdot X=X\cdot P=0$. Hence these duality
transformations change one shadow into another without changing the remaining
constraint $Q_{22}$ (which is eventually applied as $Q_{22}=0$ in each
shadow). This is the reason that the duality transformations we discuss take
the form of $\ 2\times2$ matrices as in (\ref{2TFiniteLocalTransf}), with
$\alpha\left(  x,p\right)  ,\mathcal{\gamma}\left(  x,p\right)  $ taken as
functions of 1T phase space $\left(  x\left(  \tau\right)  ,p\left(
\tau\right)  \right)  $, and with $\beta=0.$

We can, therefore, use the matrix method to find explicitly the duality
transformation constructed from phase space, with $\alpha\left(  x,p\right)
,\mathcal{\gamma}\left(  x,p\right)  $, and $\beta=0,$ given that the gauge
fixed forms of the shadows that we want to relate to each other by $2\times2$
matrices are already specified in Eqs.(\ref{g1},\ref{g2},\ref{g3}%
,\ref{g4},\ref{g5}) as $\left(
\genfrac{}{}{0pt}{}{X}{P}%
\right)  $ doublets in each direction $M$. These dualities must also be
canonical transformations since they are written only in terms of phase space
degrees of freedom and map one canonical phase space to another canonical
phase space.

\subsection{The general duality transformation}

In order to find the duality transformation we consider the general linear
form of an element of Sp$(2,R)$ given by Eq.(\ref{2TFiniteLocalTransf}) with
$\beta=0$, as explained above. We recall that the gauge fixed $X^{M}$ and
$P^{M}$ for each shadow are given explicitly as doublets in section
\ref{5gauges}. First we setup a $2\times2$ matrix transformation between two
shadows $i$ and $j$ for every direction $M$
\begin{equation}
\left(
\begin{array}
[c]{c}%
X_{j}^{M}\left(  x_{j}^{\mu},p_{j}^{\mu}\right) \\
P_{j}^{M}\left(  x_{j}^{\mu},p_{j}^{\mu}\right)
\end{array}
\right)  =\left(
\begin{array}
[c]{cc}%
\alpha\left(  \tau\right)  & 0\\
\mathcal{\gamma}\left(  \tau\right)  & \alpha^{-1}\left(  \tau\right)
\end{array}
\right)  \left(
\begin{array}
[c]{c}%
X_{i}^{M}\left(  x_{i}^{\mu},p_{i}^{\mu}\right) \\
P_{i}^{M}\left(  x_{i}^{\mu},p_{i}^{\mu}\right)
\end{array}
\right)  . \label{i-to-j-M}%
\end{equation}
Note that shadow $i$ is parametrized in terms of phase space $\left(
x_{i}^{\mu},p_{i}^{\mu}\right)  $ while shadow $j$ is parametrized in terms of
phase space $\left(  x_{j}^{\mu},p_{j}^{\mu}\right)  .$ Next, we solve for
$\alpha(\tau)$ and $\mathcal{\gamma}(\tau)$ such that the $2\times2$ matrix
corresponds to the \ gauge transformation from one fixed gauge to another.
This is done by using some doublets in convenient directions $M$ that contain
information on how the gauge was fixed in shadows $i$ and $j$. For example, in
the case of shadow \#1 in Eq.(\ref{g1}) the doublet $M=+^{\prime}$ is
convenient since it is fully fixed to $X_{1}^{+^{\prime}}=1$ and
$P_{1}^{+^{\prime}}=0.$ Finally, we express the transformation parameters
$\alpha\left(  x_{i}^{\mu},p_{i}^{\mu}\right)  ,\mathcal{\gamma}\left(
x_{i}^{\mu},p_{i}^{\mu}\right)  $ in terms of the degrees of freedom $\left(
x_{i}^{\mu},p_{i}^{\mu}\right)  $ of the shadow of origin. Having fixed the
matrix, the canonical transformation $\left(  x_{j}^{\mu},p_{j}^{\mu}\right)
\leftarrow\left(  x_{i}^{\mu},p_{i}^{\mu}\right)  $ is now obtained by taking
the $M=\mu$ direction in Eq.(\ref{i-to-j-M}) as shown in the example
(\ref{nzToZ}). Using this procedure, we found the explicit canonical
transformations in section \ref{explicit}, and listed the results of section
\ref{explicit} in the table of Eq.(\ref{canonList}).

To show that these gauge transformations are canonical transformations
(including time and Hamiltonian), we must also show that the canonical
structure holds up to a total derivative
\begin{equation}
\dot{x}_{j}^{\mu}p_{j\mu}=\dot{x}_{i}^{\mu}p_{i\mu}+\frac{d}{d\tau}%
\Lambda_{ji}\left(  \tau\right)  . \label{Lambda-ij}%
\end{equation}
When this is true, the invariance of Poisson brackets for any two quantities
$\left\{  A,B\right\}  $ is also guaranteed when they are evaluated as
derivatives in terms of either shadow. The validity of Eq.(\ref{Lambda-ij})
can be checked by using our explicit transformations in section \ref{explicit}%
. However, the result (\ref{Lambda-ij}) is already guaranteed by the canonical
structures that descended from $d+2$ dimensions. The essential observation is
that we can equate two gauge fixed forms of the same gauge invariant. That is,
the gauge invariant Lagrangian of the 2T-theory (\ref{2TAction1}) can be
equated to its five gauge fixed versions in the five shadows of section
\ref{5gauges}. Consider two shadows $i$ and $j$ which satisfy the following
relations due to the gauge invariance of the Lagrangian
\begin{align}
L  &  =\dot{X}^{M}P_{M}+\frac{1}{2}A^{kl}Q_{kl}\\
&  =\dot{x}_{i}^{\mu}p_{i\mu}+\frac{1}{2}A_{i}^{22}\left(  \tau\right)
Q_{22}\left(  x_{i}^{\mu},p_{i}^{\mu}\right)  +\frac{d\Lambda_{i}}{d\tau
},\label{Li}\\
&  =\dot{x}_{j}^{\mu}p_{j\mu}+\frac{1}{2}A_{j}^{22}\left(  \tau\right)
Q_{22}\left(  x_{j}^{\mu},p_{j}^{\mu}\right)  +\frac{d\Lambda_{j}}{d\tau}.
\label{Lj}%
\end{align}
We have shown in (\ref{A22Transf}) that under the gauge transformation
(\ref{i-to-j-M}), $A_{i}^{22}$ is related to $A_{j}^{22},$ by $A_{j}%
^{22}=\alpha^{2}A_{i}^{22},$ since $\beta=0.$ Similarly $Q_{22}\left(
x_{j}^{\mu},p_{j}^{\mu}\right)  $ must be related to $Q_{22}\left(  x_{i}%
^{\mu},p_{i}^{\mu}\right)  $ by the inverse transformation, $Q_{22}\left(
x_{j}^{\mu},p_{j}^{\mu}\right)  =\alpha^{-2}Q_{22}\left(  x_{i}^{\mu}%
,p_{i}^{\mu}\right)  ,$ so that the combination $A^{22}Q_{22}$ is invariant
under the gauge transformation (\ref{i-to-j-M})
\begin{equation}
A_{j}^{22}\left(  \tau\right)  Q_{22}\left(  x_{j}^{\mu},p_{j}^{\mu}\right)
=A_{i}^{22}\left(  \tau\right)  Q_{22}\left(  x_{i}^{\mu},p_{i}^{\mu}\right)
.
\end{equation}
This is because $A^{22}$ and $Q_{22}$ are both members of the adjoint
representation whose dot product $A^{kl}Q_{kl}$ remains invariant under the
non-derivative parts of any gauge transformation. Since $Q_{11}$ and $Q_{12}$
are already identically zero, then the above relation must hold under the
gauge transformation (\ref{i-to-j-M}). After taking into account this
identity, equating the expressions in Eqs.(\ref{Li},\ref{Lj}) establishes that
the canonical transformation described in Eq.(\ref{Lambda-ij}) is guaranteed
and predicts that the total derivative $d\Lambda_{ij}/d\tau$ must be given by%
\begin{equation}
\Lambda_{ji}\left(  \tau\right)  =\Lambda_{j}\left(  x_{j}^{\mu}\left(
\tau\right)  ,p_{j}^{\mu}\left(  \tau\right)  \right)  -\Lambda_{i}\left(
x_{i}^{\mu}\left(  \tau\right)  ,p_{i}^{\mu}\left(  \tau\right)  \right)
\end{equation}
where the $\Lambda_{i}\left(  x_{i}^{\mu}\left(  \tau\right)  ,p_{i}^{\mu
}\left(  \tau\right)  \right)  $ have been computed and given explicitly for
each $i$ in Eqs.(\ref{g1},\ref{g2},\ref{g3},\ref{lambda4},\ref{lambda5}). The
reader may verify this expression also directly from the five explicit
canonical transformations $x_{j}^{\mu}=\mathcal{X}_{j}^{\mu}\left(
x_{i},p_{i}\right)  ,\;p_{j\mu}=\mathcal{P}_{j\mu}\left(  x_{i},p_{i}\right)
$ given in the next section.

Finally, we should remark that a further consistency check for the dualities
is to verify that the different constraints of the gauge-fixed shadows
transform into each other up to overall factors as indicated in
Eqs.(\ref{dualConstraints},\ref{dualConstraints2}). This is evident from the
remarks made above on how the generator $Q_{22}$ transforms with an overall
factor $\alpha^{-2}$ under the gauge transformation (\ref{i-to-j-M}), and
noting that this factor can be written as $\alpha^{-2}=\left(  X_{i}%
^{+^{\prime}}/X_{j}^{+^{\prime}}\right)  ^{2}.$

\subsection{Explicit canonical transformations \label{explicit}}

Now we give explicitly each one of the duality relations between the five
shadows under study. They were obtained through the methods described above.

\subsubsection{Dualities $\left(  1\leftrightarrow2\right)  $}

For the duality $\left(  1\leftarrow2\right)  $ between shadows 1\&2 we first
consider the transformation (\ref{i-to-j-M}) by using Eqs.(\ref{g1},\ref{g2})
in the direction $M=0$ and obtain the relation
\begin{equation}
\left(
\begin{array}
[c]{c}%
1\\
0
\end{array}
\right)  =\left(
\begin{array}
[c]{cc}%
\alpha\left(  \tau\right)  & 0\\
\gamma\left(  \tau\right)  & \alpha^{-1}\left(  \tau\right)
\end{array}
\right)  \left(
\begin{array}
[c]{c}%
\frac{1+a}{2a}\\
\frac{-m_{2}^{2}}{2\left(  x_{2}\cdot p_{2}\right)  a}%
\end{array}
\right)
\end{equation}
From this equation we determine both $\alpha=2a/\left(  1+a\right)  $ and
$\gamma=\frac{m_{2}^{2}}{2\left(  x_{2}\cdot p_{2}\right)  a}$ as functions of
$\left(  x_{2},p_{2}\right)  .$ We insert them back in Eq.(\ref{i-to-j-M}) to
obtain the canonical transformation $\left(  1\leftarrow2\right)  $ as follows%

\begin{equation}%
\begin{tabular}
[c]{|c|}\hline
massless relativistic (1) $\leftarrow~$massive relativistic (2)\\\hline
\multicolumn{1}{|l|}{$\overset{}{%
\begin{array}
[c]{c}%
\left(
\begin{array}
[c]{c}%
x_{1}^{\mu}\\
p_{1}^{\mu}%
\end{array}
\right)  =\left(
\begin{array}
[c]{cc}%
\frac{2a}{1+a} & 0\\
\frac{m_{2}^{2}}{2\left(  x_{2}\cdot p_{2}\right)  a} & \frac{1+a}{2a}%
\end{array}
\right)  \left(
\begin{array}
[c]{c}%
x_{2}^{\mu}\\
p_{2}^{\mu}%
\end{array}
\right)  \medskip\\
a\equiv\sqrt{1+\frac{m_{2}^{2}x_{2}^{2}}{\left(  x_{2}\cdot p_{2}\right)
^{2}}}\text{ \ , \ }\ \Lambda_{12}=\left(  a-1\right)  \left(  x_{2}\cdot
p_{2}\right)
\end{array}
}$}\\\hline
\end{tabular}
\ \ \label{massivemassless1}%
\end{equation}
The inverse transformation $\left(  2\leftarrow1\right)  $ is given by the
inverse matrix, but $\alpha$ and $\gamma$ must be rewritten in terms of
$\left(  x_{1}^{\mu},p_{1}^{\mu}\right)  .$ After some algebra one gets,
$\frac{2a}{1+a}=1+\frac{m_{2}^{2}x_{1}^{2}}{4\left(  x_{1}\cdot p_{1}\right)
^{2}},$ which yields the inverse matrix as follows%

\begin{equation}%
\begin{tabular}
[c]{|c|}\hline
massive relativistic (2)$\;\leftarrow~$massless relativistic (1)\\\hline
\multicolumn{1}{|l|}{$\overset{}{%
\begin{array}
[c]{c}%
\left(
\begin{array}
[c]{c}%
x_{2}^{\mu}\\
p_{2}^{\mu}%
\end{array}
\right)  =\left(
\begin{array}
[c]{cc}%
\left(  1+\frac{m_{2}^{2}x_{1}^{2}}{4\left(  x_{1}\cdot p_{1}\right)  ^{2}%
}\right)  ^{-1} & 0\\
-\frac{m_{2}^{2}}{2\left(  x_{1}\cdot p_{1}\right)  } & \left(  1+\frac
{m_{2}^{2}x_{1}^{2}}{4\left(  x_{1}\cdot p_{1}\right)  ^{2}}\right)
\end{array}
\right)  \left(
\begin{array}
[c]{c}%
x_{1}^{\mu}\\
p_{1}^{\mu}%
\end{array}
\right)  \medskip\\
\Lambda_{21}=-\frac{2m_{2}^{2}x_{1}^{2}\left(  x_{1}\cdot p_{1}\right)
}{4\left(  x_{1}\cdot p_{1}\right)  ^{2}+m_{2}^{2}x_{1}^{2}}%
\end{array}
}$}\\\hline
\end{tabular}
\ \label{maslessmassive1}%
\end{equation}

\subsubsection{Dualities $\left(  1\leftrightarrow3\right)  $}

To determine the duality transformation $\left(  1\leftarrow3\right)  $
between shadows 1\&3 we first consider the transformation (\ref{i-to-j-M}) by
using Eqs.(\ref{g1},\ref{g3}) in the direction $M=0$ and obtain the relation
\begin{equation}
\left(
\begin{array}
[c]{c}%
1\\
0
\end{array}
\right)  =\left(
\begin{array}
[c]{cc}%
\alpha\left(  \tau\right)  & 0\\
\mathcal{\gamma}\left(  \tau\right)  & \alpha^{-1}\left(  \tau\right)
\end{array}
\right)  \left(
\begin{array}
[c]{c}%
t_{3}\\
m_{3}%
\end{array}
\right)
\end{equation}
which determines $\alpha\mathcal{=}\left(  t_{3}\right)  ^{-1}$ and
$\mathcal{\gamma=}-m_{3}.$ We insert this back in Eq.(\ref{i-to-j-M}) to
obtain the canonical transformation $\left(  1\leftarrow3\right)  $ as
follows
\begin{equation}%
\begin{tabular}
[c]{|c|}\hline
massless relativistic (1)$~\leftarrow\text{massive nonrelativistic (3)}%
$\\\hline
$\overset{}{%
\begin{array}
[c]{c}%
\left(
\begin{array}
[c]{c}%
\mathbf{x}_{1}\\
\mathbf{p}_{1}%
\end{array}
\right)  =\left(
\begin{array}
[c]{cc}%
\left(  t_{3}\right)  ^{-1} & 0\\
-m_{3} & t_{3}%
\end{array}
\right)  \left(
\begin{array}
[c]{c}%
\mathbf{x}_{3}\\
\mathbf{p}_{3}%
\end{array}
\right)  \medskip\\
\left(
\begin{array}
[c]{c}%
x_{1}^{0}\\
p_{1}^{0}%
\end{array}
\right)  =\left(
\begin{array}
[c]{cc}%
\left(  t_{3}\right)  ^{-1} & 0\\
-m_{3} & t_{3}%
\end{array}
\right)  \left(
\begin{array}
[c]{c}%
s\\
0
\end{array}
\right)  \medskip\\
s^{2}\equiv\left(  \mathbf{x}_{3}\right)  ^{2}-\frac{2t_{3}}{m_{3}}%
\mathbf{x}_{3}\cdot\mathbf{p}_{3}+\frac{2\left(  t_{3}\right)  ^{2}}{m_{3}%
}h_{3}\\
\Lambda_{13}=t_{3}h_{3}-\mathbf{x}_{3}\cdot\mathbf{p}_{3}%
\end{array}
}$\\\hline
\end{tabular}
\ \ \ \label{MassiveNRToMassless1}%
\end{equation}
The inverse transformation $\left(  3\leftarrow1\right)  $ is given by the
inverse matrix, but $\alpha$ must be rewritten in terms of $\left(  x_{1}%
^{\mu},p_{1}^{\mu}\right)  $ as $\alpha=-m_{3}x_{1}^{0}/p_{1}^{0},$ so that
the inverse transformation takes the form%
\begin{equation}%
\begin{tabular}
[c]{|c|}\hline
$\text{massive nonrelativistic (3)}\leftarrow\text{massless relativistic (1)}%
$\\\hline
$\overset{}{%
\begin{array}
[c]{c}%
\left(
\begin{array}
[c]{c}%
\mathbf{x}_{3}\\
\mathbf{p}_{3}%
\end{array}
\right)  =\left(
\begin{array}
[c]{cc}%
-\frac{p_{1}^{0}}{m_{3}x_{1}^{0}} & 0\\
m_{3} & -\frac{m_{3}x_{1}^{0}}{p_{1}^{0}}%
\end{array}
\right)  \left(
\begin{array}
[c]{c}%
\mathbf{x}_{1}\\
\mathbf{p}_{1}%
\end{array}
\right)  \medskip\\
t_{3}=-\frac{p_{1}^{0}}{m_{3}x_{1}^{0}}\\
h_{3}=\frac{m_{3}}{2}\left(  \mathbf{x}_{1}^{2}+\left(  x_{1}^{0}\right)
^{2}\right)  -\frac{m_{3}x_{1}^{0}}{p_{1}^{0}}\mathbf{x}_{1}\cdot
\mathbf{p}_{1}\medskip\\
\Lambda_{31}=-\frac{p_{1}^{0}}{2x_{1}^{0}}x_{1}^{\mu}x_{1\mu}%
\end{array}
}$\\\hline
\end{tabular}
\ \ \ \label{masslessToMassiveNR1}%
\end{equation}

\subsubsection{Dualities $\left(  2\leftrightarrow3\right)  $}

To determine the duality transformation $\left(  3\leftarrow2\right)  $
between shadows 2\&3 we can use shadow \#1 as an intermediate step since we
already know the transformations back and forth $\left(  1\leftrightarrow
2\right)  $ and $\left(  1\leftrightarrow3\right)  .$ Hence we construct
$\left(  3\leftarrow2\right)  $ via the steps $\left(  1\leftarrow2\right)  $
followed by $\left(  3\leftarrow1\right)  .$ This gives the following explicit
transformation for $\left(  3\leftarrow2\right)  $
\begin{equation}%
\begin{tabular}
[c]{|c|}\hline
$\text{massive nonrelativistic (3)}\leftarrow\text{massive relativistic (2)}%
$\\\hline
\multicolumn{1}{|l|}{$\overset{}{%
\begin{array}
[c]{c}%
\left(
\begin{array}
[c]{c}%
\mathbf{x}_{3}\\
\mathbf{p}_{3}%
\end{array}
\right)  =\left(
\begin{array}
[c]{cc}%
-\frac{1}{m_{3}}w\left(  x_{2},p_{2}\right)  & 0\\
\frac{p_{2}^{0}}{x_{2}^{0}}\frac{m_{3}}{w\left(  x_{2},p_{2}\right)  } &
-\frac{m_{3}}{w\left(  x_{2},p_{2}\right)  }%
\end{array}
\right)  \left(
\begin{array}
[c]{c}%
\mathbf{x}_{2}\\
\mathbf{p}_{2}%
\end{array}
\right)  \medskip\\
w\left(  x_{2},p_{2}\right)  \equiv\left(  \frac{m_{2}^{2}}{2\left(
x_{2}\cdot p_{2}\right)  a}+\frac{1+a}{2a}\frac{p_{2}^{0}}{x_{2}^{0}}\right)
,\;a\equiv\sqrt{1+\frac{m_{2}^{2}x_{2}^{2}}{\left(  x_{2}\cdot p_{2}\right)
^{2}}}\medskip\\
t_{3}=-\frac{1}{m_{3}}w\left(  x_{2},p_{2}\right)  \frac{1+a}{2a}\\
h_{3}=\frac{m_{3}}{w\left(  x_{2},p_{2}\right)  }\left(  \frac{p_{2}^{0}%
}{x_{2}^{0}}\frac{a~x_{2}^{2}}{1+a}-a\left(  x_{2}\cdot p_{2}\right)  \right)
~\medskip\\
\Lambda_{32}=\frac{a-1}{2}\left(  x_{2}\cdot p_{2}\right)  -\frac{p_{2}^{0}%
}{2x_{2}^{0}}x_{2}^{2}%
\end{array}
}$}\\\hline
\end{tabular}
\ \ \ \label{mToM-transform1}%
\end{equation}
The inverse transformation $\left(  2\leftarrow3\right)  $ is built in the
same manner, with the result%
\begin{equation}%
\begin{tabular}
[c]{|c|}\hline
$\text{massive relativistic (2)}\leftarrow\text{massive nonrelativistic (3)}%
$\\\hline
\multicolumn{1}{|l|}{$\overset{}{%
\begin{array}
[c]{c}%
\left(
\begin{array}
[c]{c}%
\mathbf{x}_{2}\\
\mathbf{p}_{2}%
\end{array}
\right)  =\left(
\begin{array}
[c]{cc}%
\left(  t_{3}+\frac{m_{2}^{2}}{2m_{3}}\frac{\mathbf{x}_{3}\cdot\mathbf{p}%
_{3}-t_{3}h_{3}}{\left(  2t_{3}h_{3}-\mathbf{x}_{3}\cdot\mathbf{p}_{3}\right)
^{2}}\right)  ^{-1} & 0\\
-\left(  m_{3}+\frac{m_{2}^{2}h_{3}}{2\left(  2t_{3}h_{3}-\mathbf{x}_{3}%
\cdot\mathbf{p}_{3}\right)  ^{2}}\right)  & \left(  t_{3}+\frac{m_{2}^{2}%
}{2m_{3}}\frac{\mathbf{x}_{3}\cdot\mathbf{p}_{3}-t_{3}h_{3}}{\left(
2t_{3}h_{3}-\mathbf{x}_{3}\cdot\mathbf{p}_{3}\right)  ^{2}}\right)
\end{array}
\right)  \left(
\begin{array}
[c]{c}%
\mathbf{x}_{3}\\
\mathbf{p}_{3}%
\end{array}
\right)  \medskip\\
x_{2}^{0}=\left(  t_{3}+\frac{m_{2}^{2}}{2m_{3}}\frac{\mathbf{x}_{3}%
\cdot\mathbf{p}_{3}-t_{3}h_{3}}{\left(  2t_{3}h_{3}-\mathbf{x}_{3}%
\cdot\mathbf{p}_{3}\right)  ^{2}}\right)  ^{-1}s\left(  x_{3},p_{3}\right) \\
p_{2}^{0}=-\left(  m_{3}+\frac{m_{2}^{2}h_{3}}{2\left(  2t_{3}h_{3}%
-\mathbf{x}_{3}\cdot\mathbf{p}_{3}\right)  ^{2}}\right)  s\left(  x_{3}%
,p_{3}\right)  ~\medskip\\
s\left(  x_{3},p_{3}\right)  \equiv\pm\sqrt{\mathbf{x}_{3}^{2}-\frac{2t_{3}%
}{m_{3}}\mathbf{x}_{3}\cdot\mathbf{p}_{3}+\frac{2t_{3}^{2}}{m_{3}}h_{3}}\\
\Lambda_{23}=\left(  x_{2}\cdot p_{2}\right)  \left(  a\left(  x_{2}%
,p_{2}\right)  -1\right)  -\left(  \mathbf{x}_{3}\cdot\mathbf{p}_{3}%
-t_{3}h_{3}\right)
\end{array}
}$}\\\hline
\end{tabular}
\ \ \ \label{Mtom-transform1}%
\end{equation}

\subsubsection{Dualities $\left(  1\leftrightarrow4\right)  $}

For the duality $\left(  1\leftarrow4\right)  $ between shadows 1\&4 we first
consider the transformation (\ref{i-to-j-M}) by using Eqs.(\ref{g1},\ref{g4})
in the direction $M=0$ and obtain the relation
\begin{equation}
\left(
\begin{array}
[c]{c}%
1\\
0
\end{array}
\right)  =\left(
\begin{array}
[c]{cc}%
\alpha & 0\\
\gamma & \alpha^{-1}%
\end{array}
\right)  \left(
\begin{tabular}
[c]{c}%
$\frac{1}{\sqrt{-4m_{4}h_{4}}}\left[  \left\vert \mathbf{x}_{4}\right\vert
\sqrt{-2m_{4}h_{4}}\sin u+\mathbf{x}_{4}\cdot\mathbf{p}_{4}\left(  \cos
u+1\right)  \right]  ~$\\
$\frac{1}{\sqrt{-4m_{4}h_{4}}}\left[  2m_{4}h_{4}\left\vert \mathbf{x}%
_{4}\right\vert +m_{4}e_{4}^{2}\left(  \cos u+1\right)  \right]  $%
\end{tabular}
\ \ \right)
\end{equation}
This determines both $\alpha$ and $\gamma$ as functions of $\left(
\mathbf{x}_{4},\mathbf{p}_{4},t_{4},h_{4}\right)  $%
\begin{equation}
\alpha=\frac{\sqrt{-4m_{4}h_{4}}}{\left\vert \mathbf{x}_{4}\right\vert
\sqrt{-2m_{4}h_{4}}\sin u+\mathbf{x}_{4}\cdot\mathbf{p}_{4}\left(  \cos
u+1\right)  },\;\gamma=-\frac{2m_{4}h_{4}\left\vert \mathbf{x}_{4}\right\vert
+m_{4}e_{4}^{2}\left(  \cos u+1\right)  }{\sqrt{-4m_{4}h_{4}}} \label{ac41}%
\end{equation}
We insert them back in Eq.(\ref{i-to-j-M}) to obtain the canonical
transformation $\left(  1\leftarrow4\right)  $ as follows%

\begin{equation}%
\begin{tabular}
[c]{|c|}\hline
massless relativistic (1) $\leftarrow~$H-atom (4)\\\hline
\multicolumn{1}{|l|}{$\overset{}{%
\begin{array}
[c]{c}%
\left(
\begin{array}
[c]{c}%
\mathbf{x}_{1}\\
\mathbf{p}_{1}%
\end{array}
\right)  =\left(
\begin{array}
[c]{cc}%
\alpha & 0\\
\gamma & \alpha^{-1}%
\end{array}
\right)  \left(
\begin{array}
[c]{c}%
\mathbf{x}_{4}\\
\mathbf{p}_{4}%
\end{array}
\right)  \medskip\\
\left(
\begin{array}
[c]{c}%
x_{1}^{0}\\
p_{1}^{0}%
\end{array}
\right)  =\left(
\begin{array}
[c]{cc}%
\alpha & 0\\
\gamma & \alpha^{-1}%
\end{array}
\right)  \left(
\begin{array}
[c]{c}%
\left\vert \mathbf{x}_{4}\right\vert \cos u-\frac{\mathbf{x}_{4}%
\cdot\mathbf{p}_{4}\mathbf{~}\sin u}{\sqrt{-2m_{4}h_{4}}}\\
-\frac{m_{4}e_{4}^{2}~\sin u}{\left\vert \mathbf{x}_{4}\right\vert
\sqrt{-2m_{4}h_{4}}}%
\end{array}
\right)  \medskip\\
u\equiv\frac{\sqrt{-2m_{4}h_{4}}}{m_{4}e_{4}^{2}}(\mathbf{x}_{4}%
\cdot\mathbf{p}_{4}-2h_{4}t_{4}).\\
\Lambda_{14}=3h_{4}t_{4}-2\left(  \mathbf{x}_{4}\cdot\mathbf{p}_{4}\right)
\end{array}
}$}\\\hline
\end{tabular}
\ \ \ \label{41}%
\end{equation}
To construct the inverse transformation we must rewrite the $\alpha,\gamma$ of
Eqs.(\ref{ac41}) as functions of $\left(  x_{1}^{\mu},p_{1}^{\mu}\right)  $ by
using (\ref{41})$.$ This gives%
\begin{equation}%
\begin{array}
[c]{c}%
\alpha\left(  x_{1},p_{1}\right)  =\frac{m_{4}e_{4}^{2}\left(  1+\frac{1}%
{2}\left(  x_{1}^{0}-\left\vert \mathbf{x}_{1}\right\vert \right)
^{2}\right)  }{\left(  \left\vert \mathbf{x}_{1}\right\vert p_{1}%
^{0}-\mathbf{x}_{1}\mathbf{\cdot p}_{1}\right)  \sqrt{2}L^{0^{\prime}0}}\\
\gamma\left(  x_{1},p_{1}\right)  =m_{4}e_{4}^{2}\left[  \frac{1}{\sqrt
{2}L^{0^{\prime}0}}-\frac{1}{\left\vert \mathbf{x}_{1}\right\vert \left(
\left\vert \mathbf{x}_{1}\right\vert p_{1}^{0}-\mathbf{x}_{1}\mathbf{\cdot
p}_{1}\right)  }\right] \\
\text{with }\sqrt{2}L^{0^{\prime}0}\equiv p_{1}^{0}\left(  1+\frac{1}%
{2}\left(  x_{1}^{0}-\left\vert \mathbf{x}_{1}\right\vert \right)
^{2}\right)  +x_{1}^{0}\left(  \left\vert \mathbf{x}_{1}\right\vert p_{1}%
^{0}-\mathbf{x}_{1}\mathbf{\cdot p}_{1}\right)
\end{array}
\end{equation}
The inverse transformation $\left(  4\leftarrow1\right)  $ is then%
\begin{equation}%
\begin{tabular}
[c]{|c|}\hline
H-atom (4) $\leftarrow~$massless relativistic (1)\\\hline
\multicolumn{1}{|l|}{$\overset{}{%
\begin{array}
[c]{c}%
\left(
\begin{array}
[c]{c}%
\mathbf{x}_{4}\\
\mathbf{p}_{4}%
\end{array}
\right)  =\left(
\begin{array}
[c]{cc}%
\alpha^{-1}\left(  x_{1},p_{1}\right)  & 0\\
-\gamma\left(  x_{1},p_{1}\right)  & \alpha\left(  x_{1},p_{1}\right)
\end{array}
\right)  \left(
\begin{array}
[c]{c}%
\mathbf{x}_{1}\\
\mathbf{p}_{1}%
\end{array}
\right)  \medskip\\
t_{4}=\alpha^{-1}\left(  x_{1},p_{1}\right) \\
h_{4}=-\frac{1}{2}x_{1}^{2}\gamma\left(  x_{1},p_{1}\right)  \medskip+\left(
x_{1}\cdot p_{1}\right)  \alpha\left(  x_{1},p_{1}\right) \\
\Lambda_{41}=-\Lambda_{14}%
\end{array}
}$}\\\hline
\end{tabular}
\ \label{1to4}%
\end{equation}

\subsubsection{Dualities $\left(  1\leftrightarrow5\right)  $ for general
$V\left(  x^{2}\right)  $}

For the duality $\left(  1\leftarrow5\right)  $ between shadows 1\&5 we first
consider the transformation (\ref{i-to-j-M}) by using Eqs.(\ref{g1},\ref{g5})
in the direction $M=0$ and obtain the relation
\begin{equation}%
\begin{array}
[c]{c}%
\left(
\begin{array}
[c]{c}%
1\\
0
\end{array}
\right)  =\left(
\begin{array}
[c]{cc}%
\alpha & 0\\
\gamma & \alpha^{-1}%
\end{array}
\right)  \left(
\begin{array}
[c]{c}%
A\\
\frac{A}{x^{2}}\left(  x_{5}\cdot p_{5}-\phi\right)
\end{array}
\right)  \medskip\\
\text{with }\left\{
\begin{array}
[c]{l}%
\phi\left(  x_{5},p_{5}\right)  \equiv\left(  x_{5}\cdot p_{5}\right)  \left(
1+x_{5}^{2}V\left(  x_{5}^{2}\right)  \left(  x_{5}\cdot p_{5}\right)
^{-2}\right)  ^{1/2}\\
A\left(  x_{5},p_{5}\right)  \equiv F\left(  \phi\right)  \sqrt{x_{5}^{2}}%
\exp\left[  -\frac{1}{2}\int^{x_{5}^{2}}\frac{du}{u}\left(  1-\phi
^{-2}uV\left(  u\right)  \right)  ^{-1/2}\right]
\end{array}
\right.
\end{array}
\label{A5-5}%
\end{equation}
This determines $\alpha\left(  x_{5},p_{5}\right)  =A^{-1}$ and $\gamma\left(
x_{5},p_{5}\right)  =-\frac{A}{x^{2}}\left(  x_{5}\cdot p_{5}-\phi\right)  $
as functions of $\left(  x_{5},p_{5}\right)  .$ We insert them back in
Eq.(\ref{i-to-j-M}) to obtain the canonical transformation $\left(
1\leftarrow5\right)  $ as follows%

\begin{equation}%
\begin{tabular}
[c]{|c|}\hline
massless relativistic (1) $\leftarrow~$relativistic potential
(5),\ $\text{general \ }V\left(  x^{2}\right)  $\\\hline
$\overset{}{%
\begin{array}
[c]{c}%
\left(
\begin{array}
[c]{c}%
x_{1}^{\mu}\\
p_{1}^{\mu}%
\end{array}
\right)  =\left(
\begin{array}
[c]{cc}%
A^{-1} & ~~0~~\\
-\frac{A}{x_{5}^{2}}\left(  x_{5}\cdot p_{5}-\phi\right)  & ~~A~~
\end{array}
\right)  \left(
\begin{array}
[c]{c}%
x_{5}^{\mu}\\
p_{5}^{\mu}%
\end{array}
\right)  \medskip\\
\Lambda_{15}=-\frac{1}{\phi}\int^{x_{5}^{2}}duV\left(  u\right)  \left(
1-\phi^{-2}uV\left(  u\right)  \right)  ^{-1/2}+2\int^{\phi}dz~z\frac{d}%
{dz}\ln F\left(  z\right)
\end{array}
}$\\\hline
\end{tabular}
\ \ \ \ \label{5to1}%
\end{equation}
To construct the inverse transformation we must rewrite $\alpha,\gamma,$ or
equivalently $A,\phi,$ in terms of $\left(  x_{1},p_{1}\right)  .$ To
construct this, it is useful to remember that the SO$\left(  d,2\right)  $
generator $L^{+^{\prime}-^{\prime}}$ is invariant under the Sp$\left(
2,R\right)  $ gauge transformations. It takes the form $L^{+^{\prime}%
-^{\prime}}=x_{1}\cdot p_{1}$ in shadow \#1 while it is given by
$L^{+^{\prime}-^{\prime}}=\phi\left(  x_{5},p_{5}\right)  $ in shadow \#5, but
due to the gauge invariance of $L^{+^{\prime}-^{\prime}}$ we have
\begin{equation}
L^{+^{\prime}-^{\prime}}=\phi=\left(  x_{5}\cdot p_{5}\right)  \left(
1+x_{5}^{2}V\left(  x_{5}^{2}\right)  \left(  x_{5}\cdot p_{5}\right)
^{-2}\right)  ^{1/2}=x_{1}\cdot p_{1}. \label{L+-5}%
\end{equation}
Thus, we obtain $\phi=x_{1}\cdot p_{1},$ and $x_{5}^{2}=Ax_{1}^{2},$ while
$A\left(  x_{1},p_{1}\right)  $ is determined implicitly by solving the
following algebraic equation (which is a rewriting of (\ref{A5}) after
inserting $x_{5}^{\mu}=Ax_{1}^{\mu}$ from (\ref{g5})).
\begin{equation}
\int^{x_{1}^{2}A^{2}}\frac{du}{u}\left(  1-\left(  x_{1}\cdot p_{1}\right)
^{-2}uV\left(  u\right)  \right)  ^{-1/2}=\ln\left[  x_{1}^{2}F^{2}\left(
x_{1}\cdot p_{1}\right)  \right]  ,\; \label{A5x1p1}%
\end{equation}
Inserting these results, we obtain the transformation $\left(  5\leftarrow
1\right)  $ as follows%
\begin{equation}%
\begin{tabular}
[c]{|c|}\hline
relativistic potential (5), general $V\left(  x^{2}\right)  $ $\leftarrow$
massless relativistic (1)$~$\\\hline
$\overset{}{%
\begin{array}
[c]{c}%
\left(
\begin{array}
[c]{c}%
x_{5}^{\mu}\\
p_{5}^{\mu}%
\end{array}
\right)  =\left(
\begin{array}
[c]{cc}%
A\left(  x_{1},p_{1}\right)  & ~~0~~\\
\frac{x_{1}\cdot p_{1}}{Ax_{1}^{2}}\left(  \sqrt{1-A^{2}x_{1}^{2}V\left(
A^{2}x_{1}^{2}\right)  \left(  x_{1}\cdot p_{1}\right)  ^{-2}}-1\right)  &
~~A^{-1}\left(  x_{1},p_{1}\right)  ~~
\end{array}
\right)  \left(
\begin{array}
[c]{c}%
x_{1}^{\mu}\\
p_{1}^{\mu}%
\end{array}
\right)  \medskip\\
\Lambda_{51}=\frac{1}{\phi}\int^{x_{5}^{2}}duV\left(  u\right)  \left(
1-\phi^{-2}uV\left(  u\right)  \right)  ^{-1/2}+2\int^{\phi}dz~z\frac{d}%
{dz}\ln F\left(  z\right)
\end{array}
}$\\\hline
\end{tabular}
\ \ \ \ \label{1to5}%
\end{equation}

\subsubsection{Dualities $\left(  1\leftrightarrow5\right)  $ for $V\left(
x^{2}\right)  =c\left(  x_{5}^{2}\right)  ^{b}$}

To be completely explicit, we specialize to $V\left(  x^{2}\right)  =c\left(
x^{2}\right)  ^{b}.$ Then from Eqs.(\ref{A5-5},\ref{L+-5},\ref{A5x1p1}) we
compute the explicit forms for $\phi,A$ written in terms of the phase spaces
of either shadow
\begin{equation}%
\begin{tabular}
[c]{|c|c|c|}\hline
$V=c\left(  x_{5}^{2}\right)  ^{b}$ & as function of $\left(  x_{5}%
,p_{5}\right)  $ & as function of$~\left(  x_{1},p_{1}\right)  $\\\hline
$\phi=$ & $\left(  x_{5}\cdot p_{5}\right)  \sqrt{1+c\left(  x_{5}^{2}\right)
^{1+b}\left(  x_{5}\cdot p_{5}\right)  ^{-2}}$ & $=\left(  x_{1}\cdot
p_{1}\right)  $\\\hline
$A=$ & $\left(  \left\vert x_{5}\cdot p_{5}\right\vert +\sqrt{\left(
x_{5}\cdot p_{5}\right)  ^{2}+c\left(  x_{5}^{2}\right)  ^{1+b}}\right)
^{\frac{1}{1+b}}\frac{F\left(  \phi\right)  }{c^{1/\left(  2+2b\right)  }}$ &
$\;\;=\left[  \frac{4\left(  x_{1}\cdot p_{1}\right)  ^{2}}{\left(  1+\left(
F^{2}\left(  \phi\right)  x_{1}^{2}\right)  ^{1+b}\right)  ^{2}}\right]
^{\frac{1}{2+2b}}\frac{F\left(  \phi\right)  }{c^{1/\left(  2+2b\right)  }}%
$\\\hline
\end{tabular}
\ \ \ \ \label{Aphi51}%
\end{equation}
A quick way of proving the equality of the two forms of $\phi$ is to use the
gauge invariance of the $L^{MN}$: then note that $L^{+^{\prime}-^{\prime}%
}=\phi$ when computed in shadow \#$5$ of Eq.(\ref{g5}), and $L^{+^{\prime
}-^{\prime}}=\left(  x_{1}\cdot p_{1}\right)  $ when evaluated in shadow \#1
of Eq.(\ref{g1}). Hence the result above for $\phi$ is obtained. Inserting
this in Eqs.(\ref{5to1}, \ref{1to5}) gives the duality transformations
$\left(  5\leftrightarrow1\right)  $ for the potential $V=c\left(
x^{2}\right)  ^{b}$ as follows%
\begin{equation}%
\begin{tabular}
[c]{|c|}\hline
massless relativistic (1) $\leftarrow~$relativistic potential
(5),\ $V=c\left(  x^{2}\right)  ^{b}$\\\hline
$\overset{}{\left(
\begin{array}
[c]{c}%
x_{1}^{\mu}\\
p_{1}^{\mu}%
\end{array}
\right)  =\left(
\begin{array}
[c]{cc}%
\left(  A\left(  x_{5},p_{5}\right)  \right)  ^{-1}\smallskip & ~~0~~\\
\frac{c\left(  x_{5}^{2}\right)  ^{b}~A\left(  x_{5},p_{5}\right)  }%
{\phi\left(  x_{5},p_{5}\right)  } & ~~A\left(  x_{5},p_{5}\right)  ~~
\end{array}
\right)  \left(
\begin{array}
[c]{c}%
x_{5}^{\mu}\\
p_{5}^{\mu}%
\end{array}
\right)  }\medskip$\\\hline
$\phi\equiv\left(  x_{5}\cdot p_{5}\right)  \sqrt{1+c\left(  x_{5}^{2}\right)
^{1+b}\left(  x_{5}\cdot p_{5}\right)  ^{-2}}$\\\hline
\end{tabular}
\ \ \ \ \ \ \label{5to1bc}%
\end{equation}
where the $F\left(  \phi\right)  $ that appears in $A\left(  x_{5}%
,p_{5}\right)  $ is an arbitrary function of its argument. Similarly, the
inverse transformation is
\begin{equation}%
\begin{tabular}
[c]{|c|}\hline
relativistic potential (5),\ $V=c\left(  x^{2}\right)  ^{b}$ $\leftarrow$
massless relativistic (1)$~$\\\hline
$\overset{}{%
\begin{array}
[c]{c}%
\left(
\begin{array}
[c]{c}%
x_{5}^{\mu}\\
p_{5}^{\mu}%
\end{array}
\right)  =\left(
\begin{array}
[c]{cc}%
A\left(  x_{1},p_{1}\right)  & ~~0~~\\
-\frac{2}{A\left(  x_{1},p_{1}\right)  }\frac{x_{1}\cdot p_{1}}{x_{1}^{2}%
}\frac{\left(  x_{1}^{2}F^{2}\left(  \phi\right)  \right)  ^{1+b}}{1+\left(
x_{1}^{2}F^{2}\left(  \phi\right)  \right)  ^{1+b}} & ~~\frac{1}{A\left(
x_{1},p_{1}\right)  }~~
\end{array}
\right)  \left(
\begin{array}
[c]{c}%
x_{1}^{\mu}\\
p_{1}^{\mu}%
\end{array}
\right)  ,\medskip\\
F\left(  \phi\right)  \text{ arbitrary function of its argument }\phi
=x_{1}\cdot p_{1},\;\text{ }\medskip\\
A\left(  x_{1},p_{1}\right)  \text{ given in Eq.(\ref{Aphi51}).}%
\end{array}
}$\\\hline
\end{tabular}
\ \label{1to5bc}%
\end{equation}
One can verify that these expressions satisfy
\begin{equation}
\left(  p_{5}^{2}+c\left(  x_{5}^{2}\right)  ^{b}+\cdots\right)
=A^{-2}\left(  p_{1}^{2}+\cdots\right)  ,
\end{equation}
where \textquotedblleft$\cdots$\textquotedblright\ represent the background
fields in the respective shadows. Then, by using the constraint in shadow \#1,
$p_{1}^{2}+\cdots=0,$ the constraint in shadow \#5 is automatically satisfied,
or vice-versa.

For a consistency check one may specialize to, $b=0,~c=m_{2}^{2}$ and
$F\left(  \phi\right)  =m_{2}/\left(  2\left\vert \phi\right\vert \right)  ,$
to see that the $\left(  1\leftrightarrow5\right)  $ duality expressions in
this subsection agree with the duality expressions $\left(  1\leftrightarrow
2\right)  $ given above$.$

\section{Duality invariants and hidden SO$\left(  d,2\right)  $
\label{invariants}}

In Eqs.(\ref{LMN},\ref{LMNinvar}) we argued that the SO$\left(  d,2\right)  $
generators $L^{MN}=X^{M}P^{N}-X^{N}P^{M}$ are gauge invariant under the
subgroup of Sp$\left(  2,R\right)  $ gauge transformations that correspond to
the duality transformations. Therefore, it is predicted that any function of
the $L^{MN}$ must be invariant under the duality transformations. Namely, if
one inserts the gauge fixed versions of $\left(  X_{i}^{M},P_{i}^{M}\right)  $
for $i=1,2,3,4,5,$ given in Eqs.(\ref{g1},\ref{g2},\ref{g3},\ref{g4}%
,\ref{g5}), into $L^{MN}=X^{M}P^{N}-X^{N}P^{M},$ they must equal each other%
\begin{equation}
L^{MN}=X_{1}^{[M}P_{1}^{N]}=X_{2}^{[M}P_{2}^{N]}=X_{3}^{[M}P_{3}^{N]}%
=X_{4}^{[M}P_{4}^{N]}=X_{5}^{[M}P_{5}^{N]}.
\end{equation}
We list below the different shadow-forms of the $L^{MN}=\left(  L^{+^{\prime
}-^{\prime}},L^{+^{\prime}\mu},L^{-^{\prime}\mu},L^{\mu\nu}\right)  $ for each
of the five shadows. These shadow $L^{MN}$ satisfy the Lie algebra of
SO$\left(  d,2\right)  $ under Poisson brackets computed in terms of $\left\{
x_{i}^{\mu},p_{i\mu}\right\}  $ in each shadow $i$. The closure of the
SO$\left(  d,2\right)  $ algebra in each shadow holds whether or not these
$L^{MN}$ are conserved, that is whether background fields are present or not.

Hence each shadow provides a new phase space representation of SO$\left(
d,2\right)  .$ One of these, shadow \#1, which we sometimes call the conformal
shadow, yields the familiar form of conformal transformations SO$\left(
d,2\right)  $ in $d$-dimensions. Namely $\delta_{\omega}x_{1}^{\mu}%
=\omega_{MN}\left\{  L^{MN},x_{1}^{\mu}\right\}  $, computed with the Poisson
brackets of shadow \#1, gives precisely the infinitesimal SO$\left(
d,2\right)  $ conformal transformations of $x_{1}^{\mu}.$ However, we claim
that all shadows, including the shadows with mass, also provide a
representation space for SO$\left(  d,2\right)  $, with an action of $L^{MN}$
on that phase space that is the dual of a conformal transformation in shadow \#1.

The $L^{MN}$ are not necessarily symmetry generators of the full theory, but
there are specialized forms of the theory in which they do generate the
natural SO$\left(  d,2\right)  $ rotation-type symmetry of flat $d+2$
dimensions. First, it is important to note that the $L^{MN}$ are generators of
an SO$\left(  d,2\right)  $ \textit{symmetry} of the first two constraints,
since they do commute with each other under Poisson brackets in the bulk,
$\left\{  L^{MN},\text{ }X\cdot X\right\}  =0$ and $\left\{  L^{MN},\text{
}X\cdot P\right\}  =0$. The third constraint, $\left(  P^{2}+\cdots\right)
=0,$ does not commute with $L^{MN}$ if the background fields \textquotedblleft%
$\cdots$\textquotedblright\ are present in general. But, in the case when all
background fields vanish \textquotedblleft$\cdots$\textquotedblright$=0$, the
third constraint, and indeed the full 2T Lagrangian is invariant under
SO$\left(  d,2\right)  $ transformations. Therefore, by Noether's theorem, the
$L^{MN}$ must be conserved generators $dL^{MN}/d\tau=0$ of the SO$\left(
d,2\right)  $ symmetry in that case$.$ Since the $L^{MN}$ are gauge (or
duality) invariants, then their shadows listed below, $L^{MN}=\left(
L^{+^{\prime}-^{\prime}},L^{+^{\prime}\mu},L^{-^{\prime}\mu},L^{\mu\nu
}\right)  $, must also be conserved in each shadow by virtue of being the
generators of a hidden SO$\left(  d,2\right)  $ symmetry in the case of no
background fields. Prior to the introduction of 2T-physics in 1998, the
presence of a hidden SO$\left(  d,2\right)  $ symmetry in shadows 2,3,4,5
without backgrounds had not been noticed in 1T-physics. This hidden symmetry
in the other shadows, which is a close cousin of the familiar conformal
symmetry in shadow \#1, is just as powerful and just as fundamental as
conformal symmetry. Indeed all forms of this hidden symmetry in the shadows
are the same symmetry of the bulk, which turns out to be realized \textit{in
the same irreducible unitary representation of SO}$\left(  d,2\right)  $ in
each shadow, as further discussed below at the classical or quantum levels.

In the following we use definitions for symbols given earlier in the paper,
which include,
\begin{equation}%
\begin{tabular}
[c]{|c|c|}\hline
Shadow \#\  & Definitions\\\hline
2 & $a\equiv\left[  1+m_{2}^{2}x_{2}^{2}\left(  x_{2}\cdot p_{2}\right)
^{-2}\right]  ^{1/2}~~$\\\hline
3 & $s^{2}\equiv\mathbf{x}_{3}^{2}-\frac{2t_{3}}{m_{3}}\mathbf{x}_{3}%
\cdot\mathbf{p}_{3}+\frac{2t_{3}^{2}}{m_{3}}h_{3}$\\\hline
4 & $u\equiv\frac{\sqrt{-2m_{4}h_{4}}}{m_{4}e_{4}^{2}}(\mathbf{x}%
_{4}\mathbf{\cdot p}_{4}-2h_{4}t_{4})$\\\hline
5 & $%
\begin{array}
[c]{c}%
\phi\equiv\left(  x_{5}\cdot p_{5}\right)  \left[  1+x_{5}^{2}V\left(
x_{5}^{2}\right)  \left(  x_{5}\cdot p_{5}\right)  ^{-2}\right]  ^{1/2}\\
A\equiv F\left(  \phi\right)  \sqrt{x_{5}^{2}}\exp\left[  -\frac{1}{2}%
\int^{x_{5}^{2}}\frac{du}{u}\left(  1-\frac{uV\left(  u\right)  }{\phi^{2}%
}\right)  ^{-1/2}\right]
\end{array}
$\\\hline
\end{tabular}
\end{equation}

The dual forms of the gauge invariant $L^{+^{\prime}-^{\prime}}$ and
$L^{\mu\nu}$ in the five shadows are (these are conserved if the backgrounds vanish)%

\begin{equation}%
\begin{tabular}
[c]{|c|c|c|}\hline
Shadow \#\  & $L^{+^{\prime}-^{\prime}}$ & $L^{\mu\nu}$\\\hline
1 & $x_{1}\cdot p_{1}$ & $x_{1}^{\mu}p_{1}^{\nu}-x_{1}^{\nu}p_{1}^{\mu}%
$\\\hline
2 & $\left(  x_{2}\cdot p_{2}\right)  a$ & $x_{2}^{\mu}p_{2}^{\nu}-x_{2}^{\nu
}p_{2}^{\mu}$\\\hline
3 & $2t_{3}h_{3}-\mathbf{x}_{3}\cdot\mathbf{p}_{3}$ & $%
\begin{array}
[c]{c}%
L^{0i}=s\mathbf{p}_{3}^{i}\\
L^{ij}=\mathbf{x}_{3}^{i}\mathbf{p}_{3}^{j}-\mathbf{x}_{3}^{j}\mathbf{p}%
_{3}^{i}%
\end{array}
$\\\hline
4 & $%
\begin{array}
[c]{c}%
-\left(  \mathbf{x}_{4}\cdot\mathbf{p}_{4}\right)  \cos u\\
+\frac{2m_{4}h_{4}\left\vert \mathbf{x}_{4}\right\vert +m_{4}e_{4}^{2}}%
{\sqrt{-2m_{4}h_{4}}}\sin u
\end{array}
$ & $%
\begin{array}
[c]{c}%
\overset{}{L^{0i}=\left[
\begin{array}
[c]{c}%
\left(
\begin{array}
[c]{c}%
\left\vert \mathbf{x}_{4}\right\vert \cos u\\
-\frac{\mathbf{x}_{4}\mathbf{\cdot p}_{4}}{\sqrt{-2m_{4}h_{4}}}\sin u
\end{array}
\right)  ~\mathbf{p}_{4}^{i}\\
+\frac{m_{4}e_{4}^{2}}{\left\vert \mathbf{x}_{4}\right\vert \sqrt{-2m_{4}%
h_{4}}}\sin u~\mathbf{x}_{4}^{i}%
\end{array}
\right]  }\\
\multicolumn{1}{l}{L^{ij}=\mathbf{x}_{4}^{i}\mathbf{p}_{4}^{j}-\mathbf{x}%
_{4}^{j}\mathbf{p}_{4}^{i}}%
\end{array}
$\\\hline
5 & $\left(  x_{5}\cdot p_{5}\right)  \sqrt{1+x_{5}^{2}V\left(  x_{5}%
^{2}\right)  \left(  x_{5}\cdot p_{5}\right)  ^{-2}}$ & $x_{5}^{\mu}p_{5}%
^{\nu}-x_{5}^{\nu}p_{5}^{\mu}$\\\hline
\end{tabular}
\label{L+-}%
\end{equation}

Similarly, the dual forms of the gauge invariant $L^{\pm^{\prime}\mu}$ in the
five shadows are (these are conserved if the backgrounds vanish)%
\begin{equation}%
\begin{tabular}
[c]{|c|c|c|}\hline
Shadow \#\  & $L^{+^{\prime}\mu}$ & $L^{-^{\prime}\mu}$\\\hline
1 & $p_{1}^{\mu}~$ & $\frac{1}{2}x_{1}^{2}p_{1}^{\mu}-\left(  x_{1}\cdot
p_{1}\right)  x_{1}^{\mu}~$\\\hline
2 & $\frac{1+a}{2a}p_{2}^{\mu}+\frac{m_{2}^{2}}{2\left(  x_{2}\cdot
p_{2}\right)  a}x_{2}^{\mu}$ & $\frac{a}{1+a}\left[  x_{2}^{2}p_{2}^{\mu
}-\left(  1+a\right)  \left(  x_{2}\cdot p_{2}\right)  x_{2}^{\mu}\right]
$\\\hline
3 & $%
\begin{array}
[c]{c}%
L^{+^{\prime}0}=-m_{3}s\\
L^{+^{\prime}i}=t_{3}\mathbf{p}_{3}^{i}-m_{3}\mathbf{x}_{3}^{i}%
\end{array}
$ & $%
\begin{array}
[c]{c}%
L^{-^{\prime}0}=-h_{3}s\\
L^{-^{\prime}i}=\frac{1}{m_{3}}\left(  \mathbf{x}_{3}\cdot\mathbf{p}_{3}%
-t_{3}h_{3}\right)  \mathbf{p}_{3}^{i}-h_{3}\mathbf{x}_{3}^{i}%
\end{array}
$\\\hline
$\text{4}$ & $\frac{1}{\sqrt{2}}\left(  L^{0^{\prime}\mu}+L^{1^{\prime}\mu
}\right)  ,$ see (\ref{shd4L+-m}) & $\frac{1}{\sqrt{2}}\left(  L^{0^{\prime
}\mu}-L^{1^{\prime}\mu}\right)  ,~$see (\ref{shd4L+-m}\\\hline
5 & $Ap_{5}^{\mu}-\frac{A}{x_{5}^{2}}\left(  x_{5}\cdot p_{5}-\phi\right)
x_{5}^{\mu}$ & $\frac{1}{2A}\left[  x_{5}^{2}p_{5}^{\mu}-\left(  x_{5}\cdot
p_{5}+\phi\right)  x_{5}^{\mu}\right]  $\\\hline
\end{tabular}
\label{L+m}%
\end{equation}

where for shadow \#4 it is more convenient to give $L^{0^{\prime}\mu
},L^{1^{\prime}\mu}$ instead of $L^{\pm^{\prime}\mu},$ as follows%
\begin{equation}%
\begin{tabular}
[c]{|c|}\hline
Shadow \#4,\ $L^{\pm^{\prime}\mu}\equiv\frac{1}{\sqrt{2}}\left(  L^{0^{\prime
}\mu}\pm L^{1^{\prime}\mu}\right)  ,$\\\hline%
\begin{tabular}
[c]{c}%
$L^{0^{\prime}0}=-\frac{m_{4}e_{4}^{2}}{\sqrt{-2m_{4}h_{4}}}$\\
$L^{1^{\prime}0}=\sin u~\mathbf{x}_{4}\mathbf{\cdot p}_{4}-\frac{\left\vert
\mathbf{x}_{4}\right\vert \cos u}{\sqrt{-2m_{4}h_{4}}}\left(  \frac{m_{4}%
e_{4}^{2}}{\left\vert \mathbf{x}_{4}\right\vert }+2m_{4}h_{4}\right)  $\\
$L^{0^{\prime}i}=\left(  \left\vert \mathbf{x}_{4}\right\vert \sin
u+\frac{\mathbf{x}_{4}\cdot\mathbf{p}_{4}}{\sqrt{-2m_{4}h_{4}}}\cos u\right)
\mathbf{p}_{4}^{i}-\frac{m_{4}e_{4}^{2}}{\left\vert \mathbf{x}_{4}\right\vert
\sqrt{-2m_{4}h_{4}}}\cos u~\mathbf{x}_{4}^{i}$\\
$L^{1^{\prime}i}=\frac{1}{\sqrt{-2m_{4}h_{4}}}\left[  -\left(  \mathbf{x}%
_{4}\mathbf{\cdot p}_{4}\right)  \mathbf{p}_{4}^{i}+\left(  \frac{m_{4}%
e_{4}^{2}}{\left\vert \mathbf{x}_{4}\right\vert }+2m_{4}h_{4}\right)
\mathbf{x}_{4}^{i}\right]  $%
\end{tabular}
\\\hline
\end{tabular}
\ \ \ \label{shd4L+-m}%
\end{equation}

It is interesting to point out that, in shadow \#4, the last listed
$L^{1^{\prime}i}$ is the famous Runge-Lenz vector, which is recognized as
follows. After the $\tau$-reparametrization gauge is chosen, $t_{4}\left(
\tau\right)  =\tau$ and the constraint of Eq.(\ref{gL4}) is solved when
backgrounds are absent as in Eq.(\ref{g4L}), yielding $h_{4}=\mathbf{p}%
_{4}^{2}/2m_{4}-e_{4}^{2}/\left\vert \mathbf{x}_{4}\right\vert ,$ the
$L^{1^{\prime}i}$ takes the familiar form proportional to the Runge-Lenz
vector, $L^{1^{\prime}i}=[-\left(  \mathbf{x}_{4}\mathbf{\cdot p}_{4}\right)
\mathbf{p}_{4}^{i}+\mathbf{p}_{4}^{2}\mathbf{x}_{4}^{i}-\frac{m_{4}e_{4}^{2}%
}{\left\vert \mathbf{x}_{4}\right\vert }\mathbf{x}_{4}^{i}]/\sqrt{-2m_{4}%
h_{4}}.$ This conserved vector in the H-atom or in a planetary system is
responsible for explaining the seemingly accidental systematic degeneracies of
the H-atom levels, or why planetary ellipses do not precess. In the 2T-physics
approach the explanation is because the dimension labelled by $1^{\prime}$ is
an extra hidden space dimension that, in the bulk is at the same footing as
the other usual 3 space dimensions. There is a natural rotation symmetry
SO$\left(  4\right)  $ which is part of the hidden SO$\left(  4,2\right)  $
symmetry in the H-atom shadow. The conservation of angular momentum in all 4
dimensions, not only in the first 3 dimensions, is the real explanation of the
interesting observations in the H-atom or in planetary systems. Another
familiar case of hidden symmetry is the more familiar conformal symmetry
SO$\left(  4,2\right)  $ of the conformal shadow. These are examples of the
more general hidden symmetries and conservation rules that we are advocating
in this section. For more examples of previously unknown hidden SO$\left(
d,2\right)  $ symmetries see \cite{HatomETC}\cite{ADSetc}.

Using the duality invariants $L^{MN}$ (whether the theory has background
fields or not) we make an infinite number of predictions that may be checked
both experimentally and theoretically by comparing any function of the
$L^{MN}$ in various shadows. Namely, at the classical level we predict%
\[
f\left(  L_{shadow~i}^{MN}\right)  =f\left(  L_{shadow~j}^{MN}\right)  ,\text{
any function }f,\text{ any dual pair }\left(  i\leftrightarrow j\right)  .
\]
The functions $f$ need not be SO$\left(  d,2\right)  $ invariants. For
example, we may take any function of just $L^{+^{\prime}-^{\prime}}$ and use
the corresponding expressions that are listed in the table above to make
predictions that follow from our dualities. Some such functions are the
Casimir invariants of SO$\left(  d,2\right)  ,$ for example the quadratic
Casimir is $C_{2}=\frac{1}{2}L^{MN}L_{MN}.$ At the classical level we find
$C_{2}=X^{2}P^{2}-\left(  X\cdot P\right)  ^{2}=0,$ since $X^{2}=0$ and
$X\cdot P=0$ is satisfied in every shadow \textit{for any set of background
fields} (namely without constraining $P^{2},$ leaving it off-shell). The same
vanishing result, with off-shell $P^{2},$ is found for all higher Casimirs at
the classical level, $C_{n}=\frac{1}{n!}Tr\left(  \left(  iL\right)
^{n}\right)  =0,$ $n=2,4,6,\cdots,$ where $L_{~N}^{M}$ is treated like a
matrix to evaluate the trace.

However, at the quantum level there are quantum ordering issues that must be
resolved in order to satisfy hermiticity of the $L^{MN}$ and the SO$\left(
d,2\right)  $ Lie algebra by using the quantum commutators for the operators
$\left(  X^{M},P^{M}\right)  $. Hermiticity implies that at the quantum level
we deal with unitary representations of SO$\left(  d,2\right)  .$ These
requirements lead to non-zero but definite eigenvalues for all the Casimir
operators in the physical subspace. The gauge invariant physical quantum
states are those that satisfy the vanishing of the Sp$\left(  2,R\right)  $
generators, $X^{2}|phys\rangle=0$ and $\left(  X\cdot P+P\cdot X\right)
|phys\rangle=0,$ in SO$\left(  d,2\right)  $ covariant quantization. The third
Sp$\left(  2,R\right)  $ generator $Q_{22}=\left(  P^{2}+\cdots\right)  $ is
to be imposed as well, but since the background fields $``\cdots
\textquotedblright$ are not yet specified, we consider $P^{2}$ to be
off-shell, namely so far unconstrained. This definition of physical states is
compatible with the duality transformations which do not alter the $Q_{22}$
constraint. In the physical sector as defined, we obtain a definite numerical
eigenvalue for the SO$\left(  d,2\right)  $ quadratic Casimir operator
\[
C_{2}|phys\rangle=\left(  1-\frac{d^{2}}{4}\right)  |phys\rangle
,~~P^{2}~\text{off-shell}.
\]
To see how this result is obtained we construct the Hermitian quantum
quadratic Casimir operator and re-order operator factors as follows%
\[
C_{2}=\frac{1}{2}L^{MN}L_{MN}=\left[
\begin{array}
[c]{c}%
P^{2}X^{2}+i\left(  X\cdot P+P\cdot X\right) \\
-\frac{1}{4}\left(  X\cdot P+P\cdot X\right)  ^{2}+(1-d^{2}/4)
\end{array}
\right]  ,
\]
where $X^{2}$ has been pulled to the right and $X\cdot P$ has been written in
Hermitian form. Applying this on physical states, we see that all operator
parts vanish, leaving behind a constant eigenvalue. Note that no constraint
has been imposed on $P^{2},$ hence the result $C_{2}\rightarrow\left(
1-\frac{d^{2}}{4}\right)  $ for physical states works for any set of
background fields. Similarly, for physical states we get non-zero
\textit{numerical eigenvalues} for all higher Casimirs $C_{n}$, for any set of
background fields, since $P^{2}$ is off-shell.

Due to the gauge invariance of the $L^{MN},$ the quantum theory in each shadow
must agree with the SO$\left(  d,2\right)  $ covariant quantization just
described. This requires that in each shadow the quantum ordering of the
$L^{MN}$ must be performed so that the same gauge invariant physical result is
obtained for the Casimir eigenvalues independent of the shadow and independent
of the background fields. Examples of how this quantum ordering is done in a
few shadows were given in \cite{HatomETC}\cite{ADSetc}. The numerical
eigenvalues of the Casimirs obtained in covariant quantization already
identify the specific unitary representation of SO$\left(  d,2\right)  ,$
which turns out to be the unitary \textit{singleton representation} of
$SO\left(  d,2\right)  .$ This result was known before in the absence of
background fields \cite{HatomETC}\cite{ADSetc}, and now we have established it
for \textit{any set of background fields and any shadow} since we have shown
it holds for $P^{2}$ off-shell.

We see now that the duality invariants must also hold at the quantum level in
every shadow, namely, once the $L^{MN}$ listed above are quantum ordered
properly in two dual shadows $\left(  i\leftrightarrow j\right)  $, they are
equal to each other as operators acting on a complete set of states in the
physical Hilbert space%
\[
L_{\text{quantum}}^{MN}\left(  x_{i},p_{i}\right)  =L_{\text{quantum}}%
^{MN}\left(  x_{j},p_{j}\right)  .
\]
A subset of these identities is the numerical values of the SO$\left(
d,2\right)  $ Casimirs operators being the same in every shadow, which is
already guaranteed by the correct quantum ordering. But, well beyond this, all
matrix elements between any set of quantum states for any function $f\left(
L_{\text{quantum}}^{MN}\right)  $ must also yield identical results for either
the left or the right side of this operator equation for every set of dual
shadows $\left(  i\leftrightarrow j\right)  $. This is a huge set of
\textit{quantum relations} between 1T-physics systems that can be tested as
\textit{predictions of our dualities derived from 2T-physics.}

\section{Solving problems using dualities \label{solving5}}

To illustrate the usefulness of our dualities we will solve the classical
equations of motion of the constrained system in shadow \#5 at zero background
fields. The equations of motion and constraint are given by the 1T Lagrangian%
\begin{equation}
L_{5}=\dot{x}_{5}^{\mu}p_{5\mu}-\frac{1}{2}A_{5}^{22}\left(  p_{5}%
^{2}+V\left(  x_{5}^{2}\right)  \right)  . \label{L5again}%
\end{equation}
As far as we know, the solution to the classical equations of motion of this
constrained system is not available in the literature for general $V\left(
x^{2}\right)  ,$ or even for the specialized case $V\left(  x^{2}\right)
=c\left(  x^{2}\right)  ^{b},$ except for $b=0$ (massive particle) or $b=1$
(constrained relativistic harmonic oscillator \cite{RelHarmOsc}). Furthermore,
attempting to solve it with standard methods, such as choosing a gauge, and
solving the constraint, leads to a time dependent potential that is difficult
or impossible to solve in closed form. However, by using our dualities, we
obtain the desired analytic solutions easily as follows.

The equations that determine $\left(  x_{5}^{\mu}\left(  \tau\right)
,p_{5\mu}\left(  \tau\right)  \right)  $ are
\begin{equation}
\dot{x}_{5}^{\mu}=A_{5}^{22}p_{5}^{\mu},\;\;\dot{p}_{5}^{\mu}=-A_{5}%
^{22}V^{\prime}\left(  x^{2}\right)  x_{5}^{\mu},\;\;p_{5}^{2}+V\left(
x_{5}^{2}\right)  =0. \label{eoms5}%
\end{equation}
We can make a gauge choice for $\tau$-reparametrizations by making some
convenient choice for $A_{5}^{22}\left(  \tau\right)  $ as an explicit
function of $\tau$. The solution we display below corresponds to an insightful
gauge choice for the gauge field $A_{5}^{22}\left(  \tau\right)  $ that yields
the analytic solution for any $V\left(  x^{2}\right)  .$ It would be
impossible to foresee such a gauge choice without our dualities. We proceed as follows.

First we transform the equations derived from (\ref{L5again}) to shadow \#1,
where the equations of motion and constraints are easily solved for
$x_{1}^{\mu}\left(  \tau\right)  ,p_{1\mu}\left(  \tau\right)  $ in the gauge
$A_{1}^{22}\left(  \tau\right)  =1$ as follows
\begin{equation}
x_{1}^{\mu}\left(  \tau\right)  =q_{1}^{\mu}+\tau p_{1}^{\mu},\text{ with
constant }p_{1}^{\mu},\text{ and constraint }p_{1}^{2}=0. \label{massless}%
\end{equation}
Then transforming this solution in shadow \#1 back to shadow \#5 we obtain the
desired analytic solution. Thus, we use the duality $\left(  5\leftarrow
1\right)  $ given in Eq.(\ref{1to5}) to write $\left(  x_{5}^{\mu},p_{5\mu
}\right)  $ in terms of $\left(  x_{1}^{\mu},p_{1\mu}\right)  $ and insert the
solution (\ref{massless}) to obtain the time dependence of the classical
trajectories $x_{5}^{\mu}\left(  \tau\right)  $ and $p_{5\mu}\left(
\tau\right)  $ that solve the equations of motion as well as the constraint,
$p_{5}^{2}+V\left(  x_{5}^{2}\right)  =0,$ derived from the Lagrangian $L_{5}$.

To be completely explicit, we specialize to $V\left(  x^{2}\right)  =c\left(
x^{2}\right)  ^{b}$ and use the duality in Eq.(\ref{1to5bc}) in which we
insert the explicit time dependence for $x_{1}^{\mu}\left(  \tau\right)
,p_{1\mu}\left(  \tau\right)  $ given in (\ref{massless}). To make all $\tau$
dependence evident, we note that $p_{1}^{\mu}$ is a constant that also
satisfies $p_{1}^{2}=0,$ while the other dot products have the following
explicit $\tau$ dependence
\begin{equation}
x_{1}^{2}\left(  \tau\right)  =q_{1}^{2}+2\left(  q_{1}\cdot p_{1}\right)
\tau,\;p_{1}\cdot x_{1}\left(  \tau\right)  =q_{1}\cdot p_{1}.
\end{equation}
The $F\left(  \phi\right)  $ that appears in Eq.(\ref{1to5bc}) is evaluated as
$F\left(  q_{1}\cdot p_{1}\right)  ,$ so it is another $\tau$-independent
constant that we will denote simply as a constant $F.$ \ Then the explicit
$\tau$ dependence of the solution $\left(  x_{5}^{\mu}\left(  \tau\right)
,p_{5\mu}\left(  \tau\right)  \right)  $ follows from Eq.(\ref{1to5bc}). After
some simplifications it takes the form%

\begin{equation}
x_{5}^{\mu}\left(  \tau\right)  =\left(  \frac{4\left(  q_{1}\cdot
p_{1}\right)  ^{2}}{cF^{2+2b}}\right)  ^{\frac{1}{2+2b}}\frac{q_{1}^{\mu}+\tau
p_{1}^{\mu}}{\left(  F^{-2-2b}+\left(  q_{1}^{2}+2\tau q_{1}\cdot
p_{1}\right)  ^{1+b}\right)  ^{\frac{1}{1+b}}}, \label{x5tau}%
\end{equation}
and%
\begin{equation}
p_{5}^{\mu}\left(  \tau\right)  =\left(  \frac{cF^{2+2b}}{4\left(  q_{1}\cdot
p_{1}\right)  ^{2}}\right)  ^{\frac{1}{2+2b}}\frac{p_{1}^{\mu}+\left[
q_{1}^{2}p_{1}^{\mu}-2\left(  q_{1}\cdot p_{1}\right)  q_{1}^{\mu}\right]
\left(  q_{1}^{2}+2\tau q_{1}\cdot p_{1}\right)  ^{b}}{\left(  F^{-2-2b}%
+\left(  q_{1}^{2}+2\tau q_{1}\cdot p_{1}\right)  ^{1+b}\right)  ^{\frac
{b}{1+b}}} \label{p5tau}%
\end{equation}

The $\tau$ in these expressions is the $\tau$ parameter conveniently gauge
fixed for shadow \#1 which \textit{\`{a} priori} would not occur naturally as
a gauge choice for shadow \#5$,$ although these are related to each other by
$\tau$ reparametrizations. The $\tau$-gauge in each shadow amounts to making a
choice for $A^{22}\left(  \tau\right)  .$ The gauge choice, $A_{1}^{22}\left(
\tau\right)  =1,$ was already made in shadow \#1 when writing the solution for
$\left(  x_{1}^{\mu}\left(  \tau\right)  ,p_{1\mu}\left(  \tau\right)
\right)  $ in the form (\ref{massless}). Hence, using the solution as it
stands, without further reparametrizing $\tau,$ amounts to making a definite
choice for $A_{5}^{22}\left(  \tau\right)  $ which is given by the Sp$\left(
2,R\right)  $ transformation in Eq.(\ref{A22Transf}) with $\beta=0$.
Therefore, we must take, $A_{5}^{22}\left(  \tau\right)  =\alpha^{2}\left(
\tau\right)  A_{1}^{22}\left(  \tau\right)  =\left(  A\left(  x_{1}\left(
\tau\right)  ,p_{1}\left(  \tau\right)  \right)  \right)  ^{2}\times1,$ where
$A\left(  x_{1},p_{1}\right)  $ is given in Eq.(\ref{Aphi51}),
\begin{equation}
A_{5}^{22}\left(  \tau\right)  =F^{-2b}\left[  \frac{4\left(  q_{1}\cdot
p_{1}\right)  ^{2}}{c\left(  F^{-2-2b}+\left(  q_{1}^{2}+2\tau q_{1}\cdot
p_{1}\right)  ^{1+b}\right)  ^{2}}\right]  ^{\frac{1}{1+b}}. \label{A5tau}%
\end{equation}
So, the solution for $\left(  x_{5}^{\mu}\left(  \tau\right)  ,p_{5\mu}\left(
\tau\right)  \right)  $ given in (\ref{x5tau},\ref{p5tau},\ref{A5tau}) is
expressed in terms of this choice of $\tau$-gauge applied to Eqs.(\ref{eoms5})
with $V\left(  x^{2}\right)  =c\left(  x^{2}\right)  ^{b}.$ This is a highly
non-trivial gauge choice for $A_{5}^{22}\left(  \tau\right)  $ that would be
hard to imagine without the guidance of the duality transformation. With this
understanding of the $\tau$-gauge, one may now check explicitly that the
equations of motion and the constraints in Eq.(\ref{eoms5}) are indeed
completely solved by the $\left(  x_{5}^{\mu}\left(  \tau\right)  ,p_{5\mu
}\left(  \tau\right)  ,A_{5}^{22}\left(  \tau\right)  \right)  $ given above.

Note that this solution is defined up to $\pm$ signs in different regions of
$\tau$ since it contains branch cuts in the complex $\tau$ plane (see below
for an example). This means that as $\tau$ changes, the corresponding
expressions must be continued across the branch cuts in order to get
continuously all the patches of the solution. An example of this for the case
of $b=1$ is a square-root branch cut as discussed below.

As a check, we may specialize to two cases, namely $b=0,1,$ for which we do
have a direct means of obtaining analytic solutions without using dualities,
that we may compare to the general case given above

\begin{itemize}
\item When $b=0,$ or $V\left(  x_{5}^{2}\right)  =c$, the Lagrangian $L_{5}$
reduces to the free massive relativistic particle with mass $c\equiv m^{2}$,
satisfying the equations of motion, $\dot{x}_{5}^{\mu}=A_{5}^{22}p_{5}^{\mu
},~\dot{p}_{5}^{\mu}=0,~p_{5}^{2}+c=0,$ for which a direct solution is
obtained as%
\begin{equation}
b=0:\;\left\{
\begin{array}
[c]{l}%
x_{5}^{\mu}\left(  \tau\right)  =q_{5}^{\mu}+p_{5}^{\mu}\int^{\tau}A_{5}%
^{22}\left(  \tau^{\prime}\right)  d\tau^{\prime},\\
p_{5}^{\mu}\left(  \tau\right)  =p_{5}^{\mu},\;\\
p_{5}^{2}+c=0,\text{ with }q_{5}^{\mu},p_{5}^{\mu}\text{ constants}.\text{ ~}%
\end{array}
\right.  \;~ \label{solWithAnyA}%
\end{equation}
The solution in Eqs.(\ref{x5tau},\ref{p5tau},\ref{A5tau}) is indeed of this
form when $b=0$. This is verified by noting that in the gauge (\ref{A5tau}) we
have
\begin{equation}
b=0:\;\int^{\tau}A_{5}^{22}\left(  \tau^{\prime}\right)  d\tau^{\prime}%
=\frac{-2\left(  q_{1}\cdot p_{1}\right)  F^{-2}}{c\left[  F^{-2}+\left(
q_{1}+p_{1}\tau\right)  ^{2}\right]  },
\end{equation}
and that the constrained constants $p_{5}^{\mu}$ are parametrized in terms of
the constants $q_{1}^{\mu},p_{1}^{\mu}$ with $p_{1}^{2}=0.$ The choice of
$A_{5}^{22}\left(  \tau\right)  $ in Eq.(\ref{A5tau}) as a function of $\tau$
to express the solution (\ref{x5tau},\ref{p5tau}) is clearly $\tau$-gauge
dependent, but this does not affect the gauge invariant physics. To see this
in the case $b=0$, we can write the solution \textit{for any} $A_{5}%
^{22}\left(  \tau\right)  $ given in (\ref{solWithAnyA}) in terms of the gauge
invariant variables $\left(  x_{5}^{0},\mathbf{x}_{5}^{i}\right)  $ and
$\left(  p_{5}^{0},\mathbf{p}_{5}^{i}\right)  $ as follows. First we write,
$x_{5}^{0}\left(  \tau\right)  =q_{5}^{0}+p_{5}^{0}\int^{\tau}A_{5}%
^{22}\left(  \tau^{\prime}\right)  d\tau^{\prime},$ from which we solve for
$\int^{\tau}A_{5}^{22}\left(  \tau^{\prime}\right)  d\tau^{\prime}=\left(
x_{5}^{0}\left(  \tau\right)  -q_{5}^{0}\right)  /p^{0},$ and replace it in
the solution for $\mathbf{x}_{5}^{i}\left(  \tau\right)  $ in
(\ref{solWithAnyA}) to obtain%
\[
\mathbf{x}_{5}^{i}=\mathbf{q}^{i}+\frac{\mathbf{p}_{5}^{i}}{p_{5}^{0}}\left(
x_{5}^{0}-q_{5}^{0}\right)  ,\text{ with }p_{5}^{0}=\sqrt{\mathbf{p}_{5}%
^{2}+c}.
\]
This expression, written in terms of $x_{5}^{0}$ is gauge independent because
it has the same form in terms of $\left(  x_{5}^{0}\left(  \tau\right)
,\mathbf{x}_{5}^{i}\left(  \tau\right)  \right)  $ for any choice of the
function $A_{5}^{22}\left(  \tau\right)  .$ Hence all the physical information
about the solution is encoded in any gauge choice for $A_{5}^{22}\left(
\tau\right)  $, including the choice of gauge in (\ref{A5tau}) that made the
solution (\ref{x5tau},\ref{p5tau}) possible for any $b$.

\item When $b=1,$ or $V\left(  x_{5}^{2}\right)  =cx_{5}^{2}$, the Lagrangian
$L_{5}$ reduces to the constrained relativistic harmonic oscillator with
frequency $\sqrt{c}$, satisfying the equations of motion, $\dot{x}_{5}^{\mu
}=A_{5}^{22}p_{5}^{\mu},~\dot{p}_{5}^{\mu}=-A_{5}^{22}cx_{5}^{\mu},~p_{5}%
^{2}+cx_{5}^{2}=0,$ for which a direct solution is obtained as%
\begin{equation}
b=1:\;\left\{
\begin{array}
[c]{l}%
x_{5}^{\mu}\left(  \tau\right)  =x_{0}^{\mu}\cos\left(  \sqrt{c}\int^{\tau
}A_{5}^{22}\left(  \tau^{\prime}\right)  d\tau^{\prime}\right)  +\frac
{1}{\sqrt{c}}p_{0}^{\mu}\sin\left(  \sqrt{c}\int^{\tau}A_{5}^{22}\left(
\tau^{\prime}\right)  d\tau^{\prime}\right)  ,\\
p_{5}^{\mu}\left(  \tau\right)  =p_{0}^{\mu}\cos\left(  \sqrt{c}\int^{\tau
}A_{5}^{22}\left(  \tau^{\prime}\right)  d\tau^{\prime}\right)  -\sqrt{c}%
x_{0}^{\mu}\sin\left(  \sqrt{c}\int^{\tau}A_{5}^{22}\left(  \tau^{\prime
}\right)  d\tau^{\prime}\right) \\
p_{0}^{2}+cx_{0}^{2}=0,\;\text{with }\left(  x_{0}^{\mu},p_{0}^{\mu}\right)
\text{ constants.}%
\end{array}
\right.  \;
\end{equation}
The solution in Eqs.(\ref{x5tau},\ref{p5tau},\ref{A5tau}) is indeed of this
form when $b=1.$ This is verified by noting that the constrained constants
$\left(  x_{0}^{\mu},p_{0}^{\mu}\right)  $ are parametrized in terms of the
constants $\left(  q_{1}^{\mu},p_{1}^{\mu}\right)  $ with $p_{1}^{2}=0,$ and
that the $\tau$ dependence in the gauge (\ref{A5tau}) becomes
\[
b=1:\text{\ }\sqrt{c}\int^{\tau}A_{5}^{22}\left(  \tau^{\prime}\right)
d\tau^{\prime}=\arctan\left[  F^{2}q_{1}^{2}+2F^{2}\left(  q_{1}\cdot
p_{1}\right)  \tau\right]  .
\]
Then
\begin{equation}%
\begin{array}
[c]{c}%
\cos\left(  \sqrt{c}\int^{\tau}A_{5}^{22}\left(  \tau^{\prime}\right)
d\tau^{\prime}\right)  =\pm\left(  1+F^{4}\left(  q_{1}^{2}+2\tau q_{1}\cdot
p_{1}\right)  ^{2}\right)  ^{-1/2}\\
\sin\left(  \sqrt{c}\int^{\tau}A_{5}^{22}\left(  \tau^{\prime}\right)
d\tau^{\prime}\right)  =F^{2}\left(  q_{1}^{2}+2\tau q_{1}\cdot p_{1}\right)
\left(  1+F^{4}\left(  q_{1}^{2}+2\tau q_{1}\cdot p_{1}\right)  ^{2}\right)
^{-1/2}%
\end{array}
\end{equation}
reproduces the expressions in (\ref{x5tau},\ref{p5tau}) with $b=1.$ Note that
the $\pm$ in the cosine expression is related to recovering all the patches of
the the solution; this set of signs corresponds to the continuation of the
expression across the square-root cut in the complex $\tau$ plane, as
mentioned above more generally for any $b$. Again we may argue that the gauge
choice for $A_{5}^{22}\left(  \tau\right)  $ is immaterial because the same
gauge invariant physics is reproduced when written in terms of a gauge
invariant choice of the time coordinate. For $b=1,$ a natural gauge invariant
time coordinate $t\left(  \tau\right)  $ was given in Eq.(\ref{NRharmonicOsc})
which amounts to $t\left(  \tau\right)  =\arctan\left(  \sqrt{c}x^{0}\left(
\tau\right)  /p^{0}\left(  \tau\right)  \right)  $; in that case the solution
(\ref{x5tau},\ref{p5tau}) is seen to capture all the motions of the
non-relativistic harmonic oscillator when rewritten in terms of $t$. To obtain
the gauge invariant motion it is not necessary to explicitly solve for
$t\left(  \tau\right)  $ or for the inverse $\tau\left(  t\right)  ;$ instead
one may simply do a parametric plot of $\left(  \mathbf{x}_{5}\left(
\tau\right)  ,t\left(  \tau\right)  \right)  $ and $\left(  \mathbf{p}%
_{5}\left(  \tau\right)  ,t\left(  \tau\right)  \right)  $ which is gauge
invariant. Hence, again, all the physical information about the solution is
encoded in any gauge choice for $A_{5}^{22}\left(  \tau\right)  $, including
the choice of gauge in (\ref{A5tau}) that made the solution (\ref{x5tau}%
,\ref{p5tau}) possible for any $b$.
\end{itemize}

We do not know of another approach to solve the equations (\ref{eoms5})
analytically for $V\left(  x^{2}\right)  =c\left(  x^{2}\right)  ^{b}$ with
any $b,c$, except for the duality methods discussed in this section. Even more
impressive is that we have obtained the analytic expressions for any $V\left(
x^{2}\right)  .$ This demonstrates the utility and power of our system of dualities.

\section{Outlook \label{conclude}}

The study of dualities in 1T-physics in $d$-dimensions is equivalent to
probing the properties of the underlying $d+2$ dimensions including the extra
$1+1$ dimensions. In this sense the extra dimensions are not hidden and can be
investigated both experimentally and theoretically via the dualities directly
in $3+1$ dimensions, with the guidance of 2T-physics. It is apparent that the
discussion given in this paper is just the tip of an iceberg of dualities that
will take a long time to mine.

We have seen that the underlying meaning of the dualities $\left(
i\leftrightarrow j\right)  ,$ which were realized here as canonical
transformations, is really gauge transformations from one fixed gauge to
another fixed gauge for the gauge group Sp$\left(  2,R\right)  $ acting in
phase space in $d+2$ dimensions. There definitely are gauge invariants
(equivalently duality invariants). Specifically, any function of the $L^{MN}$
is an invariant, as explained in section (\ref{invariants}), but time,
Hamiltonian or more generally space-time, as interpreted by observers in any
1T shadow, are not among the gauge invariants. This is why 1T-physics is
different in different shadows, but yet there are deep relations and
corresponding physical predictions among observers because of the underlying
gauge symmetry. This concept of gauge symmetry in phase space is more general
than the more familiar gauge symmetries, such as Yang-Mills or general
coordinate transformations, that act locally only in space-time, rather than
in phase space.

The reader may better grasp the significance of these statements by
considering the concept of observers outlined in the introduction, i.e. that a
given phase space $\left(  x_{i}^{\mu},p_{i\mu}\right)  $ in shadow $i$
defines the frame of an observer that rides along with a particle on a
worldline $\left(  x_{i}^{\mu}\left(  \tau\right)  ,p_{i\mu}\left(
\tau\right)  \right)  $ which is embedded in the bulk in $d+2$ dimensions.
Such an observer, which may be said to live on a \textquotedblleft screen
$i$\textquotedblright\ or \textquotedblleft boundary $i$\textquotedblright\ or
\textquotedblleft shadow $i$\textquotedblright\ in $1+1$ fewer dimensions,
interprets all the \textit{gauge invariant phenomena} occurring in the bulk in
$d+2$ dimensions from his/her perspective $i,$ which is totally different than
perspective $j$ defined by another observer riding along worldline $\left(
x_{j}^{\mu}\left(  \tau\right)  ,p_{j\mu}\left(  \tau\right)  \right)  $ that
defines shadow $j.$ For the 5 different shadows discussed in this paper we
have seen that these 5 perspectives are indeed very different forms of
1T-physics. Nevertheless each shadow, being just a gauge choice, captures all
the gauge invariant information in the bulk. Therefore each shadow is
holographic and hence must be dual to all other shadows. Indeed, we have shown
that all shadows are closely related to one another by explicit dualities, and
even more strongly, that all of the different 1T-physics equations in various
shadows are united and captured in a unified form of gauge invariant equations
for the phase space $\left(  X^{M},P_{M}\right)  $ in the bulk in $d+2$
dimensions, namely just $X^{2}=0,$ $X\cdot P=0,$ and $P^{2}+\cdots=0$.

These ideas resonate with Einstein's concepts of observers in his thought
experiments in various frames in special or general relativity in 1T-physics.
In our case, the analogous infinite set of frames are connected to each other
by phase space transformations. This is a much larger set as compared to the
set of frames connected to each other by only position space transformations.
Hence the unification of 1T observers is much larger in the framework of
2T-physics, while the corresponding unification of their diverse 1T equations
is a unique set of equations in $d+2$ dimensions, whose form is dictated by
gauge symmetry in phase space.

More generally, the dualities generated by 2T-physics go well beyond the realm
of canonical transformations in 1T phase space because they include additional
degrees of freedom besides $\left(  x,p\right)  $. We remind the reader that
the 2T formalism includes also the degrees of freedom of spinning systems
\cite{2Tspin1}-\cite{2Tspin4}, supersymmetric systems \cite{2TsusyParticle},
twistors \cite{twistors}, fields in local field theory \cite{2Tsm}%
\cite{2Tgravity}\cite{2Tsusy}\cite{2T-SYM-12D}\cite{ibsuper}, and fields in
phase space \cite{NC-U11}. Hence 2T-physics provides a new path to unification
of 1T systems that is not available among the familiar concepts in 1T-physics.

Extrapolating from a particle's phase space to the corresponding situation in
field theory, the shadow $i$ in field theory$,$ derived from the field theory
in the bulk in $d+2$ dimensions (such as the standard model in $4+2$
dimensions \cite{2Tsm}) describes all the physics as seen from the perspective
of observer $i,$ and similarly the shadow field theory $j$ describes all the
physical phenomena as seen from the perspective of observer $j.$ These are the
dual field theories that correspond to the duality $\left(  i\leftrightarrow
j\right)  $ in phase space. These dual field theories predict the
relationships between observers $i$ and $j$ and capture all the gauge
invariant phenomena that the observers could measure, so the duality between
1T field theories predicted by 2T field theory lead to much broader verifiable
tests of the entire approach described here. For some simple cases of $\left(
i\leftrightarrow j\right)  $ dualities discussed in the past (simpler than the
5 cases in this paper), examples of such dual field theories are developed in
\cite{dualitiesFields}. For the harder cases $\left(  i\leftrightarrow
j\right)  $ discussed in this paper it is also possible in principle to
construct the corresponding dual field theories. Our future goals include the
construction of dual versions of the standard model and their use as new tools
of investigation. The recent successful application in cosmology (involving
transformations between different fixed Weyl gauges to solve and interpret
cosmological equations) \cite{BSTconformalSM}\cite{BSThiggsCosmo}, is a simple
example of this idea involving \textquotedblleft Weyl
dualities\textquotedblright\ in $3+1$ dimensions which originated from
2T-physics gauge symmetries.

We have argued that phase space gauge symmetry in 2T-physics offers superior
unifying power than gauge symmetry in 1T-physics. Having seen that even in
simple classical mechanics systems there does exists a deeper unification, as
shown in this paper, it is natural to expect that the same must also be true
at the deepest level of physics principles. Hence 2T-physics is likely to show
the right path to the ultimate theory. Therefore, we posit that there is much
benefit in developing further this formalism and in studying its consequences,
such as the types of dualities discussed in this paper, and much more, in
order to better understand the meaning of space-time and true unification.
Along this path we should also benefit from new computational techniques in
1T-physics that emerge from 2T-physics. There is still much to be accomplished
in 2T-physics even in classical and quantum mechanics, not to mention field
theory, string theory and M theory.

\begin{acknowledgments}
IB and IJA were partially supported by the U.S. Department of Energy under
grant number DE-FG03-84ER40168. IJA was partially supported by CONICYT
(Comisi\'{o}n Nacional de Investigaci\'{o}n Cient\'{\i}fica y Tecnol\'{o}gica
- Chilean Government) and by the Fulbright commission through a joint fellowship.
\end{acknowledgments}


\begin{thebibliography}{99}                                                                                               %


\bibitem {BSTconformalSM}I. Bars, P. J. Steinhardt and N. Turok,
\textquotedblleft Local Conformal Symmetry in Physics and Cosmology
,\textquotedblright\ \ arXiv:1307.1848.

\bibitem {BSThiggsCosmo}I. Bars, P. J. Steinhardt and N. Turok,
\textquotedblleft Cyclic Cosmology, Conformal Symmetry and the Metastability
of the Higgs,\textquotedblright\ Phys. Lett. \textbf{B726} (2013) 50 [arXiv:1307.8106].

\bibitem {BSTsailing}I. Bars, P. J. Steinhardt and N. Turok, \textquotedblleft
Sailing through the big crunch-big bang transition\textquotedblright, arXiv:1312.0739.

\bibitem {2Tgravity}I. Bars, \textquotedblleft Gravity in
2T-Physics\textquotedblright, Phys. Rev. \textbf{D77 }(2008) 125027
[arXiv:0804.1585]; I. Bars, S. H. Chen \textquotedblleft Geometry and Symmetry
Structures in 2T Gravity\textquotedblright, Phys. Rev. \textbf{D79 }(2009)
085021 [arXiv:0811.2510v2].

\bibitem {2T-BDA}I. Bars, C. Deliduman, O. Andreev, \textquotedblleft Gauged
duality, conformal symmetry and space-time with two times,\textquotedblright%
\ Phys.Rev. \textbf{D58} (1998) 066004 [hep-th/9803188]

\bibitem {BarsYank-Udual}I. Bars and S. Yankielowicz, \textquotedblleft U
duality multiplets and nonperturbative superstring states\textquotedblright,
Phys.Rev. \textbf{D53} (1996) 4489 [hep-th/9511098].

\bibitem {IB-2TinMtheory}I. Bars, \textquotedblleft Supersymmetry, p-brane
duality and hidden space-time dimensions,\textquotedblright\ Phys.Rev.
\textbf{D54} (1996) 5203 [hep-th/9604139].

\bibitem {S-theory}I. Bars. \textquotedblleft S theory,\textquotedblright%
\ Phys. Rev. \textbf{D55} (1997) 2373 [hep-th/9607112]; \textit{ibid.}
\textquotedblleft Algebraic structure of S theory,\textquotedblright\ Lectures
at Strings-96 and Sakharov Conference (1996) 355-363, hep-th/9608061.

\bibitem {2Tspin1}I. Bars and C. Deliduman, \textquotedblleft Gauge symmetry
in phase space with spin: A Basis for conformal symmetry and duality among
many interaction,\textquotedblright\ Phys. Rev. \textbf{D58} (1998) 106004 [hep-th/9806085].

\bibitem {2Tspin2Fields}I. Bars, \textquotedblleft Two time physics in field
theory,\textquotedblright\ Phys. Rev. \textbf{D62} (2000) 046007 [hep-th/0003100].

\bibitem {2Tspin3GravGauge}I. Bars, \textquotedblleft Two time physics with
gravitational and gauge field backgrounds,\textquotedblright\ Phys. Rev.
\textbf{D62} (2000) 085015 [hep-th/0002140].

\bibitem {2Tspin4}I. Bars and B. Orcal, \textquotedblleft Generalized Twistor
Transform And Dualities, With A New Description of Particles With Spin, Beyond
Free and Massless,\textquotedblright\ Phys. Rev. D75 (2007) 104015 [arXiv:0704.0296].

\bibitem {2001HighSpin}I. Bars and C. Deliduman, \textquotedblleft High spin
gauge fields and two-time physics\textquotedblright, Phys. Rev. \textbf{D64}
(2001) 045004 [hep-th/0103042].

\bibitem {2TsusyParticle}I. Bars, \textquotedblleft2T physics formulation of
superconformal dynamics relating to twistors and
supertwistors,\textquotedblright\ Phys. Lett. \textbf{B483} (2000) 248 [hep-th/0004090]

\bibitem {twistors}I. Bars, \textquotedblleft Lectures on
twistors,\textquotedblright\ hep-th/0601091, I. Bars and M. Picon,
\textquotedblleft Single twistor description of massless, massive, AdS, and
other interacting particles,\textquotedblright\ Phys. Rev. \textbf{D73} (2006)
064002 [hep-th/0512091]; \textit{ibid.} \textquotedblleft Twistor transform in
d dimensions and a unifying role for twistors,\textquotedblright\ Phys. Rev.
\textbf{D73} (2006) 064033 [hep-th/0512348].

\bibitem {2TphaseSpace}I. Bars, \textquotedblleft Gauge Symmetry in Phase
Space, Consequences for Physics and Spacetime\textquotedblright, Int. J. Mod.
Phys. \textbf{A25} (2010) 5235 [arXiv:1004.0668].

\bibitem {Mtheory}I. Bars, C. Deliduman, and D. Minic, \textquotedblleft
Lifting M theory to two time physics,\textquotedblright\ Phys. Lett.
\textbf{B457} (1999) 275 [hep-th/9904063].

\bibitem {2Tsm}I. Bars, \textquotedblleft The Standard Model of Particles and
Forces in the Framework of 2T-physics\textquotedblright, Phys. Rev.
\textbf{D74} (2006) 085019 [hep-th/0606045]; I. Bars \textquotedblleft The
Standard model as a 2T- physics theory \textquotedblright, AIP Conf. Proc.
\textbf{903} (2007) 550 [hep- th/0610187].

\bibitem {2Tsusy}I. Bars and Y-C Kuo, \textquotedblleft Field Theory in
2T-physics with $\mathcal{N}$=1 Supersymmetry\textquotedblright, Phys. Rev.
Lett. \textbf{99} (2007) 041801 [hep-th/0702089]; \textquotedblleft
Supersymmetric Field Theory in 2T-physics \textquotedblright, Phys. Rev.
\textbf{D76} (2007) 105028 [ hep-th/0703002]; \textquotedblleft$\mathcal{N}%
$=2,4 Supersymmetric Gauge Field Theory in 2T-physics\textquotedblright, Phys.
Rev. \textbf{D79} (2009) 025001 [arXiv:0808.0537];

\bibitem {2T-SYM-12D}I. Bars and Y-C Kuo, \textquotedblleft Super Yang-Mills
theory in 10+2 dimensions, the 2T-physics Source for $\mathcal{N}$=4 SYM and
M(atrix) Theory\textquotedblright, arXiv:1008.4761.

\bibitem {ibsuper}I. Bars, \textquotedblleft Constraints on Interacting
Scalars in 2T Field Theory and No Scale Models in 1T Field
Theory\textquotedblright, Phys. Rev. \textbf{D82} (2010) 125025 [arXiv:1008.1540].

\bibitem {stringsBranes}I. Bars, C. Deliduman, and D. Minic, \textquotedblleft
Strings, branes and two time physics,\textquotedblright\ Phys. Lett.
\textbf{B466} (1999) 135 [hep-th/9906223].

\bibitem {stringTwistor}I. Bars, \textquotedblleft Twistor superstring in
2T-physics,\textquotedblright\ Phys. Rev. \textbf{D70} (2004) 104022
[hep-th/0407239]; \textit{ibid.} \textquotedblleft Twistors and
2T-physics,\textquotedblright\ AIP Conf. Proc. \textbf{767} (2005) 3 [hep-th/0502065].

\bibitem {NC-U11}I. Bars, \textquotedblleft U*(1,1) noncommutative gauge
theory as the foundation of 2T-physics in field theory,\textquotedblright%
\ Phys. Rev. \textbf{D64} (2001) 126001 [hep-th/0106013].

\bibitem {inflationBC}I. Bars and S-H. Chen, \textquotedblleft The Big Bang
and Inflation United by an Analytic Solution \textquotedblright, Pays.\textit{
Rev. D}\textbf{83} 043522 (2011) [arXiv:1004.0752].

\bibitem {cyclicBCT}I.~Bars, S-H.~Chen and N.~Turok, \textquotedblleft
Geodesically Complete Analytic Solutions for a Cyclic
Universe\textquotedblright, \textit{Phys. Rev. }\textbf{D84} 083513 (2011) [arXiv:1105.3606].

\bibitem {cyclic-Bars}I.~Bars, \textquotedblleft Geodesically Complete
Universe\textquotedblright, in Proceedings of the DPF- 2011 Conference,
Providence, RI, August 8-13, 2011, arXiv:1109.5872.

\bibitem {BCST1}I.~Bars, S.-H.~Chen, P.~J.~Steinhardt and N.~Turok,
\textquotedblleft Antigravity and the Big Crunch/Big Bang
Transition\textquotedblright, Phys. Lett. \textbf{B715} (2012) 278-281 [arXiv:1112.2470].

\bibitem {BCST2}I.~Bars, S.-H.~Chen, P.~J.~Steinhardt and N.~Turok,
\textquotedblleft Complete Set of Homogeneous Isotropic Analytic Solutions in
Scalar-Tensor Cosmology with Radiation and Curvature\textquotedblright, Phys.
Rev. \textbf{D86} (2012) 083542 [ arXiv:1207.1940].

\bibitem {IB-completeJourneys}I.~Bars, \textquotedblleft Traversing
Cosmological Singularities, Complete Journeys Through Spacetime Including
Antigravity,\textquotedblright\ arXiv:1209.1068.

\bibitem {Dirac}P.A.M Dirac, Ann. Math. \textbf{37} (1936) 429.

\bibitem {kastrup}H. A. Kastrup, Phys. Rev. \textbf{150} (1966) 1183.

\bibitem {salam}G. Mack and A. Salam, Ann. Phys. \textbf{53} (1969) 174.

\bibitem {adler}S. Adler, Phys. Rev. \textbf{D6} (1972) 3445; ibid.
\textbf{D8} (1973) 2400.

\bibitem {ferrara}S. Ferrara, Nucl. Phys. \textbf{B77} (1974) 73.

\bibitem {fronsdal}F. Bayen, M. Flato, C. Fronsdal and A. Haidari, Phys. Rev.
\textbf{D32} (1985) 2673.

\bibitem {siegel}W. Siegel, Int. J. Mod. Phys. \textbf{A3} (1988) 2713; Int.
Jour. Mod. Phys. \textbf{A4} (1989) 2015.

\bibitem {marnelius}R. Marnelius, Phys. Rev. D20, 2091 (1979); R. Marnelius
and B. Nilsson, Phys. Rev. \textbf{D22} (1980) 830; P. Arvidsson and R.
Marnelius, \textquotedblleft Conformal theories including conformal gravity as
gauge theories on the hypercone\textquotedblright\ [hep-th/0612060].

\bibitem {vasiliev}C. R. Preitschopf and M. A. Vasiliev, Nucl. Phys.
\textbf{B549} (1999) 450, [hep-th/9812113].

\bibitem {vasiliev2}M. A. Vasiliev, JHEP \textbf{12} (2004) 046, [hep-th/0404124].

\bibitem {taronna1}E. Joung, L. Lopez, and M. Taronna, \textquotedblleft On
the cubic interactions of massive and partially-massless higher spins in
(A)dS\textquotedblright, \ JHEP \textbf{1207} (2012) 041, [arXiv:1203.6578];
M. Taronna, \textquotedblleft Higher-Spin Interactions: three-point functions
and beyond\textquotedblright, \ arXiv:1209.5755.

\bibitem {taronna2}E. Joung, M. Taronna, A. Waldron, \textquotedblleft A
Calculus for Higher Spin Interactions\textquotedblright, JHEP \textbf{1307}
(2013) 186 [arXiv:1305.5809].

\bibitem {rivelles}J.E. Frederico and V.O. Rivelles, \textquotedblleft The
Transition Amplitude for 2T Physics\textquotedblright, Phys. Rev. \textbf{D82}
(2010) 021701, [arXiv:1002.1263].

\bibitem {weinberg}S. Weinberg, \textquotedblleft Six-dimensional Methods for
Four-dimensional Conformal Field Theories\textquotedblright, Phys.Rev. D82
(2010) 045031, [arXiv:1006.3480]; Phys. Rev. \textbf{D86} (2012) 085013 , [arXiv:1209.4659].

\bibitem {rychkov}M. S. Costa, J. Penedones, D. Poland, S. Rychkov,
\textquotedblleft Spinning Conformal Correlators\textquotedblright, JHEP
\textbf{1111} (2011) 071 [arXiv:1107.3554]; JHEP \textbf{1111} (2011) 154 [arXiv:1109.6321].

\bibitem {waldron}R. Bonezzi, E. Latini, and A. Waldron, \textquotedblleft
Gravity, Two Times, Tractors, Weyl Invariance and Six Dimensional Quantum
Mechanics\textquotedblright, Phys.Rev. \textbf{D82} (2010) 064037, [arXiv:1007.1724].

\bibitem {waldron2}R. Bonezzi, O. Corradini , and A. Waldron,
\textquotedblleft Local Unit Invariance, Back-Reacting Tractors and the
Cosmological Constant Problem\textquotedblright, J.Phys.Conf.Ser. \textbf{343}
(2012) 012128, [ arXiv:1003.3855].

\bibitem {dualitiesFields}I. Bars, S-H Chen and G. Quelin, \textquotedblleft
Dual field theories in $(d-1)+1$ emergent spacetimes from a unifying field
theory in d+2 spacetime\textquotedblright, Phys. Rev. \textbf{D76} (2007)
065016 [arXiv:0705.2834]; I. Bars and G. Quelin, \textquotedblleft Dualities
among one-time field theories with spin, emerging from a unifying two- time
field theory\textquotedblright, Phys. Rev. \textbf{D77} (2008) 125019 [arXiv:0802.1947].

\bibitem {HatomETC}I. Bars \textquotedblleft Conformal symmetry and duality
between free particle, H - atom and harmonic oscillator,\textquotedblright%
\ Phys. Rev. \textbf{D58} (1998) 066006 [hep-th/9804028].

\bibitem {ADSetc}I. Bars, \textquotedblleft Hidden symmetries, AdS$_{D}\times
$S$^{n}$, and the lifting of one time physics to two time
physics,\textquotedblright\ Phys. Rev. \textbf{D59} (1999) 045019 [hep-th/9810025].

\bibitem {RelHarmOsc}I. Bars, \textquotedblleft Relativistic Harmonic
Oscillator Revisited,\textquotedblright\ Phys. Rev. \textbf{D79} (2009) 045009 [arXiv:0810.2075].
\end{thebibliography}
\end{document}